\documentclass[10pt,journal]{IEEEtran}
\usepackage{amsmath,amsfonts}
\usepackage{algorithmic}
\usepackage{algorithm}
\usepackage{array}
\usepackage[caption=false,font=normalsize,labelfont=sf,textfont=sf]{subfig}
\usepackage{textcomp}
\usepackage{stfloats}
\usepackage{url}
\usepackage{verbatim}
\usepackage{graphicx}
\usepackage{balance}
\usepackage{booktabs}
\usepackage{colortbl}

\ifCLASSOPTIONcompsoc
  \usepackage[nocompress]{cite}
\else
  \usepackage{cite}
\fi

\usepackage{listings}

\providecommand\wenhao[1]{\textcolor{blue}{\{\textbf{wenhao:} {\em#1}\}}}
\providecommand\link[1]{\textcolor{red}{\{\textbf{link:} {\em#1}\}}}

\providecommand\ignore[1]{{}}

\providecommand\hll[1]{#1}
\providecommand\hl[1]{#1}

\newcommand{\para}[1]{\vspace{1pt}\noindent\textbf{{#1. }}}
\makeatletter
\renewcommand{\@IEEEBIOskipN}{10pt}
\makeatother

\usepackage{hyperref}
\newcolumntype{C}[1]{>{\centering\arraybackslash}m{#1}}
\usepackage{xcolor}

\DeclareRobustCommand*{\authorrefmark}[1]{\raisebox{0pt}[0pt][0pt]{\textsuperscript{\footnotesize\ensuremath{\ifcase#1\or *\or \dagger\or \ddagger\or%
    \mathsection\or \mathparagraph\or \|\or **\or \dagger\dagger%
    \or \ddagger\ddagger \else\textsuperscript{\expandafter\romannumeral#1}\fi}}}}
\usepackage{bbding} 

\usepackage{threeparttable} 

\begin{document}

\title{The Early Bird Catches the Leak: Unveiling Timing Side Channels in LLM Serving Systems}



\author{Linke Song*, Zixuan Pang*, Wenhao Wang\textsuperscript{\Envelope}, Zihao Wang, XiaoFeng Wang~\IEEEmembership{Fellow,~IEEE,} \\ Hongbo Chen, Wei Song, Yier Jin, Dan Meng, and Rui Hou
\IEEEcompsocitemizethanks{
\IEEEcompsocthanksitem The two lead authors contribute equally to the work. 
\IEEEcompsocthanksitem Corresponding author: Wenhao Wang (\href{mailto:wangwenhao@iie.ac.cn}{wangwenhao@iie.ac.cn}).
\IEEEcompsocthanksitem L. Song, W. Wang, W. Song, D. Meng and R. Hou are with State Key Laboratory of Cyberspace Security Defense, Institute of Information Engineering, Chinese Academy of Sciences, and  School of Cyber Security, University of Chinese Academy of Sciences.
\IEEEcompsocthanksitem Z. Pang and Y. Jin are with University of Science and Technology of China.
\IEEEcompsocthanksitem Z. Wang and X. Wang are with Nanyang Technological University.
\IEEEcompsocthanksitem H. Chen is with Indiana University Bloomington.
\IEEEcompsocthanksitem The authors from Institute of Information Engineering were supported by the National Natural Science Foundation of China (Grant No. 62272452), the Strategic Priority Research Program of the Chinese Academy of Sciences (Grant No. XDB0690100) and the research grant from Huawei.
}}



\maketitle

\begin{abstract}
The wide deployment of Large Language Models (LLMs) has given rise to strong demands for optimizing their inference performance. Today's techniques serving this purpose primarily focus on reducing latency and improving throughput through algorithmic and hardware enhancements, while largely overlooking their privacy side effects, particularly in a multi-user environment. In our research, for the first time, we discovered a set of new timing side channels in LLM systems, arising from shared caches and GPU memory allocations, which can be exploited to infer both confidential system prompts and those issued by other users. These vulnerabilities echo security challenges observed in traditional computing systems, highlighting an urgent need to address potential information leakage in LLM serving infrastructures.
In this paper, we report novel attack strategies designed to exploit such timing side channels inherent in LLM deployments, specifically targeting the Key-Value (KV) cache and semantic cache widely used to enhance LLM inference performance. Our approach leverages timing measurements and classification models to detect cache hits, allowing an adversary to infer private prompts with high accuracy. We also propose a token-by-token search algorithm to efficiently recover shared prompt prefixes in the caches, showing the feasibility of stealing system prompts and those produced by peer users. Our experimental studies on black-box testing of popular online LLM services demonstrate that such privacy risks are completely realistic, with significant consequences. Our findings underscore the need for robust mitigation to protect LLM systems against such emerging threats.
\end{abstract}


\section{Introduction}
\IEEEPARstart{L}{arge} Language Models (LLMs) are widely used in applications such as chatbots~\cite{chatgpt}, search engines~\cite{perplexity}, and coding assistants~\cite{copilot}. However, LLM inference is resource-intensive, requiring substantial computational power and memory due to the model's vast parameters, numerous layers, and large context sizes.  
Improving LLM inference performance has thus become essential, leading to solutions such as weight quantization~\cite{yao2022zeroquant,wei2022outlier,xiao2023smoothquant}, model compression~\cite{wang2024model}, and algorithm optimization~\cite{dao2022flashattention,kwon2023efficient}
. While these approaches reduce latency and improve inference efficiency, their privacy implications remain less clear.


In this paper, we conduct the first security analysis of performance optimization techniques employed by modern LLM systems that serve multiple users or applications concurrently. Our research reveals significant information leaks arising from distinct side channels introduced by these techniques.  
Specifically, current LLM performance optimizations use shared caches to reduce computation and storage overhead during inference. However, memory sharing, cache contention and eviction and task scheduling among different users and applications can interfere with user requests, creating noticeable timing side channels. 
Exploiting these side channels can expose private prompts from other users or applications.

\IEEEpubidadjcol
\para{LLM cache channels} We examined various caches in LLM systems, which not only reduce the computational cost of LLM inference but also improve user experience by lowering service latency. We found that these caches can be misused to infer proprietary system prompts or sensitive prompts from peer users. These prompts may contain private user information and also hold commercial value, as they enable an LLM to carry out various downstream tasks without additional fine-tuning.  
We identified two primary cache channels:

    
    \noindent$\bullet$\textit{~Leakage from the KV cache.} 
    For each inference request, the LLM maintains an in-memory state called the KV cache, which is reused throughout the request's entire service time. Due to the causal attention mask in LLMs, each token's activations are influenced only by preceding tokens in the sequence. Thus, if multiple requests share a common prefix, the key and value embeddings for those prefix tokens are identical across sequences. To optimize the KV cache's memory usage, the system identifies matching prompt prefixes across multiple requests and shares their key and value embeddings in memory at runtime~\cite{kwon2023efficient,zheng2023efficiently}. This sharing occurs when prompts include a common prefix, which frequently happens with few-shot examples, chatbot system prompts, or prompt templates~\cite{langchain-prompttemplate}. For example, it has been noted that Claude's prompt caching feature can reduce costs by up to 90\% and decrease latency by up to 85\% for long prompts~\cite{claudepromptcaching}.

   \noindent$\bullet$\textit{~Leakage from the semantic cache.} The semantic cache boosts LLM performance by caching responses based on the semantic content of the requests. For example, for the prompts ``give me suggestions for a comedy movie'' and ``recommend a comedy movie'', the LLM system can detect their semantic similarity and return similar responses without querying the LLM backend. Experiments show that when GPTCache is integrated with OpenAI's service, response speed can be improved by a factor of 2 to 10 upon a cache hit~\cite{bang2023gptcache}.




\para{Challenges and solutions}
A straightforward way to exploit these vulnerable caches is to directly search the prompt space for one that triggers a cache hit. However, this method faces multiple hurdles. First, the time difference resulting from hitting a single cache block is often minimal and can blend with GPU system noise and fluctuations in voltage and power, making it difficult to detect and exploit. Second, the KV cache only works when prompts share a common prefix, limiting attack opportunities.  
Additionally, the vastness of the prompt space makes it infeasible to systematically test every potential prompt to find a cached one. Complicating matters further, the attacker's own requests might be cached during the process, introducing additional noise and potentially causing the victim's cached data to be evicted.

To address these challenges, we developed various attack strategies to exploit LLM side channels. Specifically, we use a threshold-based classification model to detect token-level KV cache hits based on offline timing measurements. We observe that online detection accuracy can be substantially improved with only a few repeated trials.  
To reduce the search space for the KV cache channel, we propose an incremental search algorithm that capitalizes on the requirement for prompts to share a common prefix, allowing us to recover the victim's prompt token by token.  
For the semantic cache channel, we design an algorithm to select the most representative prompts as the attacker's requests, given a targeted semantic focus.  
To minimize interference from the attacker's own requests, we introduce a mechanism to clear cached data via batches of irrelevant requests. Our method also ensures the attacker's requests remain distinct by computing their semantic similarities.
\para{Experimental studies} 
In our study, we confirmed timing leakages in open-source projects, including SGLang~\cite{zheng2023efficiently}, Langchain~\cite{langchain-website}, and GPTCache~\cite{bang2023gptcache}. We further demonstrate the feasibility of deducing proprietary system prompts (i.e., \textit{prompt stealing attack}) and inferring sensitive requests from neighboring users (i.e., \textit{peeping neighbor attack}).

For the prompt stealing attack, our evaluation indicates that the accuracy of detecting per-token cache hits or misses in the KV cache is 99\%, with a false positive rate (FPR) of 0.003. Using the incremental search algorithm, we recovered the system prompt token by token, requiring an average of 112.72  queries per recovered token. This approach achieved an average recovery accuracy of 89.0\% and a corresponding FPR of 0.04.  
For the peeping neighbor attack, our measurements show an 81.4\% accuracy in distinguishing hits from misses, with an average FPR of 0.045 in a single trial. This accuracy improved to 95.4\% with a 0.056 FPR after 5 trials under GPTCache's default settings.
We further observed that it is possible to infer the documents processed by a victim user in a vulnerable LLM application, even when using standard commodity LLM services.  
Moreover, our black-box study of existing online services shows that popular LLM systems---such as Claude, DeepSeek, and Azure OpenAI---employ KV or semantic cache sharing to cut costs, rendering them susceptible to timing side-channel attacks.

Finally, we propose initial defenses against these side-channel risks. To mitigate KV cache leakage, we recommend sharing prefix caches only in batches of at least $k$ tokens ($k \ge 2$). Although this increases the prompt search space and thus the required number of guesses, the larger timing differences for sharing multiple tokens also make classifiers more robust. Consequently, attacks remain accurate but incur higher query overhead. To address semantic cache leakage, we advise anonymizing privacy-related content in user inputs before performing semantic-similarity searches. Our experiments show that this method adds modest overhead (around 4\%).  

\para{Contributions} Our paper makes the following contributions:

\noindent$\bullet$\textit{~New discovery}. We identified new timing side channels in both open-source and online LLM serving systems, arising from the sharing of KV caches and semantic caches.

\noindent$\bullet$\textit{~Novel exploit strategies}. We introduced new attack strategies to leverage the inherent side channels in LLM inference optimizations, enabling two distinctive attacks: prompt stealing attack and peeping neighbor attack.

\noindent$\bullet$\textit{~Experimental validations, real-world measurements and mitigations}. We validated the side-channel leakages locally on prominent LLM systems and conducted a black-box measurement study of popular online LLM services. We also presented preliminary mitigation measures for these risks.

\para{Responsible disclosure}
We disclosed our findings to all relevant developers (SGLang, GPTCache, etc.) and LLM service providers (OpenAI, Claude, etc.) upon identifying the side channels in September 2024.  
At the time of this manuscript's preparation, we received positive responses from the SGLang team, which noted that we were among the first two groups to report this issue, both within the same week. Moreover, we were the first to raise the topic during the SGLang development meeting, and we are now working closely with their team on a resolution. 


\begin{table*}
    \centering
    \caption{Comparisons with closely-related works.}
    \label{tab:comparison}
    \begin{tabular}{C{1.8cm} C{2.3cm} C{2.3cm}  C{1.8cm}  C{1.8cm} C{1.6cm} C{1.9cm} C{1.4cm}}
    \toprule
         & \textbf{Independent of scheduling policy?} & \textbf{Independent of prompt templates?} & \textbf{Cache eviction supported?} & \textbf{Real-world measurement?} & \textbf{KV cache channel?} & \textbf{Semantic cache channel?} & \textbf{Mitigations}  \\ \midrule
        PROMPTLEAK~\cite{wuknow} & \XSolidBrush & \Checkmark & \Checkmark  & \XSolidBrush & \Checkmark & \XSolidBrush & \XSolidBrush \\ 
        InputSnatch~\cite{zheng2024inputsnatch} & \Checkmark  & \XSolidBrush & \XSolidBrush & \XSolidBrush & \Checkmark & \Checkmark & \XSolidBrush \\ 
        \cellcolor{lightgray} This work & \cellcolor{lightgray}\Checkmark  & \cellcolor{lightgray}\Checkmark  & \cellcolor{lightgray}\Checkmark  & \cellcolor{lightgray}\Checkmark & \cellcolor{lightgray}\Checkmark & \cellcolor{lightgray}\Checkmark & \cellcolor{lightgray}\Checkmark \\ \bottomrule
    \end{tabular}
\end{table*}

\para{Comparison with concurrent and follow-up works (\autoref{tab:comparison})} 
Concurrently and independently to our research, Wu et al. proposed PROMPTLEAK~\cite{wuknow}, an attack that exploits the Longest Prefix Match (LPM) scheduling policy in SGLang, which prioritizes requests with longer prefix matches, to leak user prompts. 
Their method performs collision attacks by sending carefully crafted batched prompts. When a request shares more KV cache entries than others, SGLang prioritizes it, enabling adversaries to extract victim tokens.
In contrast, our work identifies a more general KV cache side channel that does not depend on LPM-based scheduling and thus remains effective even without prefix-based prioritization. Moreover, we are the first to reveal a semantic cache side channel, propose practical mitigations, and conduct real-world measurements to assess leakage risks in commercial LLMs.




Inspired by our findings, Gu et al. conducted a subsequent large-scale measurement study on prompt caching in real-world LLM services~\cite{gustanford}, identifying its presence in 8 out of 17 evaluated providers. 
More recently, Zheng et al. investigated timing side channels in LLMs through the InputSnatch attack~\cite{zheng2024inputsnatch}. However, their work lacks the optimized search strategy we introduce for efficient request recovery and suffers from practical limitations. Specifically, their KV cache attacks on vLLM can extract blocks of 16 tokens but are limited to template-dependent scenarios, requiring attackers to have prior knowledge of fixed input structures. 
Additionally, their semantic cache attacks on GPTCache aim to infer patterns from retrieved documents but do not address self-induced interference from the attackers' own queries and lack eviction strategies to mitigate adversarial noise. In contrast, our approach enables template-free extraction, independent of application-specific formats, and incorporates eviction-controlled probing to systematically eliminate interference. Furthermore, we validate the effectiveness of our method in a realistic setting: a document summarization service powered by a commercial LLM API.

\hl{In general, PROMPTLEAK and InputSnatch achieve higher prompt recovery rates and lower per-token recovery costs than our attack, but at the expense of much stronger assumptions. Specifically, PROMPTLEAK depends on the LPM scheduling policy in SGLang, while InputSnatch relies on predefined template structures and fixed sentence blocks known to the attacker. By contrast, our method only assumes cache sharing and does not require such constraints. Even under this weaker and more realistic model, our attack achieves a reasonable single-token recovery accuracy and can recover a meaningful number of tokens --- 81 tokens with only 519 guesses (}\autoref{subsec:evalpsa}\hl{). In practice, attackers could recover even more tokens by employing a stronger next-token predictor and increasing the maximum number of attack queries.}

\ignore{Concurrently and independently to our research, Wu et al. proposed PROMPTLEAK~\cite{wuknow}, an attack that exploits the Longest Prefix Match (LPM) scheduling policy in SGLang, which prioritizes requests with longer prefix matches, to leak user prompts. Their method performs collision attacks by sending carefully crafted batched prompts. When a request shares more KV cache entries than others, SGLang prioritizes it, enabling adversaries to extract victim tokens. More recently, Zheng et al. investigated timing side channels in LLMs through the InputSnatch attack~\cite{zheng2024inputsnatch}. However, their work suffers from practical limitations. Specifically, their KV cache attacks on vLLM can extract blocks of 16 tokens but are limited to template-dependent scenarios, requiring attackers to have prior knowledge of fixed input structures. Additionally, their semantic cache attacks on GPTCache aim to infer patterns from retrieved documents but do not address self-induced interference from the attackers' own queries and lack eviction strategies to mitigate adversarial noise.

Our approach achieves 89\% accuracy in single token recovery with a maximum of fewer than 100 tokens recovered, whereas PROMPTLEAK and InputSnatch can achieve complete prompt recovery in certain scenarios with high success rates for specific prompts. Although our recovery performance may not match theirs in terms of completeness, our approach relies on the weakest assumptions among existing methods. In our PSA threat model, we do not depend on any behavioral patterns of victim request transmission. Unlike InputSnatch, which relies solely on predefined template structures and specific sentence blocks known by attackers, our method does not require such constraints. Furthermore, our approach is inherently immune to the challenges faced by PROMPTLEAK, where attack windows can disappear due to constant victim requests and require reconstruction, nor do we depend on the priority scheduling of SGLang that PROMPTLEAK focuses on.

In contrast, our work identifies a more general KV cache side channel that does not depend on LPM-based scheduling and thus remains effective even without prefix-based prioritization. Our approach enables template-free extraction, independent of application-specific formats, and incorporates eviction-controlled probing to systematically eliminate interference. Furthermore, we validate the effectiveness of our method in a realistic setting: a document summarization service powered by a commercial LLM API. Our attack demonstrates that by relying solely on widely accepted OpenAI API usage patterns and sharing mechanisms of prefix KV cache, we can extract information of sufficient value under the most basic and universally present conditions in modern LLM serving systems.

Moreover, we are the first to reveal a semantic cache side channel, propose practical mitigations, and conduct real-world measurements to assess leakage risks in commercial LLMs. Inspired by our findings, Gu et al. conducted a subsequent large-scale measurement study on prompt caching in real-world LLM services [17], identifying its presence in 8 out of 17 evaluated providers, validating the real-world relevance of our discoveries.
}


\para{Availability}
All the code and datasets necessary to reproduce our experiments are publicly available at: \url{https://github.com/Maxppddcsz/llm-sidechannel}. The demos for our attacks are available at: \url{https://sites.google.com/view/early-bird-catches-the-leak/}.


\ignore{Large Language Models (LLMs) have found widespread deployment in various applications such as chatbots~\cite{chatgpt,gemini}, search engines~\cite{perplexity}, and coding assistants~\cite{copilot}. However, LLM inference is a resource-intensive task that requires significant computational power and memory capacity. This is due to the model's vast number of parameters, numerous layers, and large context sizes.
As a result, optimizing LLM inference has become crucial, prompting researchers and engineers to explore avenues like weight quantization~\cite{yao2022zeroquant,wei2022outlier,xiao2023smoothquant,frantar2022gptq,liu2023llm}, model compression~\cite{wang2024model}, algorithm  optimization~\cite{dao2022flashattention,kwon2023efficient,xiao2023efficient}, hardware infrastructure improvements~\cite{zhao2024alisa,song2023powerinfer}, and parallel processing techniques~\cite{yu2022orca}. These advancements aim to reduce latency and enhance overall performance~\cite{miao2023towards}.


In this paper, our primary focus is on the deployment of LLMs in shared computing systems, serving multiple users or applications simultaneously. By drawing parallels to traditional computing systems like CPUs, GPUs, and accelerators, we argue that existing research predominantly emphasizes performance improvements, such as throughput and latency, while neglecting important security considerations. This oversight mirrors the vulnerabilities observed in hardware computing systems, which can be compromised through micro-architectural side channels.

Specifically, previous studies have aimed to minimize computation requirements and optimize memory utilization. Strategies such as GPU kernel scheduling, iteration-level batching, and token-level batching have been explored. 
However memory sharing, cache contentions and evictions, and GPU scheduling among different users and applications can introduce interference among multiple requests, potentially leading to observable timing side channels. Exploiting these side channels allows for the inference of other users' or applications' private prompts.

\para{Identifying the leakage sources}
In recent years, ``prompt as a service'' has greatly enhanced the utility of LLMs by enabling them to perform various downstream tasks efficiently without fine-tuning. This has also increased the commercial value of prompts.
This paper takes the first step in exploring this new type of side channels to leak the system prompt or user prompt. We focus on various caches in LLM systems, which not only reduce the computation cost of LLM inference but also enhance the user experience by reducing the service latency. Specifically, we have identified the following leakages.
\begin{packeditemize}
    
    \item \textit{Leakage of sharing the KV cache.} Efficient inference requires careful allocation of GPU memory. For each request, an LLM maintains an in-memory state known as the KV cache and reuses it in every iteration for the duration of the request. 
    The causal attention mask in LLMs results in each token's activations being influenced only by previous tokens in the sequence. Therefore, if multiple requests share a common prefix, the keys and values corresponding to the prefix tokens will be identical across sequences. To optimize the memory utilization of the KV cache, the system can identify matching prompt prefixes across multiple requests and share their key/value tensors in memory at runtime~\cite{kwon2023efficient,zheng2023efficiently}. This sharing occurs when prompts have a common prefix, which is often the case when using few-shot examples~\cite{reynolds2021prompt} or a chatbot system prompt~\cite{giray2023prompt} or prompt templates~\cite{langchain-prompttemplate}. For instance, it is reported that  Claude's prompt caching feature can cut costs by as much as 90\% and decrease latency by up to 85\% for long prompts~\cite{claudepromptcaching}.

   \item \textit{Leakage of sharing the semantic cache.} The semantic cache~\cite{bang2023gptcache} improves the performance of LLM applications by caching responses based on the semantic meaning or context within the requests themselves. For example, for two requests such as ``give me suggestions for a comedy movie'' and ``recommend a comedy movie'', the LLM system detects their semantics are similar and returns similar responses without querying the LLM. It has been shown that when GPTCache is integrated with OpenAI's GPT service, response speed can be boosted by a factor of 2 to 10 when the cache is hit~\cite{bang2023gptcache}.


\end{packeditemize}


\para{Challenges and solutions}
One straightforward strategy is to brute-force requests and attempt to identify the specific request that triggers a cache hit. However, It can be challenging to exploit the above leakages effectively. 
Firstly, the input space of the requests is vast, making it difficult to systematically explore all possibilities. Additionally, during this process, the attacker's own requests can also be cached, potentially adding noise to the following-up requests and leading to the eviction of the targeted victim's requests. 
Secondly, LLMs typically employ positional encoding to maintain the word order before inputting the data into the transformer block for predicting the next word. As a result, the KV cache can only be effectively reused when the prompts have a common prefix, limiting the opportunities for KV cache based attacks. Lastly, the timing difference resulting from hitting a single cache block is typically minimal and can easily blend in with the natural variations in network latencies, making it harder to detect and exploit.

To tackle these challenges, we propose several attack strategies specifically tailored to exploit the side channels inherent in LLMs.
Firstly, we developed a classification model to detect token-level cache hits based on timing measurements in an offline phase (for the KV cache). By adjusting the timing threshold, we found that online detection accuracy could be greatly improved with just a few repeated trials.
For the semantic cache, we identified the most representative trials during an offline phase and devised attacker strategies aimed at maximizing accuracy while minimizing false alarms.
Secondly, in order to mitigate the noise caused by the attacker's own attempts, we leverage the caching system's capabilities to efficiently trigger cache flushes (for the KV cache) or ensure that the attacker's trials do not interfere with each other (for the semantic cache).
Moreover, we devised an efficient token-by-token search algorithm to reduce the search space, that capitalizes on the observation that the KV cache can only be reused when requests share a common prefix.







\para{Experimental validations}
We have verified the vulnerabilities in open-source projects, including GPTCache~\cite{gptcache-github}, Langchain~\cite{langchain-website}, and SGLang~\cite{sglang-github} within our local environment. The result shows that the accuracy of detecting a per-token hit or miss in KV cache stands at 86\%. Furthermore, the accuracy of identifying semantic cache and hit exceeds 95\% under the default parameters in GPTCache.

Building upon these vulnerabilities, we demonstrate the potential to deduce proprietary system prompts (i.e., a \textit{prompt stealing attack}) or infer sensitive requests of neighbouring users (i.e., a \textit{peeping neighbour attack}). For the prompt stealing attack, searching for each token of the system prompt requires only 194 queries with an average accuracy of 92.3\%. For the peeping neighbour attack, we can infer sensitive information with an accuracy exceeding 95\%. Furthermore, we find that common deployment practices of LLMs, such as Retrieval-Augmented Generation (RAG), introduce timing vulnerabilities that can be exploited to infer the document and website processed by the LLM.
Based on our black-box measurement study on existing online services. We further find that popular online LLM systems, such as OpenAI's GPT4 and Google's Gemini, use cache sharing to reduce costs and are thus vulnerable to these timing side-channel attacks. Finally, we present evaluations on preliminary mitigation of the side-channel risks we identified, through obfuscating the sensitive information or increasing the minimum number of tokens shared in KV cache.

\para{Contributions} The paper makes the following contributions.

\begin{packeditemize}
    \item \textit{New discovery}. We have identified timing side channels in both open-source and online LLM serving systems, arising from the sharing of the semantic cache and KV cache to minimize costs.
    \item \textit{Novel exploit strategies}. We proposed attack strategies specifically tailored to exploit the side channels inherent in LLMs, and demonstrate 2 novel attacks: a prompt stealing attack and a peeping neighbour attack.
    \item \textit{Experimental validations, real-world measurements and mitigation}. We verified the leakages in our local environment, and conducted a black-box measurement study of popular online LLM services. We also provided preliminary solutions to mitigate our identified vulnerabilities.
\end{packeditemize}
}
\section{Background}
\label{sec:background}

\subsection{LLM Serving Systems}


In this paper, we explore shared LLM deployments serving multiple users or applications within computing systems. This setup is common in commercial public services (e.g., ChatGPT) and  local LLMs tailored to specific tasks. 
In these scenarios, LLMs optimize latency and throughput via memory usage optimization, efficient batching, and GPU scheduling. These factors can introduce interference among concurrent requests, potentially leading to observable timing side channels. 




\noindent\textbf{LLM serving modes.}
The LLM service offers two operation modes. In non-streaming mode, the response is fully generated and then delivered once the request has been processed. However, this approach can result in an extended waiting period for long completions.  
To achieve faster responses, the streaming mode is available. In this mode, the LLM emits tokens sequentially, allowing users to view the beginning of the completion while the remaining tokens are still being generated. Streaming is the preferred method for interacting with LLMs, especially in chatbot scenarios where real-time conversation is essential.
Popular LLM applications (e.g., ChatGPT) use a system prompt containing task definitions, examples, and safety rules to guide their behavior. This prompt is typically static and shared among all users.

\para{Metrics}
Latency measures how long it takes for an LLM to respond to a user's query.
Low latency is particularly important for real-time interactions, such as chatbots. Time to First Token (TTFT) is the interval from the moment a user submits a prompt until receiving the first token of the response. It reflects the initial processing delay and serves as a crucial indicator of user-perceived responsiveness.
This paper examines the risks arising from optimizing an LLM's serving latency and employs TTFT as the primary metric for side-channel observations.




\subsection{Serving Backend}
Most LLMs rely on the Transformer architecture, which uses \textit{Attention}~\cite{vaswani2017attention} to pinpoint the most relevant parts of the input. Core to this mechanism are Query (Q), Key (K), and Value (V) embeddings: Q specifies what to seek at each position, K determines how to match relevant information across the sequence, and V provides the data retrieved upon a match.
LLM inference consists of two stages: the \textit{prefill phase} and the \textit{decoding phase}. 
In the prefill phase, the LLM processes the request prompt by converting it into a sequence of tokens, each mapped to a numerical representation called embedding. It then computes the K and V embeddings for each token across all attention layers, enabling the generation of \textit{the first token} in a single step.
In the decoding phase, the LLM generates each subsequent token by using the prefilled context and the previously generated token. For each layer, it computes the Q, K and V embeddings for the new token and attends to all existing context tokens. Unlike the prefill phase, decoding processes only one token at a time.


\para{Memory management of KV cache}
The attention mechanism in LLMs requires computing pairwise similarities among tokens, which leads to quadratic complexity with respect to sequence length. 
To address this, KV caching stores the K and V embeddings in GPU memory, avoiding redundant computations and reducing the computation cost to scale linearly with sequence length.
Early LLM systems statically allocated large memory regions for KV caches to accommodate unpredictable output lengths, leading to significant internal and external fragmentation. vLLM mitigated this with PagedAttention, which divides the KV cache into blocks accessed via a lookup table~\cite{kwon2023efficient}. This enables efficient memory sharing across requests. Modern frameworks like Nvidia’s TensorRT-LLM and Huggingface’s TGI adopt similar techniques, but the security implications of KV cache sharing remain underexplored, leaving a critical gap in existing research.


\ignore{
\para{RAG}
\wenhao{we may remove RAG related terms}
Retrieval-Augmented Generation (RAG) is a technique that combines retrieval and generation to enhance natural language generation tasks. First, a retrieval system searches an external knowledge source for relevant information, which is then passed to a generative model to produce a more accurate and contextually appropriate response. RAG is particularly useful in knowledge-intensive tasks like question answering and summarization. The retrieval step adds real-time, external knowledge, improving the quality of generated output.However, the process often involves repeated use of long texts, which can cause efficiency issues due to the need for repeated retrieval from large datasets. To improve efficiency, techniques such as caching, parallel processing, or pruning can help reduce redundant computations, ensuring the system remains fast and scalable. This allows RAG to generate high-quality responses while maintaining system efficiency.
}

\ignore{
\para{Semantic Cache}
\link{This para might not be in urgent need. Because Semantic cache is simpler and less interesting.}
}
\ignore{
\para{Continuous batching}
State-of-the-art inference engines adopts continuous batching to handle the varying sequence lengths and dynamic arrivals of requests.
Continuous batching is first proposed by Orca~\cite{yu2022orca}. A new or completed request can join or leave the running batch immediately, without waiting for all the ongoing requests to finish. For this purpose, Orca employs an iteration-level scheduling algorithm, ensuring that new sequences can be processed as soon as the last one is done. This approach leads to higher GPU utilization compared to static batching. In reality, continuous batching has attracted the attention from both industry and academia, following works like vLLM~\cite{kwon2023efficient}, FastGen~\cite{holmes2024deepspeed} have fulfilled this idea, making it a consensus in LLM serving field. 
}

\subsection{Threat Model and Assumptions}
\label{subsec:threat_model}

\para{Threat model}
In this paper, we examine the security risks of deploying a shared LLM to serve multiple users or applications within a single computing system.  
Specifically, we consider two main scenarios.  
First, an LLM service provider offers public APIs that registered users can employ to send requests, all of which are processed by the same underlying serving system. A victim user may establish a proprietary system prompt to power a popular LLM application, while an attacker could leverage the same LLM APIs to infer this system prompt, thereby gaining financial benefits or bypassing embedded safety instructions.
Second, public LLM applications---such as chatbots (e.g., GPT-4) or document analysis services (e.g., AnythingLLM~\cite{anythingllm}, etc.)---handle concurrent user requests via the same LLM serving system. If the application itself relies on a public LLM API, these requests are generally routed through the same developer's API key.  
An attacker could register as a user of the application to detect whether specific requests have been submitted by others and infer, for instance, users’ interest in certain topics or uploaded files. They could also monitor requests over time to detect private attributes or sensitive personally identifiable information (PII) (Table~\ref{tab:templates}).
In both scenarios, the attacker's and the victim's requests share the same platform and thus same cache resources. This shared environment can produce interference and create observable timing side channels---the core subject of our investigation.   
The attacker needs only black-box access to the underlying model, without knowledge of its architecture or weight parameters. However, the attacker must first examine the system's leakage profile in an offline phase to understand how different inputs affect timing, enabling them to craft queries that exploit the cache-induced timing discrepancies.

{In this paper, we explore the side-channel risks in both local and remote LLM services. For local settings, we assume a stable network connection between client and server. For remote services, previous works such as NetCAT~\cite{kurth2020netcat} and NetSpectre~\cite{schwarz2019netspectre} have addressed mitigating noise caused by unstable connections and jitters, particularly in CPU cache side-channel attacks. Extending such noise-reduction strategies to remote LLM scenarios remains an avenue for future research.}
We do not consider hardware side channels tied to GPU micro-architectures~\cite{naghibijouybari2018rendered,zhou2016vulnerable,taneja2023hot}.
Instead, our focus lies on software caches maintained by the LLM serving system, making our attacks applicable across various hardware platforms (CPUs, GPUs, ASICs, etc.).

\para{Assumptions}
\hl{For end-to-end attacks, we assume that the adversary has certain prior knowledge of the target user or system prompt. Specifically:}

\begin{itemize}
    \item \textbf{Prompt Stealing Attacks} (\autoref{subsec:evalpsa}): 
    \hl{The attacker is assumed to have access to a public dataset of system prompts that can be used to fine-tune the next-token predictor. Such datasets are semantically diverse and representative of real-world prompt engineering practices. In our evaluation, the victim randomly selects a small subset (e.g., 200 prompts) as its proprietary system prompt, while the attacker uses the remaining prompts as training data. This reflects realistic scenarios where production systems employ a limited number of carefully crafted prompts, while public repositories offer abundant similar data for adversarial training.}

    \item \textbf{Peeping Neighbor Attacks} (\autoref{subsec:evalpna}): 
    \hl{We assume the attacker is focused on a particular topic (e.g., healthcare, travel planning) and has a public set of prompt templates containing sensitive private attributes such as personal names, medical conditions, travel destinations, or business details. These templates can be obtained from public datasets or generated using paraphrasing tools. The attacker uses them to probe semantic caches, where timing differences between cache hits and misses may reveal whether the victim’s request contains the targeted attributes.}

    \item \textbf{Document Inference Attacks} (\autoref{subsec:psarag}): 
    \hl{Because KV-cache sharing typically requires requests to share a common prefix, we assume the attacker knows either the full document or at least its prefix (i.e., the initial portion of the content). Even if the document’s content is known, detecting its submission to an LLM service may disclose an organization’s interests, posing serious privacy and competitive risks across different sectors. 
    }
\end{itemize}

\section{Attacks}
\label{sec:attacks}





\subsection{Overview}
Creating effective prompts is a challenging task that requires substantial effort, particularly in scenarios like in-context learning where extensive data is needed to optimize LLM performance. Furthermore, prompts can include personal or sensitive information, making them valuable assets that must be safeguarded. For instance, Samsung Electronics has prohibited employees from using Chatgpt to prevent accidental disclosure of confidential data to OpenAI~\cite{samsung}.

In our research, we investigated two types of attacks. The first is the \textit{prompt stealing attack (PSA)}, which targets system prompts. A system prompt defines the model's operational behavior and may incorporate carefully crafted business logic, private data, or safety-related instructions. Consequently, LLM application developers treat it as confidential intellectual property~\cite{syspromptvalue}. Moreover, once exposed, the system prompt could facilitate other attacks, such as jailbreaking.  
The second is the \textit{peeping neighbor attack (PNA)}, which focuses on uncovering the semantics of another user's prompt. Since these prompts may contain personally identifiable information (PII) or other sensitive data, any disclosure poses a substantial risk to user privacy.
There are three entities involved in these attacks:
the server ($\mathcal{S}$), the victim user ($\mathcal{C}$), and the attacker ($\mathcal{A}$). The attacker's goal is to infer the prompt submitted by the victim user. The attack proceeds in two phases. In the \textit{offline phase}, the attacker studies how a request alters the server's state and how these modifications manifest in the latency of subsequent requests. Critically, these timing profiles stem primarily from the system's optimization techniques rather than from a specific model or parameter set.




\ignore{
\begin{figure}
    \centering
    \includegraphics[width=0.99\linewidth]{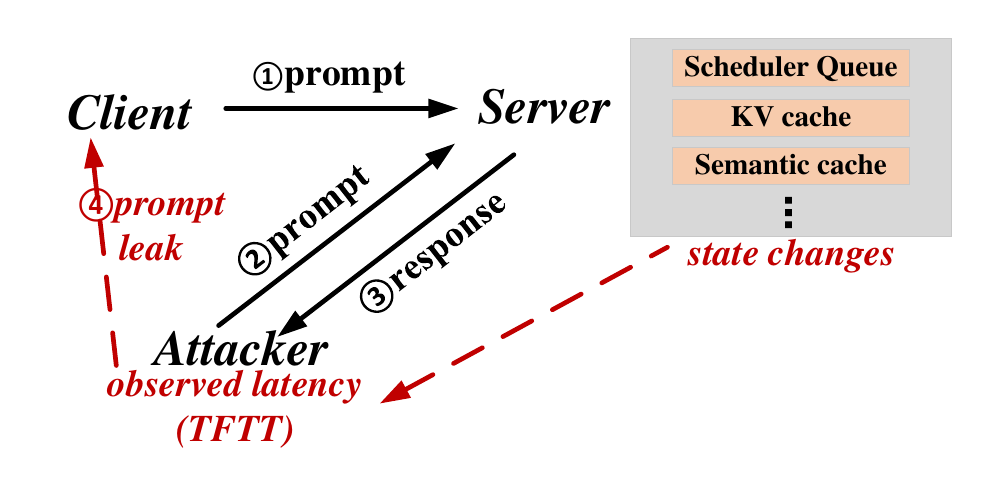}
    \caption{Attack steps\wenhao{todo: offline/online phase}.}
    \link{offline_online.pdf is ready to be used.}
    \label{fig:attacksteps}
\end{figure}
}

In the \textit{online phase}, the attacker leverages insights gained during the offline phase to craft requests that exploit the identified timing properties. Initially, $\mathcal{S}$ is in state $State_0$. When $\mathcal{C}$ issues a request, the state changes to $State_1$, reflecting updates like modifications to the KV or semantic cache. These state transitions can affect the performance of later requests.  
To track the system state, $\mathcal{A}$ regularly sends a request $r$ at intervals starting from time $t_{start}$, measuring the resulting latency $l = t_{end} - t_{start}$, where $t_{end}$ denotes the time point when the first token in the response arrives.  
By analyzing these latency readings, $\mathcal{A}$ can infer the prompt submitted by $\mathcal{C}$.

\ignore{
For the system prompt recovery, we assume that the attacker can trigger a request both with and without system prompt. We believe the assumption is reasonable, considering platforms like OpenAI support modes (e.g., API mode and web mode) that handle requests with preset, custom, or no system prompts~\cite{chatgptapi}.
To recover the system prompt,  $\mathcal{A}$ initiates an initial request $r_0$ to trigger the system prompt within $\mathcal{S}$, resulting in the actual request $r = r_0 + {system\_prompt}$. Immediately after that, $\mathcal{A}$ sends another request $r_1$ and measures the latency $(t_{end} - t_{start})$ to infer the state changes, thereby deducing the content of the $system\_prompt$. 
}


\subsection{Prompt Stealing Attacks (PSA)}


\noindent\textbf{Background.}
KV caching is a widely adopted optimization in LLM which avoids redundant computation during autoregressive generation by storing key and value embeddings. 
Representative implementations include PagedAttention in vLLM~\cite{kwon2023efficient}, which detects shared prefix segments at runtime for KV cache reuse, and RadixAttention in SGLang~\cite{zheng2023efficiently}, which efficiently manages shared KV caches in prefix tree.

\para{The side channel}
Sharing KV caches can introduce a timing side channel. In the prefill phase, if a request's prefix matches one already stored in the KV cache, it will be processed more quickly. Since most LLMs stream their outputs (i.e., token by token), it is possible to measure the fine-grained TTFT and detect timing discrepancies associated with cache hits versus misses.
Consider a model with 7 billion parameters (e.g., Llama-7B) running on an A100 GPU with 312 TFLOPS of computational power and a memory bandwidth of 1.5 TB/s. In the best case, with full GPU utilization, a cache miss for a single token during the prefill phase may take:

{\small
\begin{align*}
\text{prefill time (miss)} 
&= \#tokens \times \frac{\#parameters}{\text{GPU compute bandwidth}}\\
&= \frac{1 \times (2 \times 7B)~\text{FLOP/token}}{312~\text{TFLOP/s}} \\ 
& \approx 0.045~\text{ms}.
\end{align*}}

By contrast, if the token hits the cache, the time is dominated by loading the precomputed KV cache from HBM (assuming 16-bit precision parameters):

{\small
\begin{align*}
\text{prefill time (hit)}
&= \#tokens \times \frac{\text{KV cache per token}}{\text{GPU memory bandwidth}} \\
&= \frac{1 \times (2 \times 4096 \times 32) \times 2~\text{Bytes/token}}{1.5~\text{TB/s}} \\
& \approx 0.35~\mu\text{s}.
\end{align*}}

These gaps become larger for more complex models or when serving multiple requests concurrently. For instance, on a \path{Llama-3.1-70B-Instruct-GPTQ-INT4} model (70 billion parameters at 4-bit precision), the prefill time for a token miss is about 0.45~ms, while a hit is roughly 0.22~\(\mu\)s.

These timing differences underpin the prompt stealing attack, which leverages KV cache sharing with the same prefix. After sending a request to the LLM, an attacker can observe TTFT to infer prefix matches. Building on this observation, we devised an incremental search algorithm to recover prompts on a token-by-token basis in \autoref{subsec:evalpsa}.


\subsection{Peeping Neighbor Attacks (PNA)}
\label{subsec:pna_intro}

\noindent\textbf{Background.}
Semantic caching (e.g., GPTCache~\cite{bang2023gptcache}) stores prior requests and their corresponding responses. Upon receiving a new request, it measures semantic similarity with cached requests. If the similarity surpasses a certain threshold, the system returns the cached response; otherwise, it queries the LLM again. This technique reduces costs, improves performance, and is widely adopted in frameworks like LangChain~\cite{langchain-website} and LlamaIndex~\cite{lammaindex-github}.

\para{The side channel} 
Unlike KV caching, which reuses data only for identical prompt prefixes, semantic caching allows reuse based on semantic similarity beyond a predefined threshold.  
However, sharing a semantic cache among multiple users can inadvertently reveal their requests. Cache hits provide responses in mere milliseconds, whereas cache misses can take several seconds—creating exploitable timing differences.
This discrepancy enables the attacker to infer the semantics of concurrent requests issued by nearby users, a scenario we refer to as the \textit{peeping neighbor attack}.
Despite this, when the attacker tries to match a victim's request semantically, the attacker's own requests may also be cached, introducing noise into subsequent attempts. To address this challenge, we propose an efficient search algorithm that both minimizes the caching effects of the attacker's own requests and improves the detection rate for the victim's request, detailed in \autoref{subsec:evalpna}. 


\section{Side-channel Analysis and Evaluation}
\label{sec:eval}



In this section, we present our empirical analysis of the identified side channels and describe strategies for their efficient exploitation.  
All experiments were conducted on a Lenovo server equipped with two Intel Xeon E5-2678 v3 CPUs (12 cores at 2.50 GHz each), 100 GB DDR4 memory, and two NVIDIA A100 PCIE GPUs (80 GB memory each). The system ran Ubuntu 22.04 (kernel 5.15.0-125-generic) with GPU driver 550.127.05, CUDA 12.4, and PyTorch 2.4.0.  
We used open-source models from the Llama family as the underlying LLM, adhering to their default hardware and software configurations for all evaluations.

\subsection{Analysis on PSA}
\label{subsec:evalpsa}

\begin{figure}
    \centering
    \includegraphics[width=0.8\linewidth]{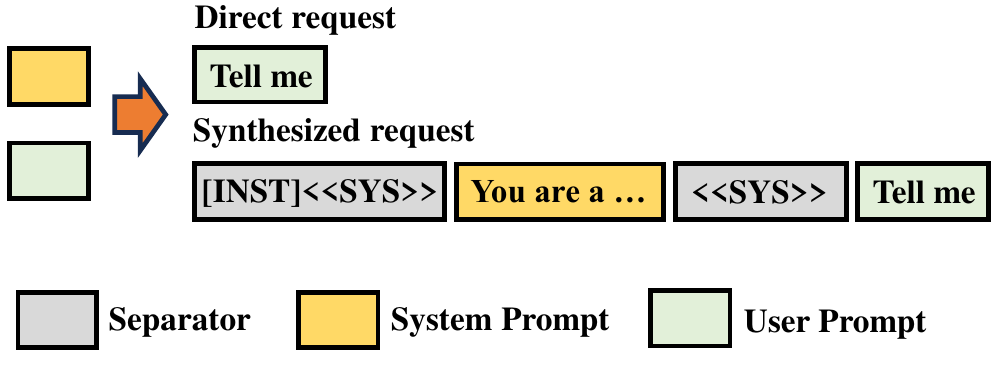}
    \caption{LLM API servers like OpenAI allow user input through both direct requests (top) and synthesized requests via a template (bottom).}
    \label{fig:system_prompts}
    \vspace{-8pt}
\end{figure}

\noindent\textbf{Attack setup.}
{With increasing concerns about client data leakage in public LLM services, local enterprise LLMs have become increasingly popular~\cite{jpmorgan}. In our study, we focus on a use case where the LLM API service is constructed from open-source projects within a \textit{local network environment}.}  
As described in \autoref{subsec:threat_model}, we consider a scenario in which a victim develops a popular LLM application (e.g., a chatbot) using a proprietary system prompt via the LLM service. The attacker interacts with this LLM application over the local network, measures the TTFT, and attempts to uncover the system prompt based on timing discrepancies.
Specifically, the LLM service uses the SGLang backend API server~\cite{zheng2023efficiently}, which supports KV cache sharing for common prefixes. Notably, LLM API servers including OpenAI allow users to define various roles, such as ``system'' and ``user'', within their requests \cite{openaiapi}, or send requests directly. Based on this capability, we categorize requests into two modes: synthesized and direct. The victim's LLM chatbot, built on the FastChat framework~\cite{zheng2023judging}, explicitly supports both, as shown in \autoref{fig:system_prompts}. In direct mode, the user sends requests directly to the SGLang backend, whereas in synthesized mode, the full prompt is created by concatenating messages from each role according to predefined templates.

\begin{figure*}[!ht]
    \centering
    \includegraphics[width=0.95\textwidth]{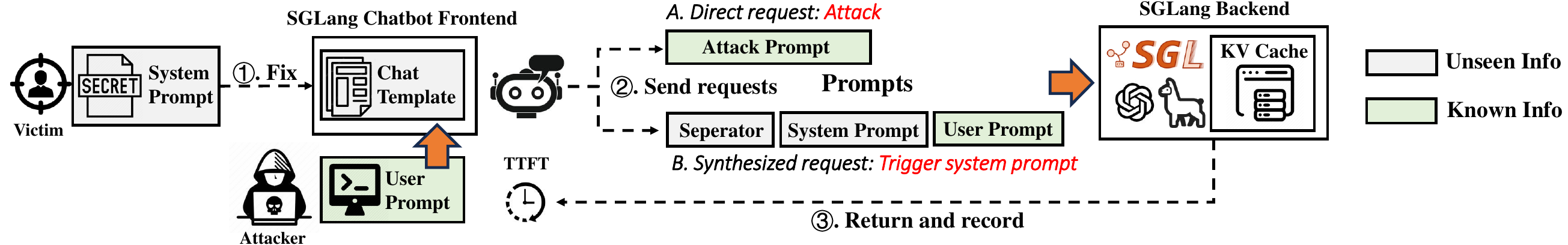}
    \caption{Overview of prompt stealing attacks.
    }
    \label{fig:model_PSA}
\end{figure*}

We consider that the victim's chatbot employs a proprietary system prompt for the ``system'' role in the synthesized mode, while the user's inputs fall under the ``user'' role. In the synthesized mode, this system prompt is prepended at every conversational turn to form the complete prompt sent to the SGLang backend.
{Notably, BloombergGPT~\cite{DBLP:journals/corr/abs-2303-17564} serves as a real-world example of a purposely built LLM for financial use, deployed for internal use at Bloomberg. It leverages 3-shot prompting to handle domain-specific tasks more effectively, safeguarding sensitive data and maintaining control over proprietary financial processes.}
As illustrated in \autoref{fig:model_PSA}, an attacker can masquerade as a chatbot user, submitting either a direct request or a synthesized request that includes the static system prompt. When the LLM processes these synthesized prompts, it retains the separator and system prompt in the KV cache used by the SGLang backend, expediting subsequent requests that share partial prefixes of the system prompt. In this attack, the attacker first submits a synthesized request to cache the system prompt, then employs direct queries to reveal it through timing leakages.





\ignore{
Open-sourced chatbots typically input the system prompt once at the start or repeatedly throughout each conversation. While inputting it once at the begining may be efficient, when token count increases or malicious input occurs, system prompt data may be evicted, leading to disruptions or security risks like jailbreaking~\cite{liu2023jailbreaking}. To mitigate this, services like FastChat rely on a fixed system prompt for gpt-4-turbo or claude-3-sonnet~\cite{fastchatsrc}, which is repeatedly prepended at each conversational turn to maintain stability. Notably, these chatbots allow users to either send prompts directly to the LLM or construct complete requests using the \textbf{LEGO} shown in \autoref{fig:system_prompts}, in an OpenAI-compatible API manner~\cite{Send_request}.

Based on these findings, we devised an end-to-end attack model in \autoref{fig:model_PSA}. We assume the victim deploys a chatbot via SGLang with a fixed prompt and default chat template of selected model. An attacker, acting as a user, can either send a direct query or a synthesized request containing the fixed system prompt. Upon processing the synthesis prompts, the LLM will store the KV cache of seperator and system prompt in the backend, enabling subsequent direct queries with partial prefixes of the separator and system prompt to benefit from accelerated responses, as reflected in TTFT. Thus, attackers may use synthesized requests to keep the system prompt in the KV cache and infer it through direct queries.

While the separator under a default chat template of the selected model is finite and can be revealed, reconstructing the victim's secret system prompt remains challenging due to its infinite possible variations. We assume the initial tokens can often be statistically guessed, such as starting with "You are...". Attackers focus on system prompts closely aligned with their interests, testing their relevance to their objectives before proceeding further, while disregarding prompts with minimal semantic similarity to their goals.
}

\begin{figure}
    \centering
    \subfloat[(a). Llama-3.1-8B-Instruct\label{fig:llama-3.1-8B-distribution}]{
        \includegraphics[width=0.46\linewidth]{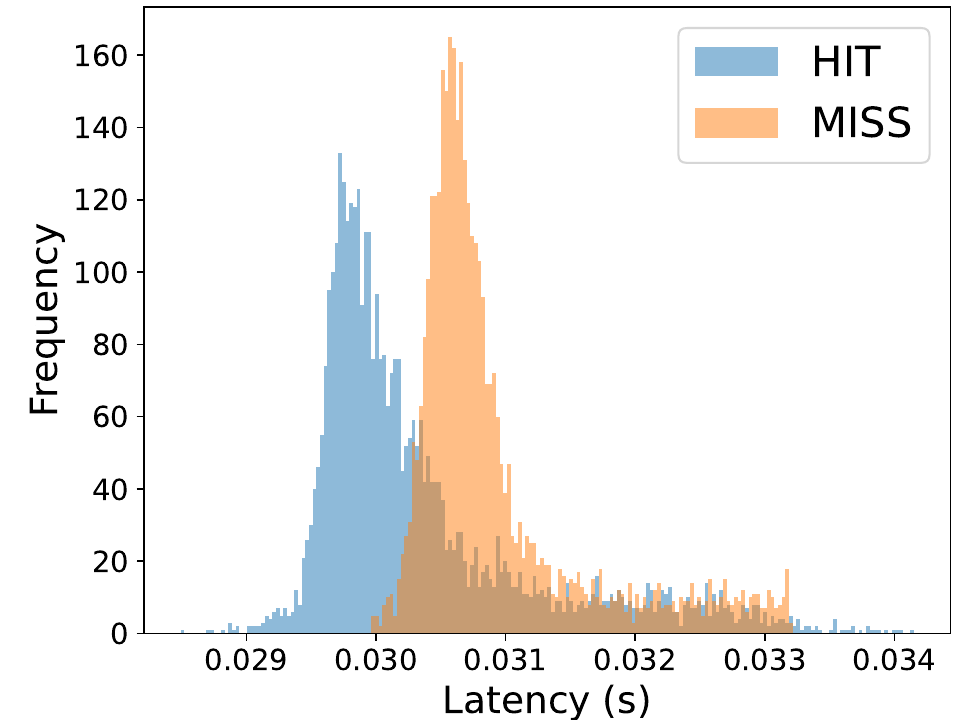}
    }
    \hfill
    \subfloat[(b). Llama-3.1-70B-Instruct-GPTQ-INT4
    \label{fig:Llama-3.1-70B-Instruct-GPTQ-INT4}]{
        \includegraphics[width=0.46\linewidth]{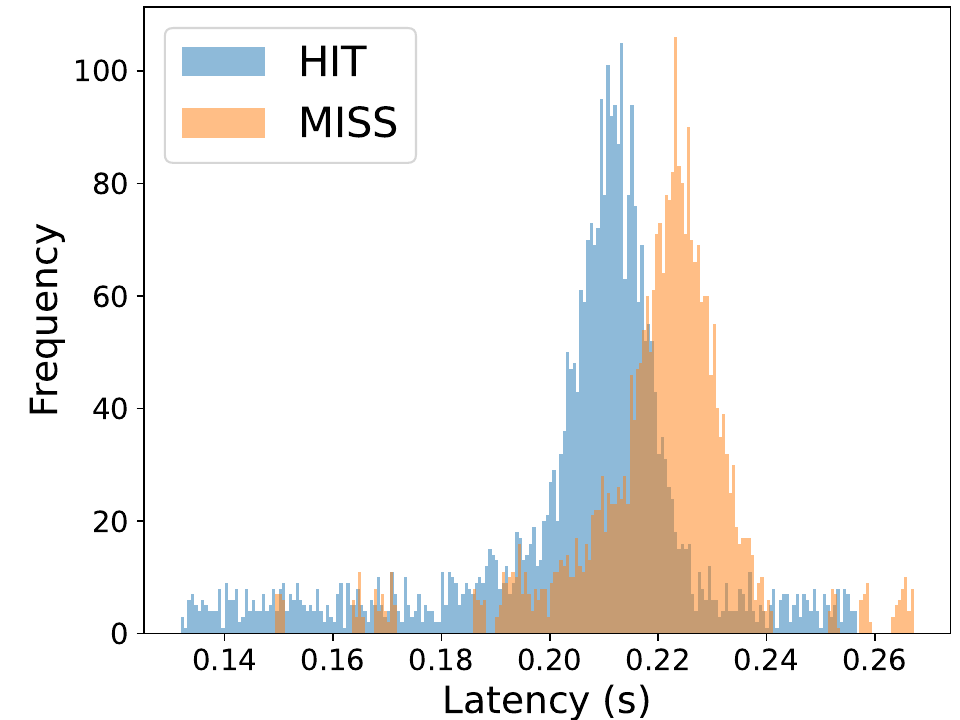}
    }
    \caption{Latency distribution of one token hit and miss. 
    }
    \label{fig:distribution}
\end{figure}

\ignore{
\textcolor{red}{However, latency distributions for hit and miss cases might still overlap, making classifier of simple thresholds ineffective. Based on the experimental conditions outlined in \autoref{sec:eval}, we present latency distributions for cases where the shared-prefix token count differs by one or two, corresponding to cache hits and misses, with 1,000 tests per sample. The \autoref{fig:distribution} shows that single-token hits and misses are indistinguishable without special handling, while two-tokens differences, though overlapping, can be distinguished.}
}

\para{Characterizing the leakage}
We began by examining the timing difference between a cache hit and a miss for a single token across different model sizes. Specifically, we tested SGLang v0.3.0, which uses a radix tree to manage KV cache blocks for efficient prefix matching. Two models were evaluated: \path{Llama-3.1-8B-Instruct} and \path{Llama-3.1-70B-Instruct-GPTQ-INT4}. 
System prompts were derived from the \path{gabrielchua/system-prompt-leakage}~\cite{systemprompt-huggingface} dataset. 
Using 15-token prompts differing only in the last token—ensuring a shared-prefix difference of exactly one—we collected TTFT measurements over 4,000 runs. 
\autoref{fig:distribution} illustrates the resulting time distributions, revealing a pronounced distinction between hit and miss scenarios for both models.

Based on these observations, a simple classifier can be built to categorize the last token in the prompt as a hit if its latency is below a predefined threshold. However, as prompts grow longer, the latency also tends to rise due to increased computational demands, necessitating different thresholds for each prompt length. To address this, \hl{instead of the absolute latency}, we measured the relative latency difference between hit and miss cases for last tokens across prompts of varying lengths (1–200 tokens). Our findings reveal that \hl{even though the absolute latency varies}, this difference remains stable across prompt length, and significantly larger than the variance within miss cases, enabling a single unified threshold for reliable hit detection. 

\ignore{
To assess the impact of the shared prefix token length on the response latency, 
we generated five groups of prompts of the same length as $T$, and the groups were denoted as $T_0$, $T_1$, $T_2$, $T_4$,and $T_8$, which shared exactly the initial $k$ tokens with $T$ for $k = 0, 1, 2, 4,$ and $8$, by replacing the $(k+1)$-th token with different tokens in each group. \textcolor{red}{Each group consisted of 2000 prompts, and TTFT was measured for each prompt immediately after $T$ was processed and cached by the LLM. Prompts from $T_0$ (no shared prefix with $T$) were labeled as $0$, while those from groups with shared prefixes (lengths $1, 2, 4, 8$) were labeled as $1$. }

\begin{figure*}[!ht]
    \centering
    \begin{subfigure}{0.33\linewidth} 
        \centering
        \includegraphics[width=\textwidth]{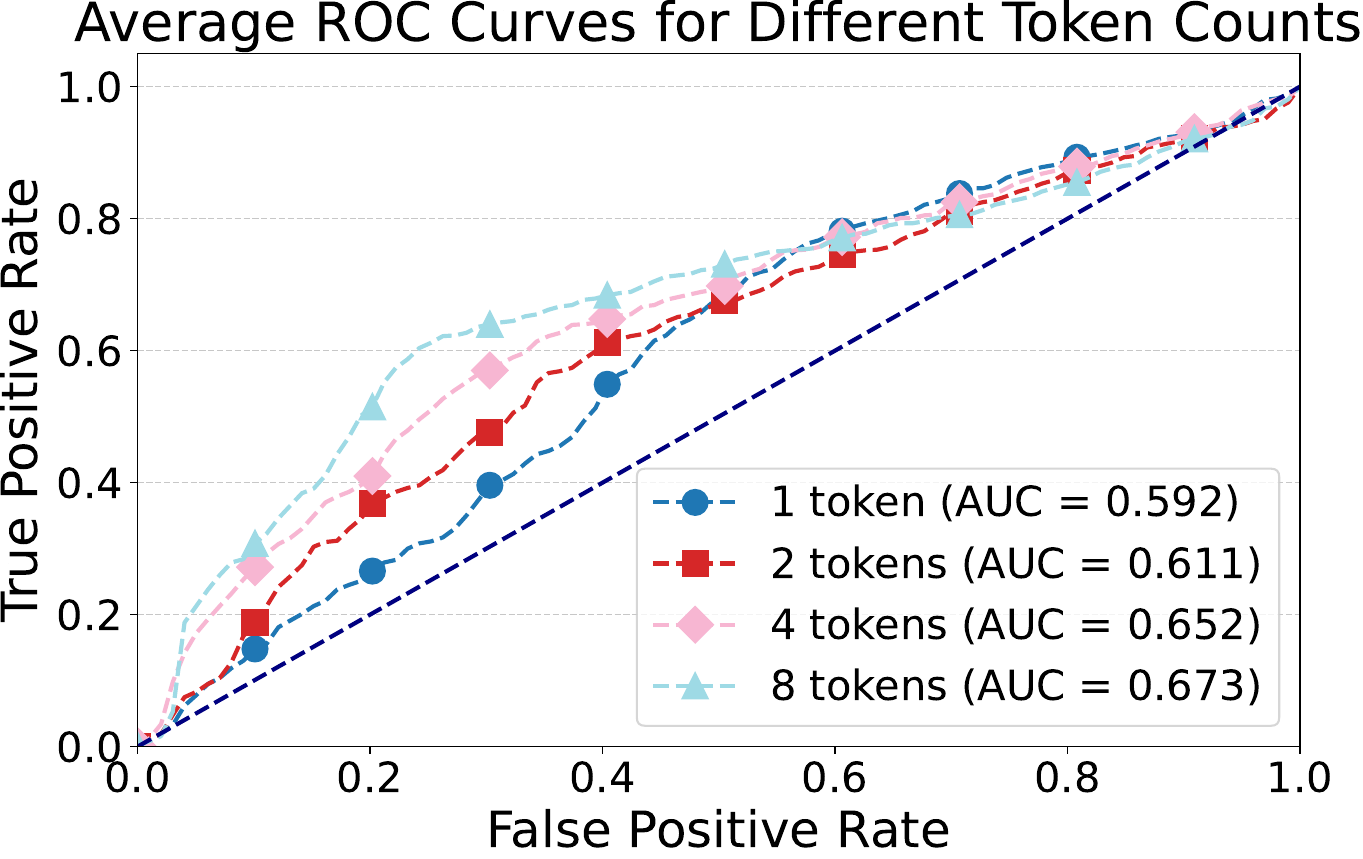}
        \caption{llama3.1-8B-Instruct}
        \label{fig:llama13b}
    \end{subfigure}
    \hfill
    \begin{subfigure}{0.33\linewidth}
        \centering
        \includegraphics[width=\textwidth]{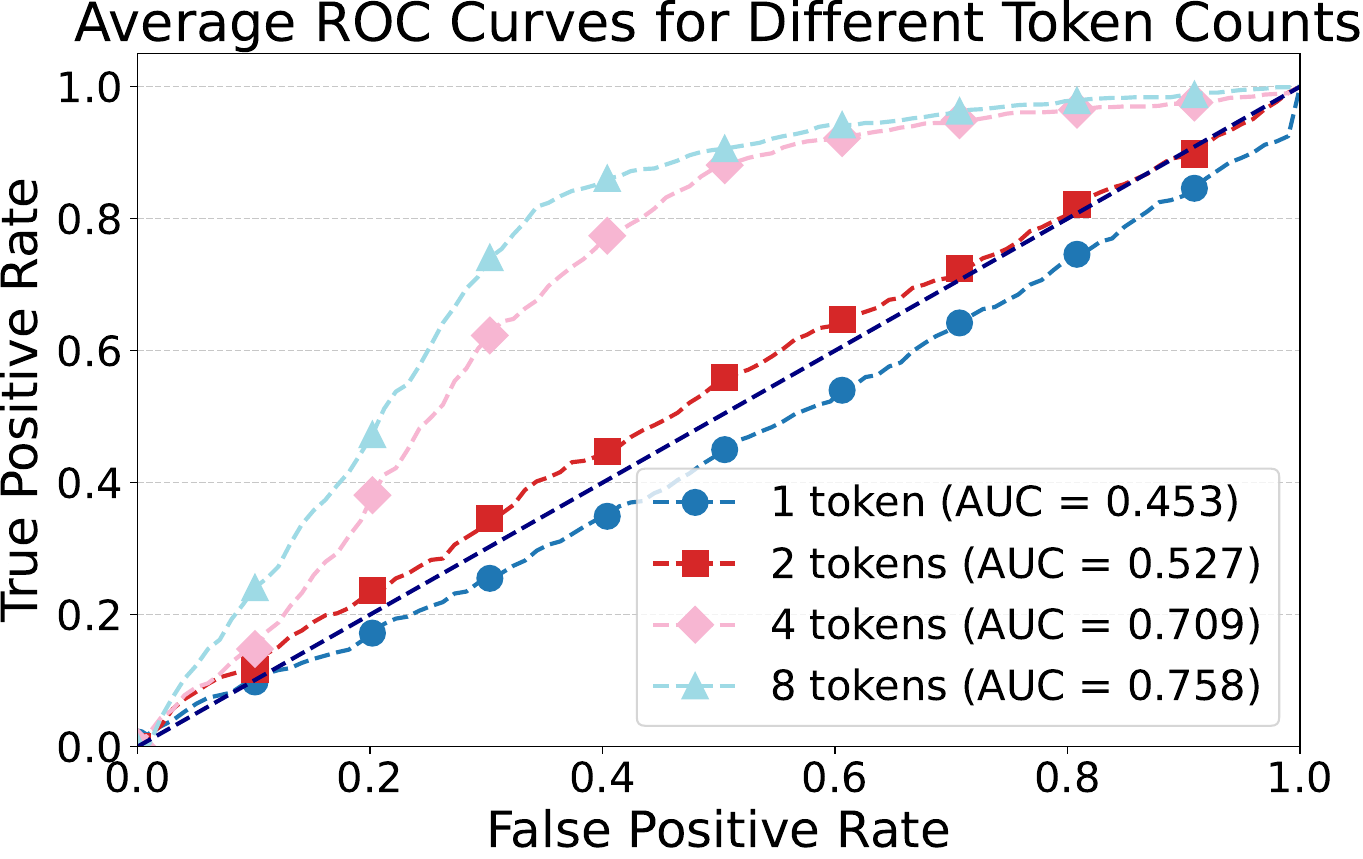}
        \caption{llama2-13B}
        \label{fig:llama70b}
    \end{subfigure}
    \hfill
    \begin{subfigure}{0.33\linewidth}
        \centering
        \includegraphics[width=\textwidth]{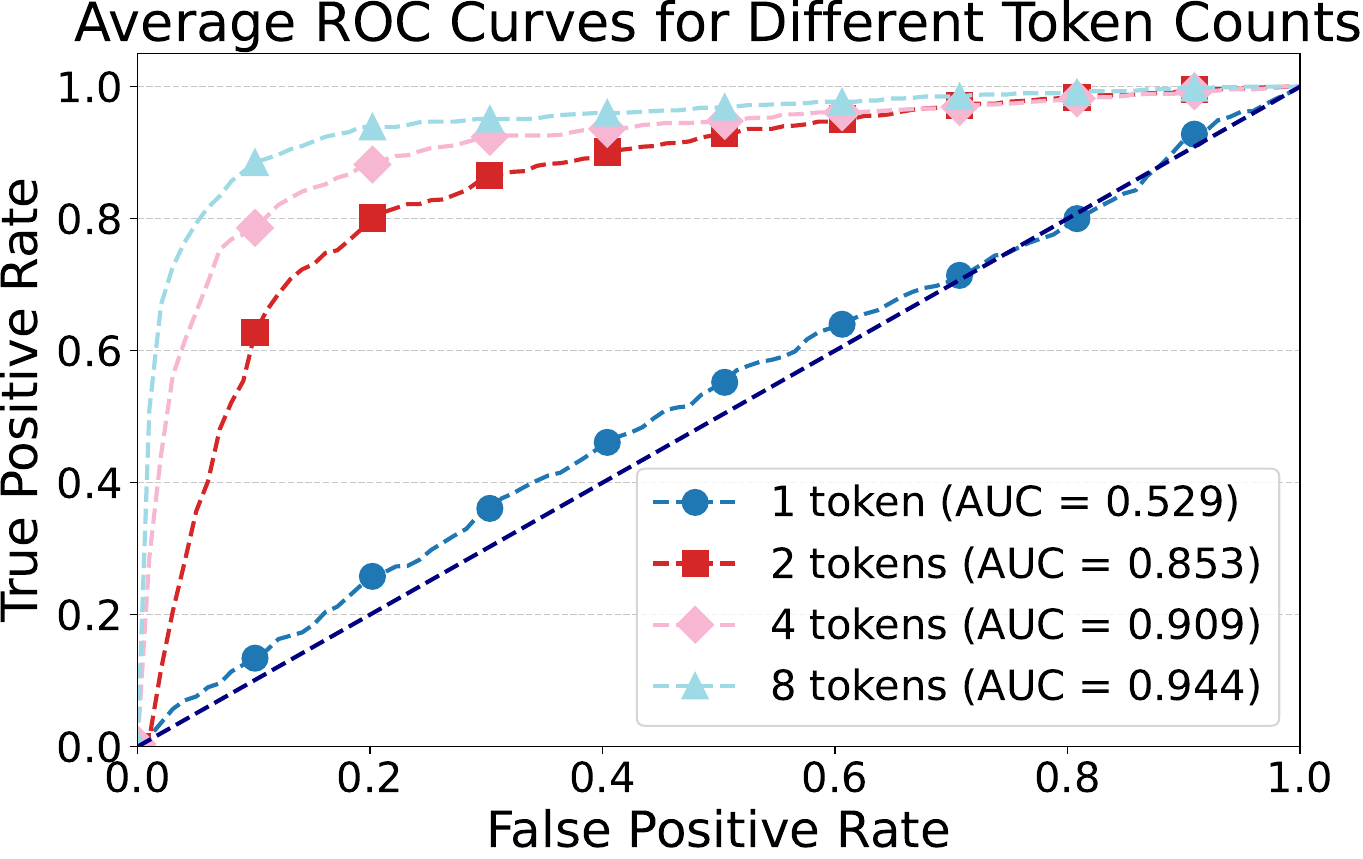}
        \caption{llama2-70B-GPTQ}
        \label{fig:llama70b_2}
    \end{subfigure}

    \caption{Leakage profile of sharing the KV cache. We used three models with different sizes: \texttt{llama3.1-8B-Instruct}, \texttt{llama2-13B} and \texttt{llama2-70B-GPTQ}, and plotted the ROC curves to fingerprint the timing difference when the prompts share the prefix of 1, 2, 4, and 8 tokens respectively.\link{These three graph Maybe all need to be updated. I am afraid we might use other form instead of ROC curves.}}
    \label{fig:leakage_kv}
\end{figure*}

\textcolor{red}{\autoref{fig:distribution} reveals timing differences from a few tokens are negligible, with overlapping latencies in distribution. To separate $T_0$ from other prompt sets, we used ROC curves, which illustrate the trade-off between FPR and TPR, to effectively capture timing differences. \autoref{fig:leakage_kv} demonstrates that longer shared prefixes significantly enhance TTFT caching, with larger models showing a stronger and more stable effect.} 
}

In real-world evaluations of our classifier, we noted that TTFT varies due to factors such as GPU system noise and power fluctuations, weakening the effectiveness of a fixed classification threshold, as shown in \autoref{fig:distribution}. To mitigate this, we first construct a ``miss prompt'' by appending a rare token to the partially recovered prompt (i.e., requests without predicted tokens) and designate the prompt being evaluated as hit or miss as the ``candidate prompt''. Then, within a brief time window, we simultaneously collect TTFT data for both the miss and candidate prompts. By using the difference between these measurements rather than absolute values, we establish a threshold-based classification that remains robust against temporal system variations.

\ignore{To address this challenge, we introduce a dynamic calibration scheme that continuously adjusts the threshold in real time, enhancing the classifier's robustness. The primary mechanism involves simultaneously collecting TTFT data from known requests (i.e., requests without appended predicted tokens) within a brief time window alongside the targeted request's TTFT. These concurrent measurements establish a baseline threshold representing real-time system performance. Based on this baseline, the threshold is dynamically updated at runtime to boost accuracy.}

To illustrate the classifier's effectiveness, consider an example using the \path{Llama-3.1-70B-Instruct-GPTQ-INT4} model under our evaluation settings. We built a classifier to determine whether the last token of prompts hit the KV cache. 
We tested the classifier using 4,000 hits and 4,000 misses drawn from randomly selected system prompts, achieving a TPR of 0.88 and an FPR of 0.10. \hl{To further mitigate noise, we employ multi-sampling by collecting $n$ TTFT samples per token. A token is classified as a hit only if the number of detections exceeds a threshold $k$. This leverages binomial statistics to amplify the TPR while suppressing the FPR.} With $n=10$ and $k=5$, the final classifier attains a TPR of 0.99 and an FPR of 0.003.


\ignore{
\para{Recovering one token}
We previously explained how attackers preserve the system prompt in the KV cache. \autoref{fig:process_PSA} provides a detailed overview of the PSA attack process, focusing on recovering a single token of system prompt. During the offline phase, the attackers make two key preparations. First, they collect system prompts related to their target topic, ensuring semantic similarity, and fine-tune a language model to improve next-token prediction. Second, the attackers deploy a chatbot with a fixed system prompt in the KV cache and send user prompts to the SGLang backend, varying the shared prefix lengths with the system prompt. The returned TTFT variations are used to train a voting classifier, combining XGBoost, random forest, and gradient boosting, as simple thresholding is insufficient to handle system noise such as network latency and GPU load.

During the online phase, after the attackers employ statistical methods to obtain the system prompt's initial tokens and verify them using TTFT, they send prompts to the fine-tuned language model to generate a probability distribution for the next token. Based on this distribution, the attackers randomly sample a candidate token, append it to the current prompt, and construct a new query. As \autoref{fig:distribution} shows, the prefill acceleration for a few tokens creates subtle and overlapping timing differences, increasing false positives when relying on single-sample evaluations. To address this, the attackers send up to $n$ identical requests to the SGLang backend which has the KV cache of the system prompt, records TTFTs. The voting classifier checks if the "Hit" count surpasses the threshold $k$; if so, the token is appended to the user prompt. Otherwise, the token's probability is penalized by halving the sampling probability to reduce the chance of incorrect predictions.


\textcolor{red}{We embedded over 350k diverse system prompts, 80\% of which are 500--1500 characters~\cite{systemprompt-huggingface}, using SBERT \path{all-mpnet-base-v2}. Then we calculated and extracted the largest cluster with cosine similarity over 0.9. From this, 200 prompts were randomly selected as the victim's system prompts, while the rest were provided to the attacker for fine-tuning. We fine-tuned LLaMa-3.1-8B-Instruct on an A100 GPU using LLAMA-Factory's LoRA option~\cite{zheng2024llamafactory} for 20 epochs.}


\textcolor{red}{
The voting classifier relies on maintaining the system prompt's KV cache in the backend. Positive and negative samples are crafted with identical token lengths but differ in their shared prefix length with seperator and system prompt by $x$ tokens. Requests for both sample types are sent to the SGLang Chatbot system powered by the llama2-70B-GPTQ model, with their TTFT values recorded. TTFT differences between these samples guide classifier training, with 10,000 samples per category collected to avoid system noise. }

\link{We will use filter to eliminate some strange data, ensuring that these data are in a stable range. This can be categorized as challenge, please think about this question.}

\link{Optional tasks: warmup the GPU.}

\link{It seems we didn't have a closer look at the case that one-token hit and one-token miss, even we don't know why this amplify the performance, we can always predict for two tokens while recovering one single token. But this is not good, because we can't persuade the reviewers.}
\begin{figure*}[!ht]
    \centering
    \includegraphics[width=0.95\textwidth]{figures/PSA_Design.pdf}
    \caption{Process of PSA.}
    \label{fig:process_PSA}
\end{figure*}

\para{Building the classifier} 
In the \textit{offline phase}, we trained a cluster of classifiers, $classifier_0$ to $classifier_{n-1}$, where each $classifier_i$ ($0 \le i < n$) detects if the $i$-th token hits the cache, given that all preceding tokens have also hit the cache.

We used the llama2-70B-GPTQ model for evaluation. As shown in \autoref{fig:leakage_kv} (b), the timing difference between a cache hit and miss of 1 token is small (AUC=0.52), making it hard to build a robust classifier. Fortunately, we found that the timing difference can be enlarged by appending one dummy token right after the target token, i.e., adding the dummy token as the $i+1$-th token when building $classifier_i$. This may be because the dummy token takes up additional GPU computational resources, thus increasing the timing difference.

\ignore{
\begin{figure}[!ht]
    \centering
    \includegraphics[width=\linewidth]{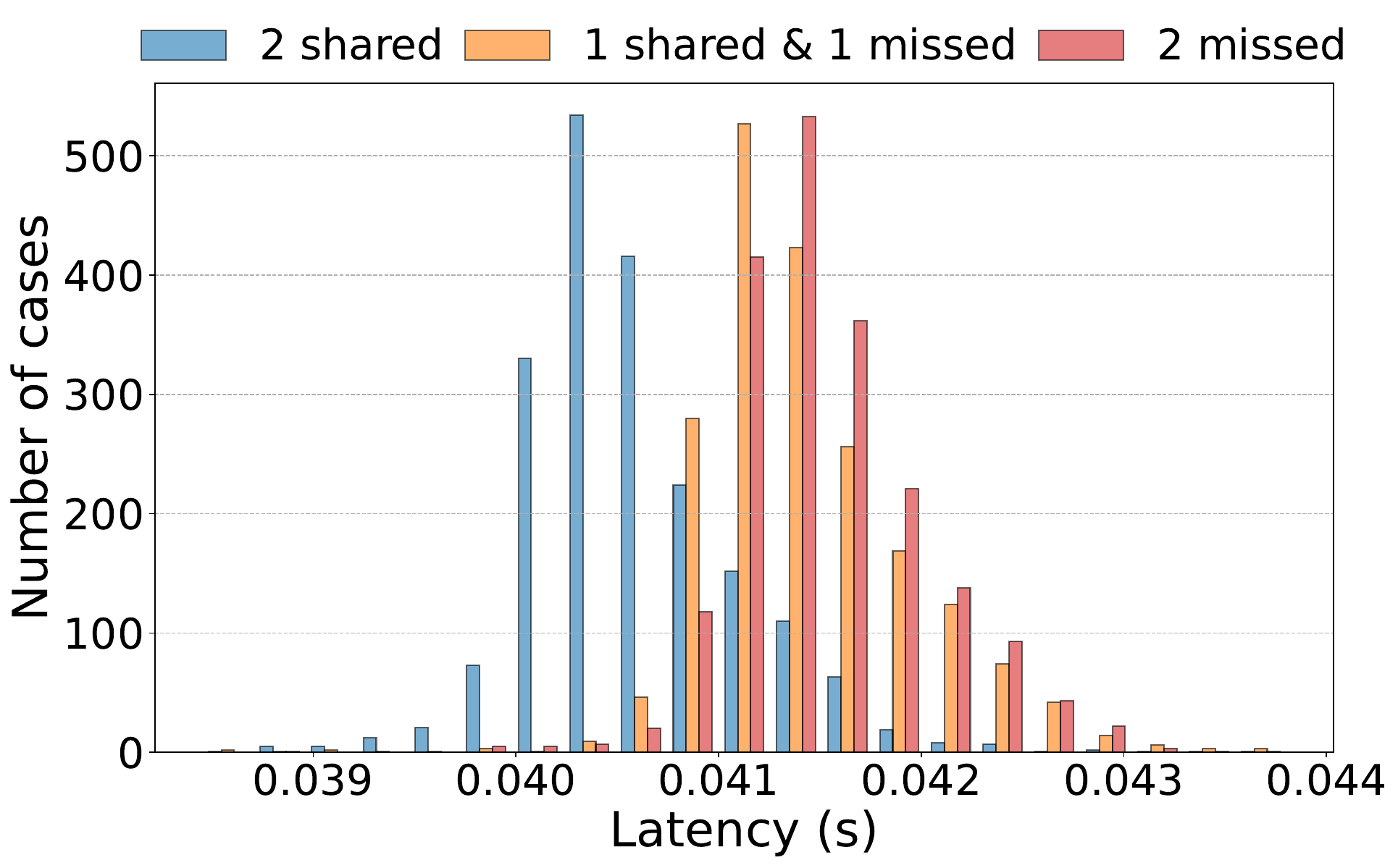}
    \caption{Histogram for matching 0, 1, and 2 tokens when the lengths of the prompts are 2. The cases of matching 2 tokens (i.e., 2 shared) are distinguishable with other 2 cases.}
    \label{fig:histogram}
\end{figure}
}


We now describe the process of generating the dataset used to train $classifier_i$.
We first randomly sampled a system prompt $P$ from the dataset. We then generated 2000 unique sentences of length $i+1$ as negative samples, which all have the first $i-1$ tokens in common with $P$, but differ on the last two tokens. We also generated 2000 unique positive samples, which all have the first $i$ tokens in common with $P$ and a random dummy $(i+1)$-th token.
To collect the latency of the positive and negative samples, we first fed $P$ to the LLM server (SGLang with the llama2-70B-GPTQ model, then sent the sample to the LLM and measured the latency of each sample. Before collecting the latency of next sample, we flushed the KV cache to avoid noises.
Based on the collected latency for both positive and negative samples, 
we utilized the XGBoost algorithm to train the classifier and optimized its hyper parameters using Bayesian optimization. To enhance the performance of the classifier, we adjusted the decision threshold to maximize the difference between the TPR and FPR, while ensuring that the TPR remains high. In our evaluation, the classifiers have a TPR of 55\% and FPR of 42\%.

}

\begin{figure}
    \centering
    \subfloat[(a). Random sample of 10k prompts from the dataset\label{fig:umap_visualization_smaller}]{
        \includegraphics[width=0.46\linewidth]{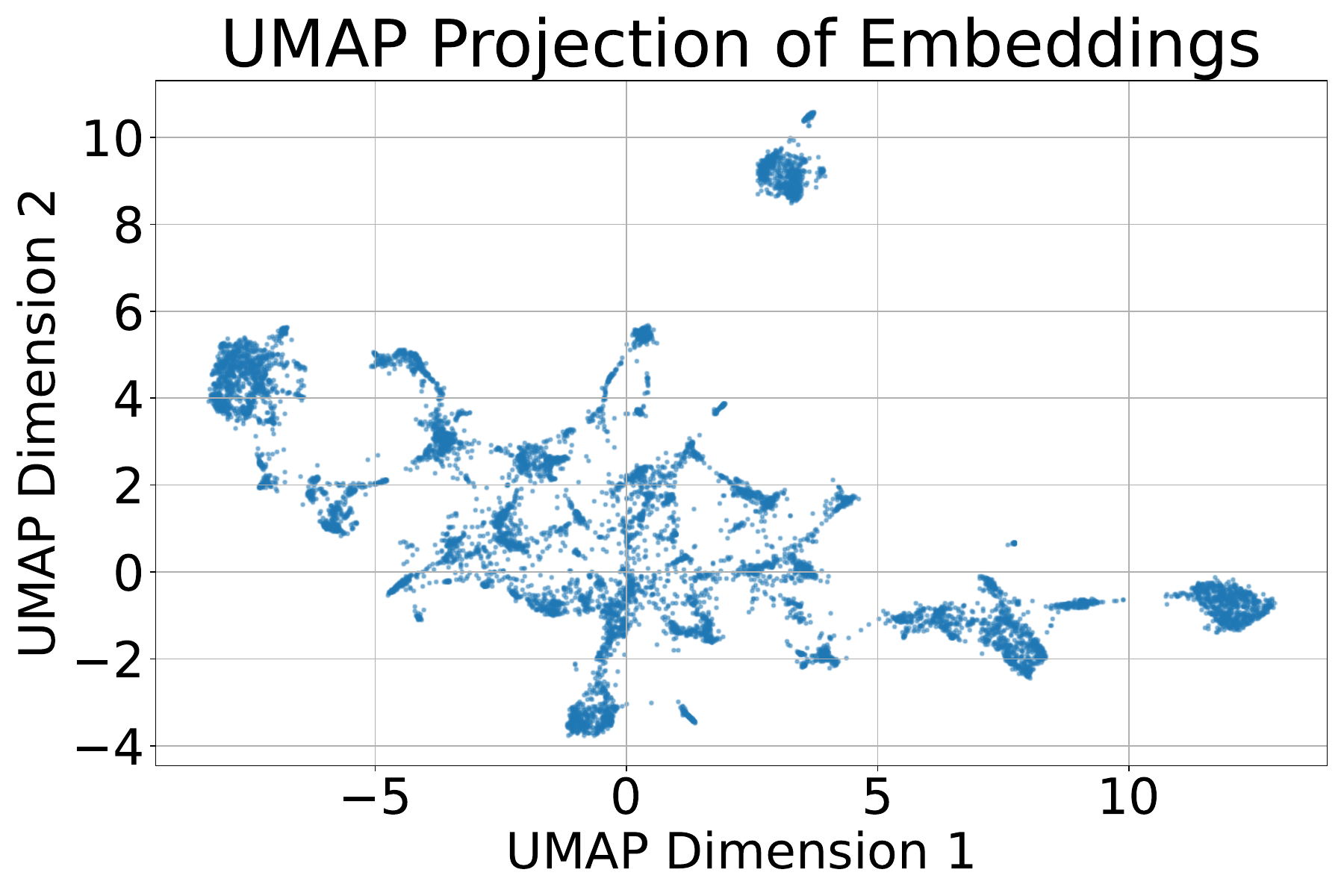}
    }
    \hfill
    \subfloat[(b). Complete dataset (over 350k prompts)
    \label{fig:umap_visualization_bigger}]{
        \includegraphics[width=0.46\linewidth]{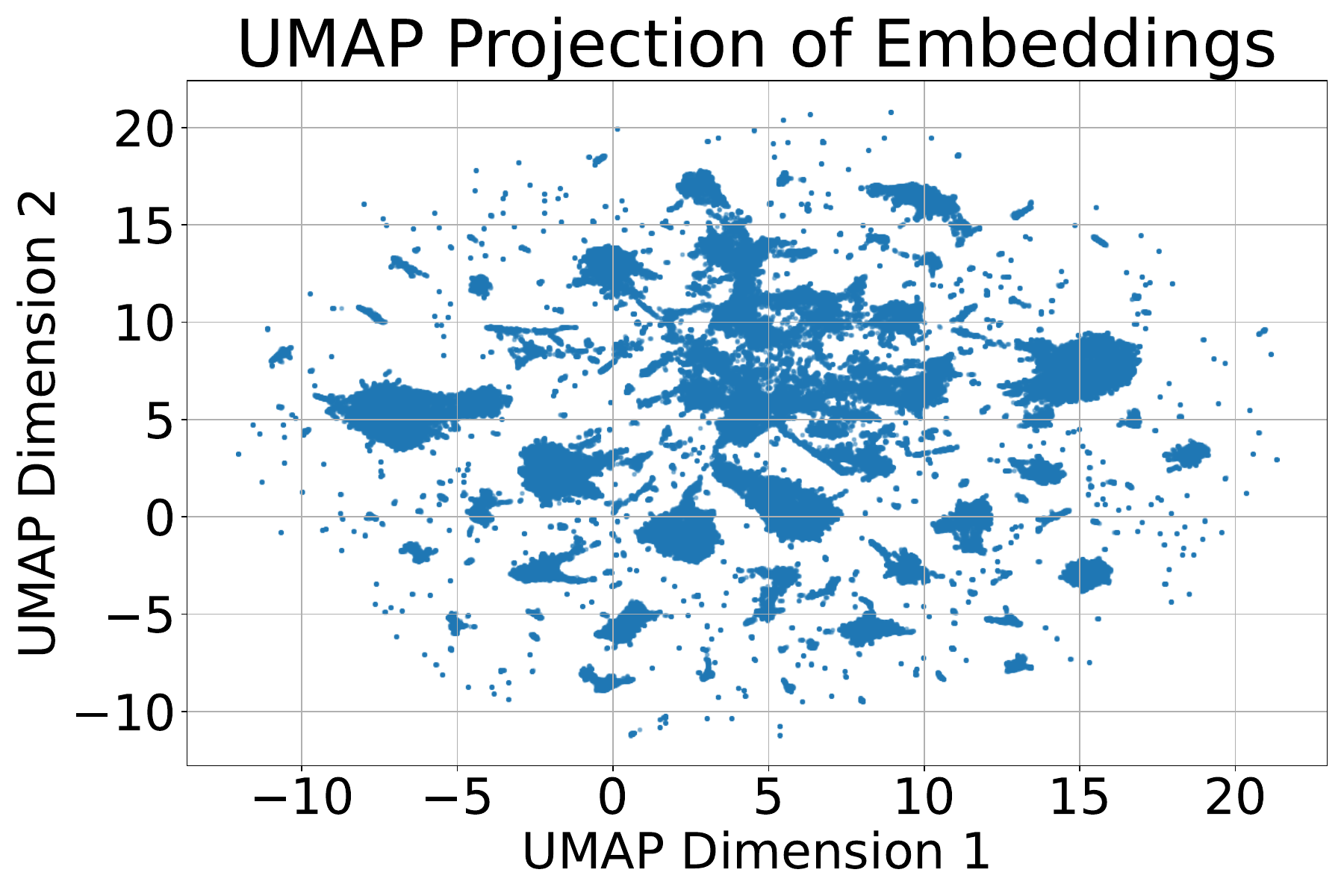}
    }
    \caption{UMAP projection demonstrating the dataset's heterogeneity: As sampling density increases, the visualization reveals expanded dimensional ranges with distinct semantic clusters and isolated points, highlighting the complex multidimensional structure of the prompt dataset.}
    \label{fig:data_heter}
    \vspace{-8pt}
\end{figure}

\para{End-to-end attacks}
We built a local chatbot using the FastChat as the victim application.  
Its backend API server is configured with SGLang v0.3.0, and the model deployed by victim is \texttt{Llama-3.1-70B-Instruct-GPTQ-INT4}. For evaluation, we used the \path{gabrielchua/system-prompt-leakage} dataset, containing over 350,000 synthetic system prompts. We applied \path{Llama-3.1-8B-Instruct} to compute embeddings for each prompt and visualized the results using Uniform Manifold Approximation and Projection (UMAP), which reduces the dimensions of datasets and enables effective visualization of data distribution~\cite{mcinnes2018umap}. UMAP visualization confirmed substantial semantic diversity across this dataset, reflecting the varied topics and styles found in real-world prompt engineering. As shown in \autoref{fig:data_heter}, this semantic heterogeneity increases as the dataset expands, significantly complicating predictive modeling. Consistent with the threat model detailed in \autoref{subsec:threat_model}, we assume the attacker has a public dataset of system prompts that mirrors those of the victim. To reflect real-world conditions—where production systems typically implement a small number of carefully crafted system prompts—we randomly select 200 prompts as the victim's configuration while reserving the remainder as the attacker's training corpus. This experimental design deliberately simulates the ``needle in a haystack'' challenge faced in realistic attack scenarios, where adversaries must identify specific target prompts within a substantially larger heterogeneous corpus. 

\begin{figure}
    \centering
    \includegraphics[width=0.9\linewidth]{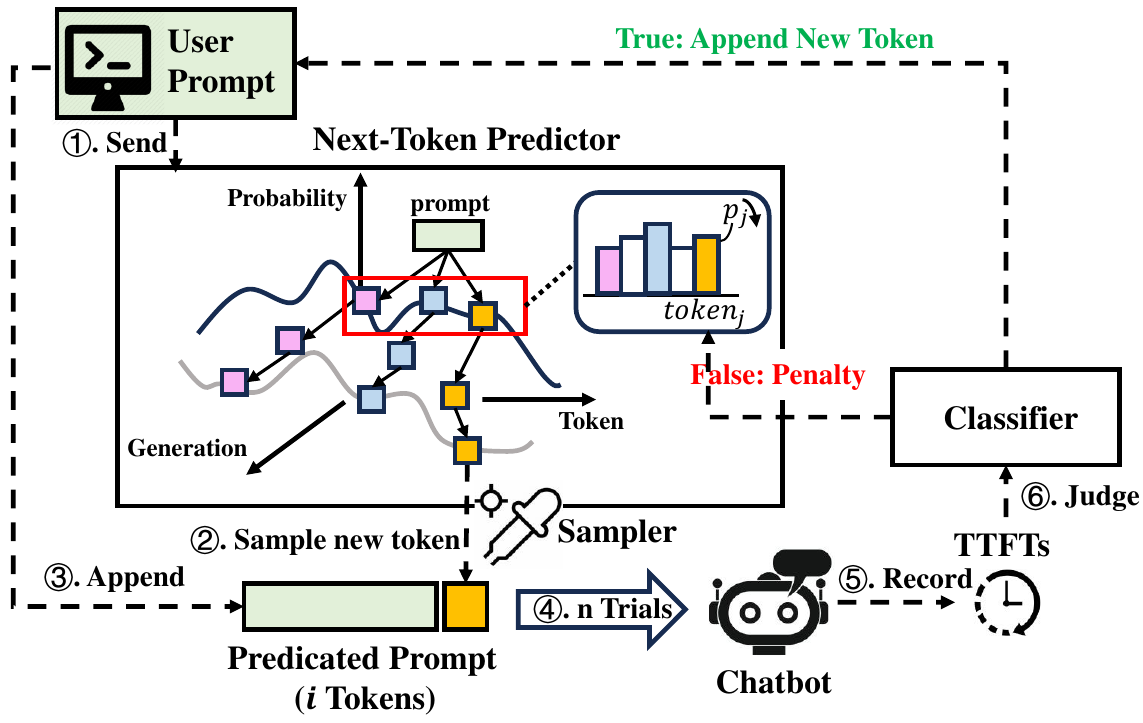}
    \caption{Efficient token-by-token request recovery. 
    }
    \label{fig:searchprefix}
    \vspace{-8pt}
\end{figure}

To streamline the search process, as illustrated in \autoref{fig:searchprefix}, we propose an incremental token-by-token approach to recover the target prompt.  
This approach relies on multiple components: a \textit{classifier} for validation, a \textit{next-token predictor} to estimate token probabilities, and a \textit{sampler} that selects candidate tokens based on the predictor's temperature setting. The \textit{next-token predictor} is fine-tuned on the public system prompt dataset available to the attacker, allowing it to forecast the next token given the already retrieved tokens.  
For each candidate token at position $i$, we feed the partially reconstructed prompt and miss prompt into the LLM to obtain TTFT difference (same method in \autoref{subsec:evalpsa}), which then serves as the input to the $classifier$. If the classifier identifies this token as a cache hit, the token is appended to the prompt; otherwise, a \textit{repetition penalty} is applied by adjusting the probability distribution of the current token (in our implementation, this penalty is applied by halving the sampling probability for an incorrect token in the next round).




In constructing the \textit{next-token predictor}, we adopt \path{Llama-3.1-8B-Instruct} as the base model, chosen as it shares an identical tokenizer with the victim LLM. We fine-tune this model using the attacker's training dataset. Given the partially recovered prompt and the chosen temperature, the predictor generates a probability distribution for the next token, while temperature scaling introduces variability into the prediction. Our predictor is not heavily optimized; it was fine-tuned on a single NVIDIA A100 PCIE GPU for only a few hours with limited training resources. With larger models, more epochs, and bigger batch sizes, the predictor's performance could be further improved.



To obtain TTFT values, the attacker first clears both the victim's and the attacker's requests from the cache. Next, the attacker measures the TTFT for their own request after the victim's system prompt has been cached. Below, we describe how the TTFT is gathered and how caches are flushed.


\noindent\textit{$\bullet$ Timing measurement}.
As shown in \autoref{lst:time}, the attacker begins by issuing a synthesized request containing the targeted system prompt. Once the end-of-sequence token is received in the POST response, a short delay is introduced, ensuring the system prompt resides in the KV cache. The attacker then sends a direct request via a POST call using the predicted prompt, configured to generate only 1 output token. The TTFT is computed as the interval between sending the request and detecting the first non-blank token in the streamed response. 

\begin{figure}
\centering
{\scriptsize
\begin{lstlisting}
def get_ttft(text):
    start_time = time.perf_counter()
    response = requests.post(
        {..., max_tokens=1,}, 
        stream = True
    )
    for line in response.iter_lines():
        if line:
            end_time = time.perf_counter()
            break
    return end_time - start_time
    
def complete(text):
    response = requests.post(..)
    for line in response.iter_lines():
        if line:
            data = json.loads(line)
            if data.get("end_of_sequence", 
                False):  
                break

complete(triggering_prompt) 
time.sleep(0.2)
ttft = get_ttft(predicted_prompt)
\end{lstlisting}
\caption{Code for measuring response latency in PSA.}
\label{lst:time}
}
\end{figure}

\noindent\textit{$\bullet$ Flushing the caches through eviction}.
We observed SGLang provides a $flush\_cache$ API~\cite{flush_cache} that efficiently clears the cache. However, for our end-to-end attack scenario, we chose not to use this API, as it is unlikely to be accessible to attackers in real-world environments. \hl{Instead, we employed a more robust method of evicting the KV cache by issuing batches of irrelevant requests. These irrelevant requests consist of diverse texts with rare prefixes to rapidly fill the GPU memory and trigger cache eviction.} Under default SGLang settings, sending 15 such requests (each containing about 200 tokens) was sufficient to trigger eviction in about 5 seconds. This approach proved successful in 100\% of our 10,000 tests.

\ignore{
To improve the classification accuracy, we need to collect a set of TTFT values for the classifier using the same candidate prompts in $n$ repetitive trials. For this purpose, we explicitly flush the cache and begin our timing measurements after the victim prompt is served in the LLM server.
We consider the candidate prompt to have hit the cache if the number of hits returned by the classifier exceeds $k$ out of $n$ trials. In our evaluation, we set $n = 100$ and $k = 50$, and the accuracy of correctly identifying a prefix cache hit is more than 86\%, and the FPR is about 5\%. 
}



We evaluated both the token recovery accuracy and the average number of queries required per token. The results indicate a success rate of 89.0\%, an FPR of 0.04, and an average of 5.64 guesses with 112.72 attack queries needed per recovered token when $n=10$. The corresponding cost per token recovery is approximately \$0.16 using OpenAI o1, \$0.012 with OpenAI o3-mini, and \$0.001 with Deepseek-Chat (as of March 5, 2025).
{Out of 200 victim prompts, we successfully recovered an average of 11.32 tokens for the top 100 prompts, and 17.84 tokens on average for the top 50 prompts. The maximum number of tokens recovered for a single prompt was 81, achieved with just 519 total guesses.}
Notably, we limited the token prediction attempts to a maximum of 80 per position, ceasing further attempts if the correct token was not identified within these trials. Consequently, the primary constraint in reconstructing entire system prompts stems from the precision of the predictor rather than the accuracy of the classifier---an aspect that falls outside the scope of this study.
In real-world attacks, the attacker can recover additional tokens by deploying a more advanced next-token predictor and increasing the maximum number of attack queries. {A demo for the attack is presented on our website~\cite{Demo}.}

\para{\hl{Discussions}}
\hl{Currently, we assume that the attacker has access to a public dataset of system prompts that can be used to fine-tune the next-token predictor. When a generic (i.e., untuned) predictor is employed, it tends to generate more false predictions, which in turn requires additional verification queries to distinguish genuine cache hits from prediction errors (i.e., to reduce false positives). This added overhead increases the effort and resources needed for an attacker to recover tokens. However, these challenges raise the cost of the attack rather than eliminating the underlying vulnerability. The core timing side channels we identified in KV-cache mechanisms remain exploitable regardless of predictor quality. With sufficient resources and persistence, attackers can still leverage these timing differences to recover tokens, albeit at a higher cost. To evaluate how predictor quality affects attack efficiency, we conducted the attack using the untuned} \path{Llama-3.1-8B-Instruct} \hl{model as the generic predictor. 
Our results show a recovery rate of 69\%, with an average of 163.48 guesses and 5,231.30 attack queries per token (here we set $n=16$ and $k=8$ to reduce false positives, and limit the token prediction attempts to a maximum of 1,000 per token), and an FPR of 1.5\%. Further analysis reveals that, compared with fine-tuned predictors, generic predictors struggle to correctly predict certain rare tokens, leading to a lower recovery rate. Moreover, because generic predictors are less likely to quickly guess the target token, the average number of guesses rises sharply (from 5.64 to 163.48). This further demonstrates that our side-channel-based classifier is robust to false positives, as the attack maintains a low overall false positive rate (1.5\%) even under significantly increased guessing attempts.}

\ignore{
\begin{table*}
    \centering
    \caption{Examples of recovered system prompts, including the number of attack queries and recovered tokens. Travel planning-related prompts are selected to demonstrate the ability to recover diverse expressions.}
    \label{tab:systemprompt}
    {\small
    \begin{tabular}{C{0.5cm} | C{7.1cm} | C{6.2cm} | C{2.3cm}}
      \hline
   \textbf{No.} & \textbf{Targeted system prompts} & \textbf{Recovered system prompts} & \textbf{\#queries, \#recovered tokens}\\ \hline
   1 & In your role as a dedicated travel itinerary assistant, you will craft well-organized travel plans based on user preferences and interests. Gather information from the user's inputs such as destinations, travel dates, and preferred activities to formulate a comprehensive itinerary. The output should include: dates, activities planned for each day, estimated costs, and important local information such as culture or tips. Emphasize clear, organized, and insightful suggestions for leisure and significant stops based on the given inputs about destination and user interest. &  In your role as a dedicated travel itinerary assistant, you will craft well-organized travel plans based on user preferences and interests. Gather information from the user's inputs such as destinations, travel dates, and preferred activities to formulate a comprehensive itinerary. The output should include: dates, activities planned for each day, estimated costs, and important local information such as culture or tips. Emphasize clear, organized,  $\cdots$ & 10,260/81 \\ \hline
   2 & You are programmed to function as a travel itinerary planner focusing exclusively on creating unique travel experiences. Provide tailored itineraries for destinations worldwide. Include major attractions $\cdots$ & You are programmed to function as a travel itinerary planner focusing exclusively on creating unique travel experiences. Provide tailored itineraries for destinations worldwide. $\cdots$ & 3,760/26 \\ \hline
   
   3 & Imagine you are a travel itinerary planner specializing in creating unique and personalized travel experiences. Your role is to craft itineraries that cater to the diverse interests and needs of travelers. These itineraries should $\cdots$ & Imagine you are a travel itinerary planner specializing in creating unique and personalized travel experiences. Your role is to craft itineraries that cater to the diverse interests and needs of travelers. $\cdots$ & 1,320/32\\ \hline
    \end{tabular}
    }
\end{table*}
}

\ignore{
\begin{table*}
    \centering
    \caption{Examples of recovered system prompts, including the number of attack queries and recovered tokens. We only listed travel planning-related prompts to demonstrate the PSA's ability to recover diverse expressions.}
    \label{tab:systemprompt}
    {
    \begin{tabular}{C{0.4cm} | C{13.4cm} | C{2.5cm}}
      \hline
   \textbf{No.} &  \textbf{Recovered system prompts} & \textbf{\#queries, \#recovered tokens}\\ \hline
   1 &  In your role as a dedicated travel itinerary assistant, you will craft well-organized travel plans based on user preferences and interests. Gather information from the user's inputs such as destinations, travel dates, and preferred activities to formulate a comprehensive itinerary. The output should include: dates, activities planned for each day, estimated costs, and important local information such as culture or tips. Emphasize clear, organized,  $\cdots$ & 10,380/81 \\ \hline
   2 &  You are programmed to function as a travel itinerary planner focusing exclusively on creating unique travel experiences. Provide tailored itineraries for destinations worldwide. $\cdots$ & 3,760/26 \\ \hline
   
   3 &  Imagine you are a travel itinerary planner specializing in creating unique and personalized travel experiences. Your role is to craft itineraries that cater to the diverse interests and needs of travelers. $\cdots$ & 1,360/32\\ \hline
    \end{tabular}
    }
\end{table*}
}

\ignore{
for the system\_prompt dataset: 72 \%

Reason: Uncertainty of the tokenizer time for different token and different templates of sentences. So some 'different' senteces show a very bad result in prediction.

Defense: Diversity => against both the fine-tuned GPT predictor, and the Classifier. We can based on our case.
}




\subsection{Analysis on PNA}
\label{subsec:evalpna}

\noindent\textbf{Attack setup.}
In the PNA attack, we observe that not all user queries contain sensitive data. An attacker typically ignores generic requests like ``What's the weather today?'' and focuses on those specific revealing private information. For example, a prompt like ``Draft a travel plan for me to Rome in September for a 3-day stay at the Hilton hotel'' could expose personal and location details. We assume both the victim and the attacker are normal users of a chatbot service that uses a semantic cache in its backend. By identifying high-risk queries, the attacker can exploit timing side channels to recover private information from other users' requests.  
For this purpose, the attacker could compile a list of privacy-related prompts from online sources (\autoref{tab:templates}). The attacker has obtained a list of private attributes, and aims to discover connections among these private attributes, e.g., whether a specific user is linked to a particular medical condition.

\begin{figure}
    \centering
    \includegraphics[width=0.9\linewidth]{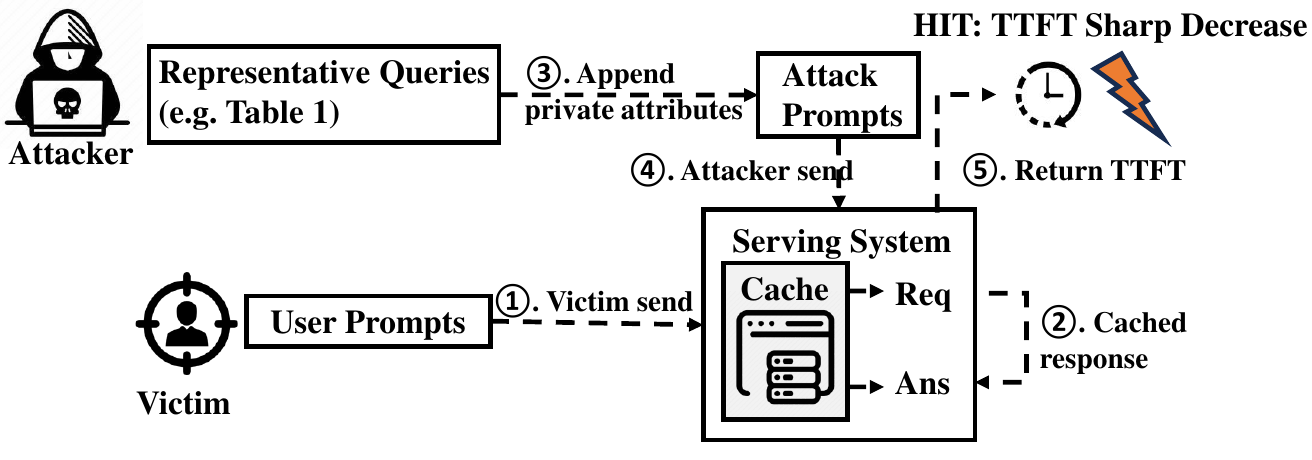}
    \caption{Peeping Neighbor Attacks.}
    \label{fig:semantic_attack}
    \vspace{-10pt}
\end{figure}

\autoref{fig:semantic_attack} illustrates the PNA steps.  
When semantic caching is used, the LLM stores victim requests and serves cached responses for similar queries. \hl{To exploit this channel, the attacker creates requests containing private attributes and monitors TTFT. Since cache hits retrieve responses directly from local storage and bypass LLM inference, they produce a sharp TTFT reduction (from 5 seconds to under 1 second), which far exceeds the timing variations caused by network jitter. Consequently, by monitoring such timing differences, an attacker can infer semantic matches with a victim’s cached request and potentially expose private data.}



In practice, users may express the same intent through varied phrasing and may embed diverse private attributes---such as personal names or medical conditions---within their queries. Our objective is to determine how these different attributes affect semantic similarity and whether the resulting variations exceed a threshold described in \autoref{subsec:pna_intro}. If so, such differences can be detected, enabling the attacker to determine whether the victim's request carries particular attributes, even when the victim rephrases the content.

\begin{table*}
    \caption{Examples of user prompts that contain private attributes.}\label{tab:templates}
    \centering
    {  
    \begin{tabular}{C{2.3cm} | C{14.6cm}}
      \hline
        \textbf{Use cases} & \textbf{Prompts} \\ \hline
       Healthcare  & Compose a meeting agenda for an interdisciplinary team discussing the treatment plan for \hll{[Name]} with \hll{[medical condition]}. \\ \hline
        Travel planning & I'm \hll{[flying/driving]} to \hll{[destination]} with \hll{[Name]} for a leisurely trip, and we'll be staying at \hll{[hotel]} for \hll{[number of days]}. Can you create a comprehensive packing list organized by category?\\ \hline
        Business planning & Act as an expert business plan writer, and help me generate a product and services section of my \hll{[business type]} business plan. My company is \hll{[business]} called \hll{[business name]} that specializes in \hll{[USP or specialization]}. \\ \hline
        Performance review & I'm preparing for the annual performance review of an employee named \hll{[Name]}. \hll{[Name]}'s role involves \hll{[roles]}. Draft a performance review for \hll{[Name]} and suggesting improvements in \hll{[area of improvement]}. \\ \hline
        E-mails & Draft an e-mail to the \hll{[company]} on \hll{[subject]}. \\ \hline
        Cover letter & Write a conversational cover letter for \hll{[Name]} for a job application as a \hll{[position]} at \hll{[company]}. \\ \hline
        Out-of-office message & Write a short out-of-office message. Reason: \hll{[vacation]}. Dates: \hll{[month and dates]}. Person to contact in an emergency or for immediate help: \hll{[name]} at \hll{[email address]}. \\ \hline
    \end{tabular}
    }
\end{table*}

In our experimental study, we chose LangChain as the LLM serving framework, given its widespread adoption and extensive application ecosystem. The local chatbot system was built using LangChain and integrated with GPTCache for semantic caching. We used \path{gpt-3.5-turbo} API as the backend LLM, and MedQuAD~\cite{MedQuAD-website} as the evaluation dataset. For evaluating semantic similarity, we employed  $OnnxModelEvaluation$ method in GPTCache, which leverages the \path{albert-duplicate-onnx} model.


\ignore{
\textcolor{red}{
We used the albert-base-v2~\cite{onnx_albert} model in ONNX~\cite{onnx-github} to compute semantic similarity, striking a balance between accuracy and efficiency for robust prompt analysis. While the default SearchDistanceEvaluation ~\cite{searchdistance} method is faster and maintains attack efficacy, it is more likely to misclassify unrelated sentences as similar, leading to irrelevant responses that could compromise the system's usability.}
}

\begin{figure*}
    \centering
    \includegraphics[width=\textwidth]{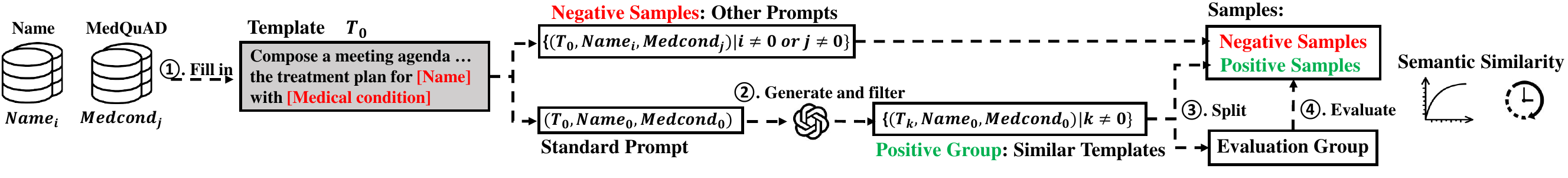}
    \caption{Evaluating semantic leakage  of private attributes.}
    \label{fig:Semantic}
\end{figure*}

\para{Characterizing the leakage}
\autoref{fig:Semantic} outlines our evaluation steps. We illustrate the process with a query template ``Compose a meeting agenda ... for \hll{[Name]} with \hll{[medical condition]}'', denoted as \(T_0\). Here, both the name and medical condition are treated as private attributes.

\textit{Step 1.} We randomly sampled 10 names from the Python package \path{names-dataset} (\(Names\)) and extracted 10 medical conditions (\(Medconds\)) from semantically unrelated Q\&A pairs in MedQuAD dataset using GPT-3.5-turbo.
We then randomly chose 1 name (\(Name_0\)) and 1 medical condition (\(Medcond_0\)) as the private attributes to be recovered; the remaining pairs serve as the negative group (described below).


\textit{Step 2.} Since real-world users may phrase the same content differently, we generated multiple sentences that share the same semantics as \(T_0\). Specifically, we asked GPT-3.5-turbo to paraphrase \(T_0\) into \(n\) variations \(\{T_1, \dots, T_n\}\), each filled with \(Name_0\) and \(Medcond_0\).
Following these steps, we created two sample sets:
(1) \textit{Positive Group}: Sentences that are semantically similar to \(T_0\), all containing \(Name_0\) and \(Medcond_0\). Formally, \(\{(T_i, Name_0, Medcond_0)\mid i=1,\dots,n\}\).
(2) \textit{Negative Group}: The original template \(T_0\) populated with other names or medical conditions. Formally, \(\{(T_0, Name_i, \\ Medcond_j)\mid i\neq 0 \text{ or } j\neq 0\}\).

\textit{Step 3.} We split the positive group into two subsets. The first (20\% of the samples) is designated as the ``positive'' reference set, while the remaining 80\% forms the evaluation set. We also ensure that the evaluation set's positive and negative samples are equal in size. Finally, we compute semantic similarities between the evaluation samples and both the positive and negative groups, yielding two respective similarity distributions.





\begin{figure}
    \centering
    \subfloat[(a). The ``name'' and ``medical condition'' in the negative group are both different from the positive group.\label{fig:semantic_roc}]{
        \includegraphics[width=0.45\linewidth]{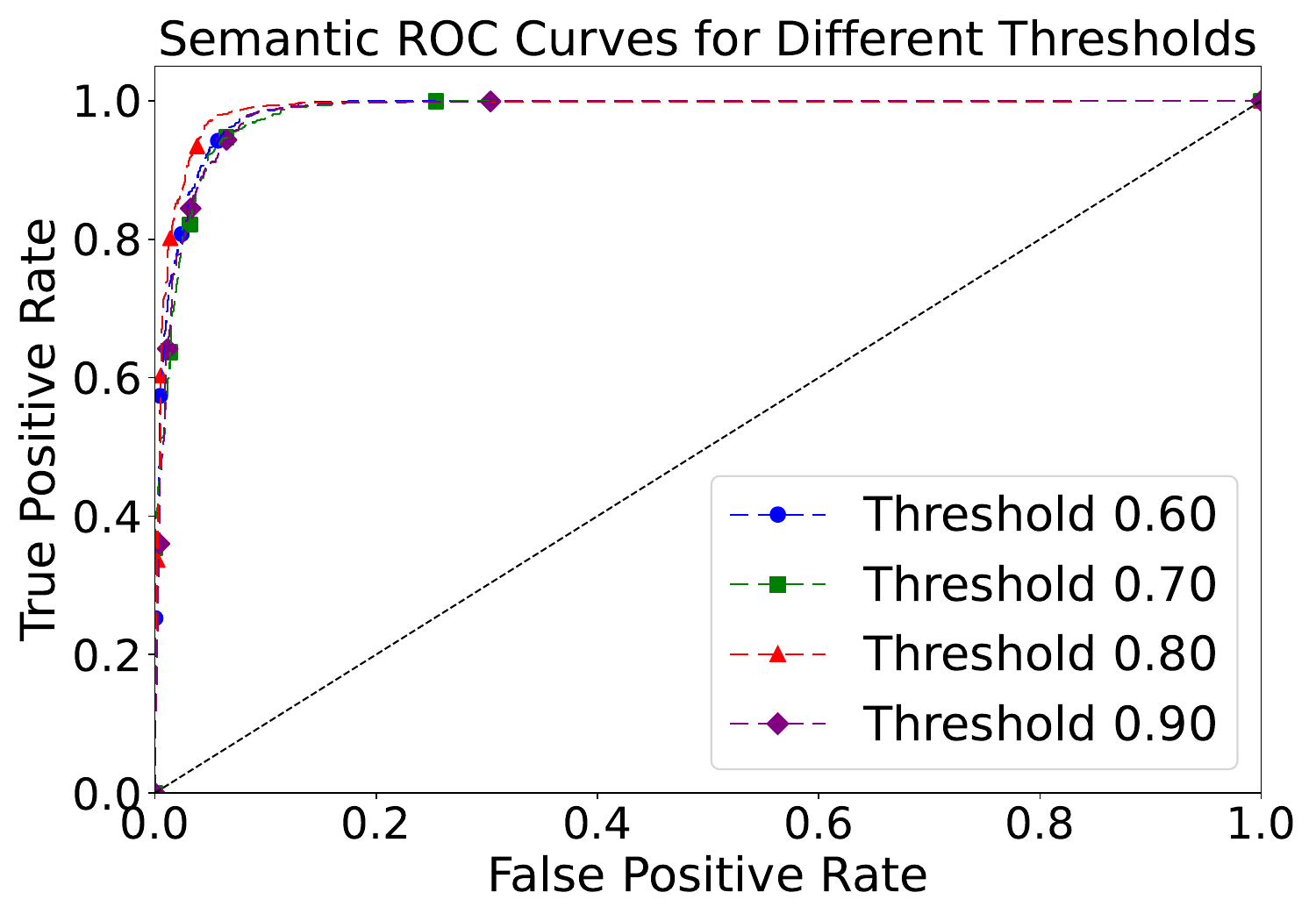}
    }
    \hfill
    \subfloat[(b). Either the ``name'' or ``medical condition'' in the negative group is different from the positive group.\label{fig:semanticone_roc}]{
        \includegraphics[width=0.45\linewidth]{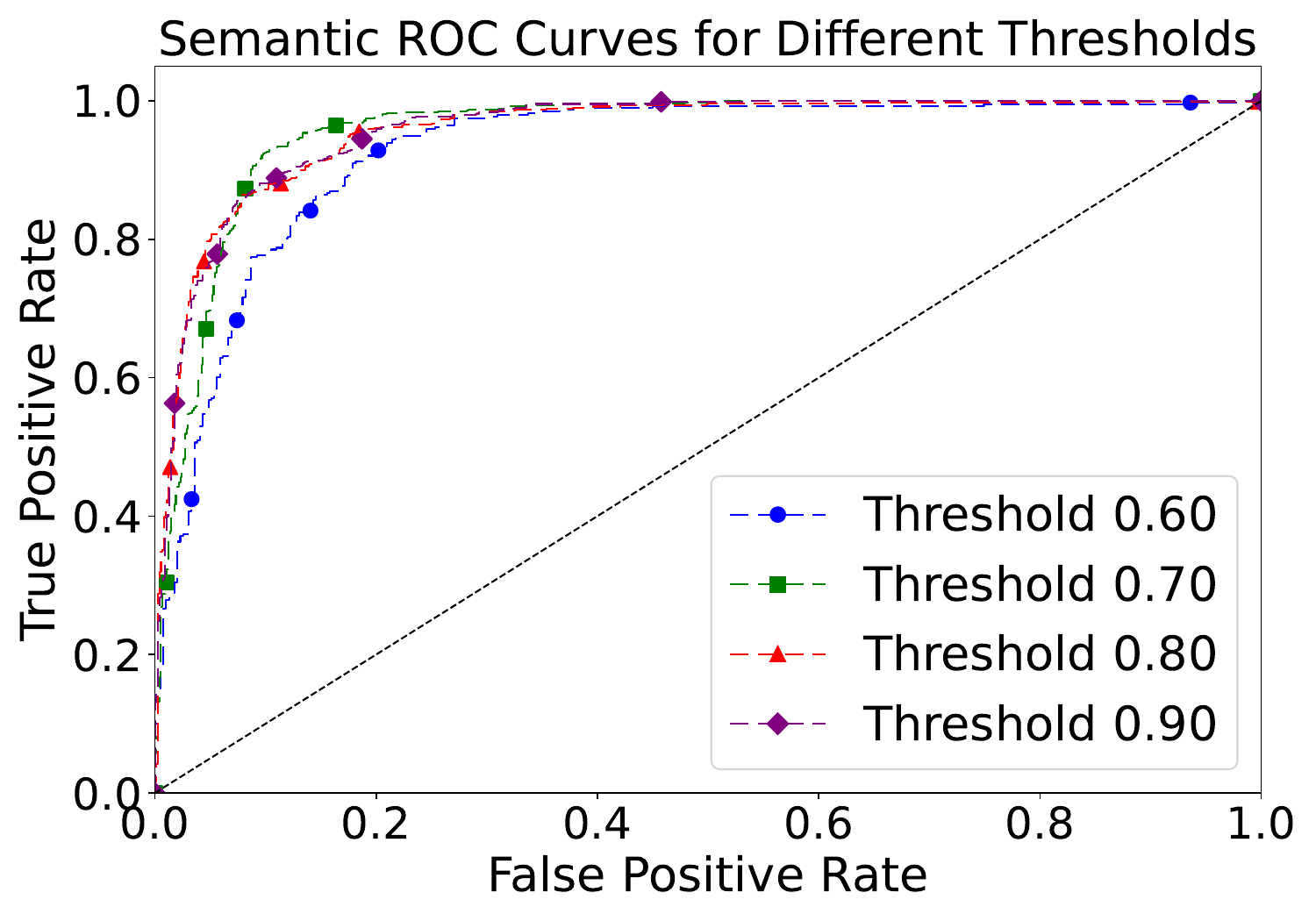}
    }
    \caption{Leakage profile of semantic cache sharing. We plotted the ROC curve to fingerprint the relationship between the similarity vectors of the positive and negative groups.}
    \label{fig:semanticsim}
    \vspace{-8pt}
\end{figure}

We tested similarity thresholds from 0.6 to 0.9 (the default value in GPTCache is 0.8). Figure\autoref{fig:semantic_roc} shows the Receiver Operating Characteristic (ROC) curves for the positive and negative similarity distributions, revealing a clear separation between the two. At the default threshold of 0.8, a TPR of 0.95 can be achieved with an FPR below 0.1.
We also examined cases where only one private attribute (either \(Name_0\) or \(Medcond_0\)) matched. Here, the negative group consists of sentences with only one correct private attribute, while the positive group remains the same. Figure\autoref{fig:semanticone_roc} shows the semantic distinctions remain substantial: at the default threshold of 0.8, a TPR of 0.85 corresponds to an FPR under 0.1.

\ignore{
\begin{itemize}
    \item First, we select a template $T_0$ to be filled in later.
    \item Second, we select 10 different Names from the \textit{NameDataset} package in python. And randomly select 10 semantic unrelated Q\&A pairs from the dataset. Besides, we use \path{gpt-3.5-turbo} to summarize the medical condition from the question. We then choose $Name_0$ and $Medical_0$ as our privacy, and form the seed $(T_0, Name_0, Medical_0)$, while other pairs will be stored in keywords.
    \item Third, to obtain the template cluster, we paraphrase the template ($T_0$) without changing the keywords using \path{gpt-3.5-turbo}. After filling in new template $\{T_1,..,T_n\}$ with $Name_0$ and $Medical_0$, we utilize the threshold of GPTcache to filter in semantic similar sentences "seeds". We randomly choose 80\% as our positive samples, the remaining 20\% as our evaluation data.
    \item Next, we use the original template $T_0$, fill in with different names and medical conditions as the negative samples. The number of the negative samples equals to the positive samples, and can be represented as $NS$, which is the subset of $\{(T_0, Name_i, Medical_j)|i \neq 0 \lor j \neq 0\}$.
    \item Finally, using the same threshold, we calculate the similarity between our seeds and the samples, and obtain the similarity matrix at the end. In order to enhance the generalizability of the results and increase the sample size, we conducted the experiment 50 times and plotted ROC curves based on the relevant data.
\end{itemize}
}






\para{End-to-end attacks}
We consider a typical open-world scenario in which a victim user requests healthcare assistance from an LLM. For instance, the user might submit a query with semantics similar to the template ``compose a meeting agenda...'' shown in \autoref{tab:templates}, but with various names and medical conditions. The user may also send queries unrelated to the targeted request. To simulate this, we model the user's queries as follows:

\vspace{2pt}\noindent$\bullet$\textbf{~Type-1 (true samples)}: Queries with the specific name (e.g., ``Alice'') and specific medical condition (e.g., ``heart disease'').

\vspace{2pt}\noindent$\bullet$\textbf{~Type-2 (false samples)}: Queries that use the same name as the true samples (e.g., ``Alice'') but feature different medical conditions, such as ``diabetes'', ``hypertension'', or ``asthma''.

\vspace{2pt}\noindent$\bullet$\textbf{~Type-3 (false samples)}: Queries with the same medical condition as the true samples (e.g., ``heart disease'') but with different names.

\vspace{2pt}\noindent$\bullet$\textbf{~Type-4 (false samples)}: Unrelated queries.

We assume that the attacker targets the private attribute associations in Type-1 queries. The victim can freely choose different paraphrases while preserving the same underlying semantics. To simulate this, we configure the victim to send five random requests per round: one Type-1 query (true sample) and four false samples (one Type-2, one Type-3, and two Type-4 requests).  
To measure the effectiveness of the attack, we use the TPR to assess how successfully the attacker retrieves private attributes from Type-1 requests. We also measure separate FPRs to capture how often the attacker misclassifies each of the three false sample types as positive.



To perform effective end-to-end attacks on the semantic cache, we must eliminate noise from the attacker's own cached requests.
We address this by clearing the cache after each attack round: under GPTCache’s default settings, sending 1,000 semantically unrelated requests suffices to remove any leftover cache entries. \hl{These requests were pre-extracted from multi-domain articles. Their semantic irrelevance was validated using SBERT}~\cite{sbert_model}\hl{, considering pairs with a similarity score below 0.3 as semantically unrelated.}
In this scenario, we assume an attacker aims to determine whether a particular user (e.g., ``Alice'') is associated with a specific medical condition (e.g., ``heart disease''). Since the victim may use different phrases, the attacker issues multiple requests to enhance coverage and boost the TPR. However, increasing the total number of requests also raises the risk of false positives, especially if new requests strongly resemble earlier ones. To mitigate this issue, the attacker prioritizes \textit{representative requests} in the request space, thus increasing the overall number of queries while minimizing interference among them.


\begin{figure}
    \centering
    \includegraphics[width=0.75\linewidth]{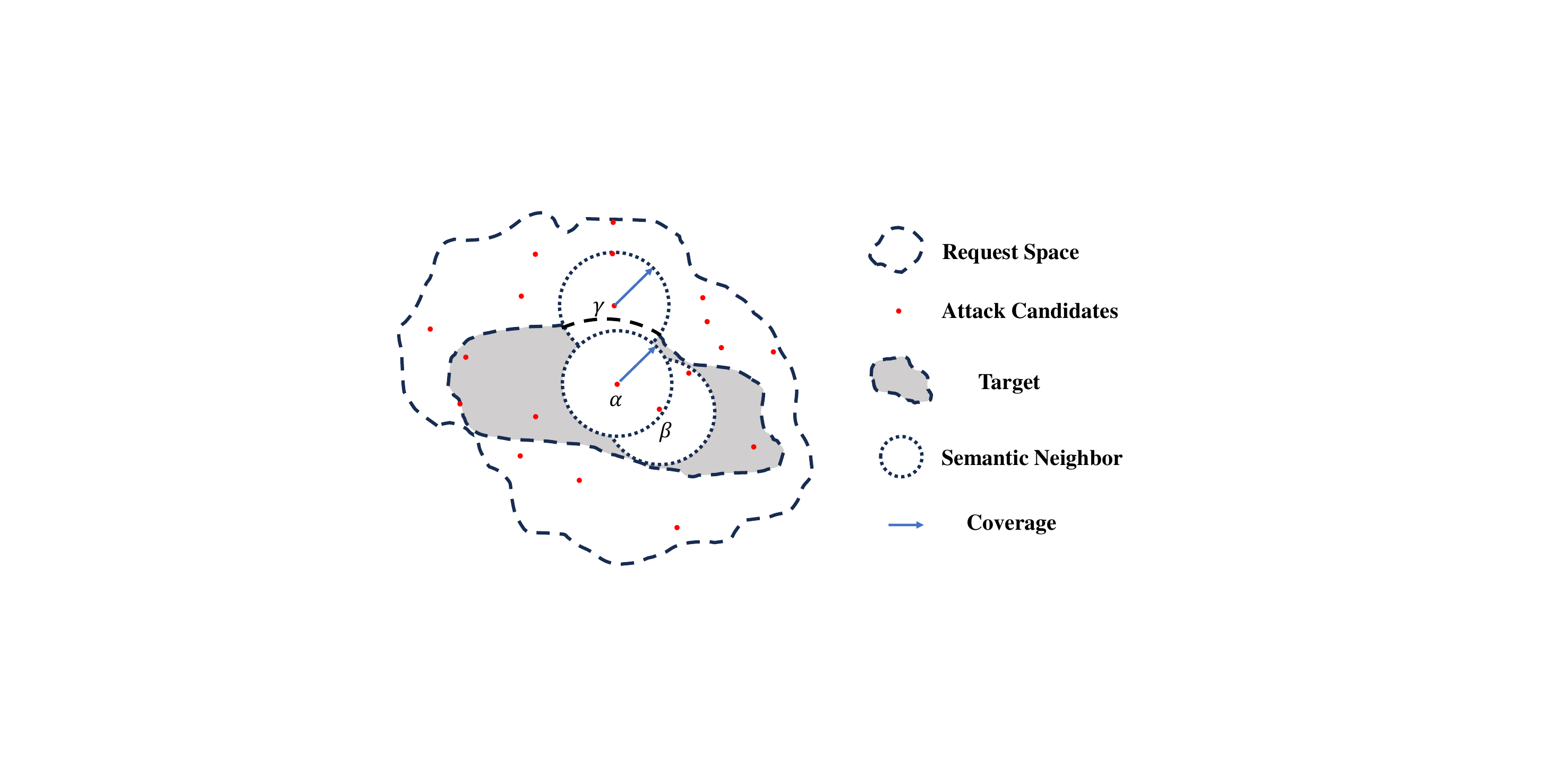}
    \caption{The greedy search strategy for PNA.
    }
    \label{fig:semantic_space}
\end{figure}

\textit{Representative} requests are those that most closely approximate the rest of the request space. To identify them, we use the \path{distilbert-base-uncased} model by default to generate embeddings and then compute the L2 distance between these embeddings. We then sort the requests by their L2 distances; those with the smallest distances are deemed the most representative and selected as attack requests to maximize coverage.
To further expand coverage, we incorporate \textit{orthogonal requests}—requests that are semantically distinct from one another. This reduces the chance that semantic overlaps among the attacker's own requests degrade accuracy in identifying victim requests.  
We classify a cache access as a hit if at least one of the attacker's requests triggers a cache hit in the timing channel. Although this strategy boosts coverage, it also raises the FPR, necessitating a careful balance.


\autoref{fig:semantic_space} depicts our greedy search algorithm, which iteratively selects the \textit{most representative} candidates (e.g., \(\alpha\)) within a semantically similar target space, unless they are overly similar to previously selected ones (e.g., \(\beta\)). This process continues until no suitable candidates remain or when the FPR exceeds a predefined threshold \(\sigma\).

\ignore{
According to \autoref{fig:semantic_space}, while selecting more prompts can increase coverage, it does not guarantee improved accuracy. Thus, we select several parameters to judge whether we should halt or continue our experiment.
\begin{itemize}
    \item $\Delta TPR$ and $\Delta FPR$: represent the change of TPR and FPR after selecting one more attack candidate.
    \item $\Delta Times$: represent the increment of test times after selecting one more attack candidate.
\end{itemize}
}



In the evaluation, each of the 4 victim request types was tested 500 times. In each iteration, a random pair of private attributes in the Type-1 request was selected. We set \(\sigma = 0.06\) and identified 5 \textit{orthogonal} attack requests using the proposed greedy search strategy. \autoref{tab:semantic_attack} summarizes the TPR for true samples, and the separate FPRs for each false sample type when increasing the attack requests from 1 to 5.
{Specifically, we recovered 407 victim requests out of the 500 true samples with only 1 attack request, achieving a recovery accuracy of 81.4\% with an average FPR of 0.045. With 5 attack requests, 477 victim requests were recovered, demonstrating a recovery accuracy of 95.4\% with an average FPR of 0.056. A demo of this attack is available on our website~\cite{Demo}.}

\begin{table}
 \centering
    \caption{Attack accuracy for the 4 types of victim requests with different number of attack trials.}
    \label{tab:semantic_attack}
     {
    \begin{tabular}{C{1.0 cm} | C{1.0 cm}| C{1.0  cm} | C{1.0  cm} | C{1.0  cm} }
    \hline 
         \textbf{\#Trials}    & \textbf{Type 1 \newline (TPR)} & \textbf{Type 2 \newline (FPR)} & \textbf{ Type 3 \newline (FPR)} & \textbf{ Type 3 \newline (FPR)} \\ \hline  
        1 & 0.814 & 0.116 & 0.054 & 0.004\\ \hline
        2 & 0.884 & 0.142 & 0.056 & 0.005\\ \hline
        3 & 0.930 & 0.146 & 0.060 & 0.005\\ \hline
        4 & 0.946 & 0.150 & 0.062 & 0.005\\ \hline
        5 & 0.954 & 0.152 & 0.062 & 0.005\\ \hline

    \end{tabular}
    }
    \vspace{-8pt}
\end{table}


\subsection{Inferring Documents on Commodity LLM}
\label{subsec:psarag}

\begin{figure}
    \centering
    \includegraphics[width=0.8\linewidth]{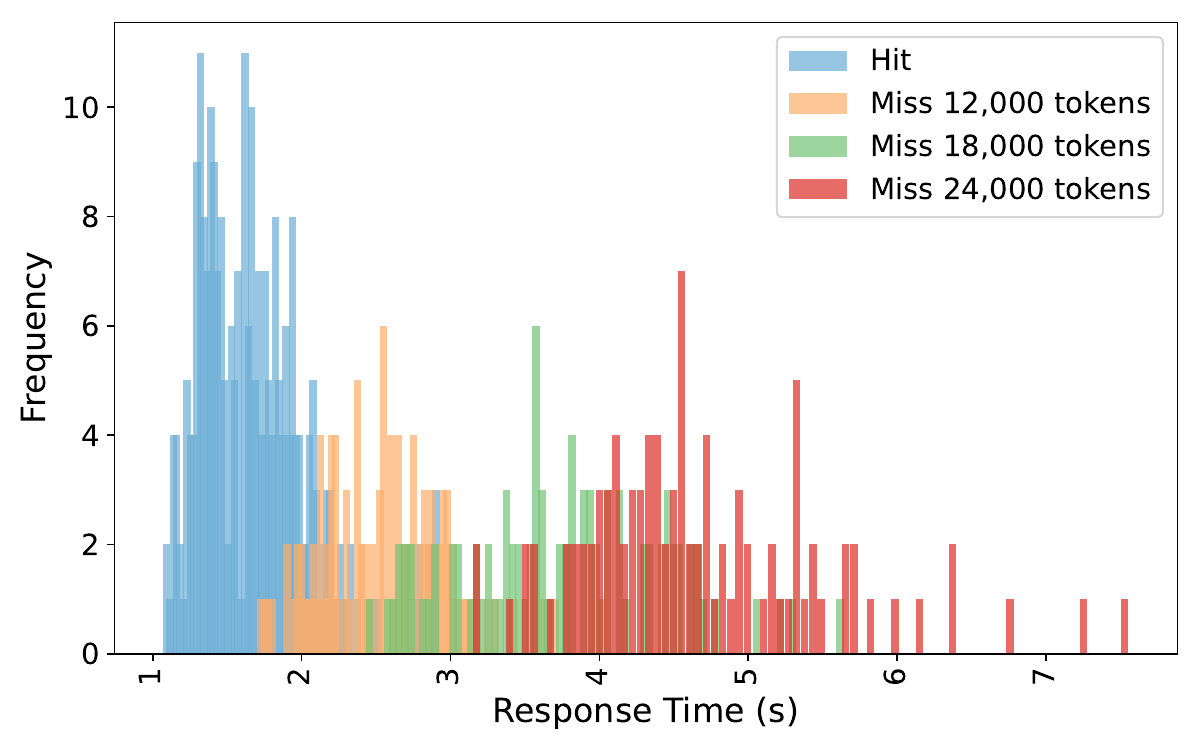}
    \caption{Timing distribution for hits and misses of processed documents with 12,000, 18,000, and 24,000 tokens.}
    \label{fig:infer_documents}
    \vspace{-8pt}
\end{figure}


{The KV cache can be shared among the same user, within an organization, or even across organizations~\cite{gustanford}, creating the potential for cross-user information leaks. In our research, we discovered that such leaks are feasible even in remote attack scenarios, particularly when the target LLM processes documents. To demonstrate this, we utilized a document summarization application powered by a commodity LLM API service, where all user requests are processed using the same developer’s API key, and both the victim and the attacker are standard users of the application. We provide an end-to-end demonstration showing how an adversary could infer processed documents from the application through the KV cache side channels. Importantly, such cross-user attacks fundamentally do not require application-specific vulnerabilities, as the KV cache can be shared across organizations~\cite{gustanford}. However, our use of the application highlights the privacy risks inherent to LLM-powered applications, even when they fully comply with the guidelines provided by the LLM service.}

{Note that the observation of the document uploaded to an LLM service, even when the content of the document is known, can expose an organization's interest, with substantial privacy and competitive ramifications across various domains. For example, a law firm relying on an LLM-based document processor could unknowingly disclose its involvement in complex litigation or pivotal mergers, tipping off opposing parties about strategic decisions before they become public. 








\para{The victim application}
{We implemented the document summarization application with direct summarization (also known as the stuff approach~\cite{stuffsummary}), using the public \path{Deepseek-chat} model as
the backend LLM API server.}
The application was built in accordance with Deepseek's guideline and operates by first extracting text from user-uploaded PDF files using the \path{pdfplumber} package. 
This text is then formatted into a request and sent to the LLM for summarization.
In this setup, documents uploaded by users are included in messages under the ``user'' role and sent to the Deepseek model, which returns summaries of the documents. Notably, all user inputs are processed under a single Deepseek account, which is typical in enterprise LLM deployments. However, this poses a privacy concern because Deepseek's prompt caching can inadvertently allow cached content to be reused across different users. As a result, a malicious user could potentially infer which documents other users have processed by exploiting timing side channels.

\para{Characterizing the leakage}
We conducted experiments with 200 documents of various lengths (approximately 12,000, 18,000, and 24,000 tokens), each saved in PDF format and derived from segments of the \path{zero_scrolls} dataset~\cite{shaham2023zeroscrolls}. Our results revealed distinct latencies between cache hits (where documents had been cached) and cache misses (where documents had not been cached), as shown in \autoref{fig:infer_documents}. In particular, responses to cache hits remained consistently fast across different document lengths, while cache misses grew noticeably slower with larger documents.

\para{End-to-end attacks}
In this evaluation, we assume an attacker aims to determine whether a specific document is uploaded by the victim for summarization. The attacker prepares a set of 200 documents of varying lengths (the ``interested'' documents). Meanwhile, the victim submits a total of 200 documents, half of which come from the interested set and half from outside it. This sequence is repeated 5 times, and each time the attacker attempts to distinguish which of the victim's documents belong to the interested set.

Specifically, the victim first submits 100 documents from the interested set. The attacker then probes each of the 200 interested documents once, recording response latencies. Based on a predefined threshold (2.0 seconds in our experiments), the TPR is computed as the fraction of probed documents correctly identified as cache hits (i.e., with latencies below the threshold). Next, the victim uploads 100 additional documents 
outside the interested set, and the attacker probes the entire set of 200 documents again. The FPR is then calculated as the fraction of documents incorrectly labeled as hits. Our evaluation shows that this attack achieves an average accuracy of 89\%, with an average FPR of 0.05. {A demo for the attack is presented on our website~\cite{Demo}.}


\para{\hl{Discussions}}
\hl{Since KV-cache sharing typically requires requests to have a common prefix, we assume that the attacker knows either the full document or at least its prefix (i.e., the initial portion of the content). However, recent work has introduced more flexible cache-sharing mechanisms that extend beyond strict prefix matching. For example, CacheBlend~}\cite{yao2024cacheblend} \hl{and LMCache~}\cite{lmcache} \hl{enable sharing of arbitrary reused text segments, while KVShare~}\cite{yang2025kvshare} \hl{and SemShareKV~}\cite{anonymous2025semsharekv} \hl{support semantic-aware key-value cache sharing. These techniques are already being integrated into popular LLM serving engines such as vLLM, llm-d, and KServe. In such systems, an attacker could probe whether a target document has been processed by the LLM using only partial common text (not necessarily a prefix) or even general topical knowledge. We leave deeper exploration of these attack scenarios to future work.}

\ignore{
In the current Retrieval-Augmented Generation (RAG) scenario, LLMs generate more specific and accurate responses by knowledge from existing documents or user-uploaded content. In such a scenario, the model will repeatedly use previously accessed documents on one hand, while on the other hand, a small subset of documents will be frequently used in certain specific contexts. Therefore, the KV cache reuse techniques, such as CacheBlend~\cite{yao2024cacheblend} and RAGCache~\cite{jin2024ragcache}, and semantic cache reuse techniques, such as Ollama~\cite{ollama-website} can significantly reduce the response time of the LLM system and increase its throughput. This section presents preliminary findings concerning the timing leakage in the RAG scenario.

\para{Inferring the queried document through KV cache reuse}
We conducted our evaluation on the CacheBlend inference framework~\cite{cacheblend-github}, which is built upon vLLM and supports KV cache reuse. Using the Multi-news dataset~\cite{multinews-github,Fabbri2019MultiNewsAL}, we selected documents with different content at various lengths and obtained ROC curves of response times when users queried a specific document, both with and without the document being cached. We selected 500 documents of different lengths and contents from the dataset through simple data processing. When testing the caching effect of documents of different lengths, we first uploaded one of these documents to the LLM. Then, we repeatedly used this document to query the LLM to obtain a dataset of response times for document cache hits. Additionally, we used the remaining documents, which differ from the target document, to query and obtain a dataset of response times for document cache misses. During the AUC calculation, we labeled the response times for document matches as 1 and those for document misses as 0, and plotted the ROC curves for both. The results, as shown in \autoref{fig:ragroc}, indicate that since document lengths in RAG scenarios are typically in the range of a few hundred words, it becomes apparent whether a document is cached. This could potentially allow for the identification of private documents used by other users.

\begin{figure}[!ht]
    \centering
    \includegraphics[width=0.95\linewidth]{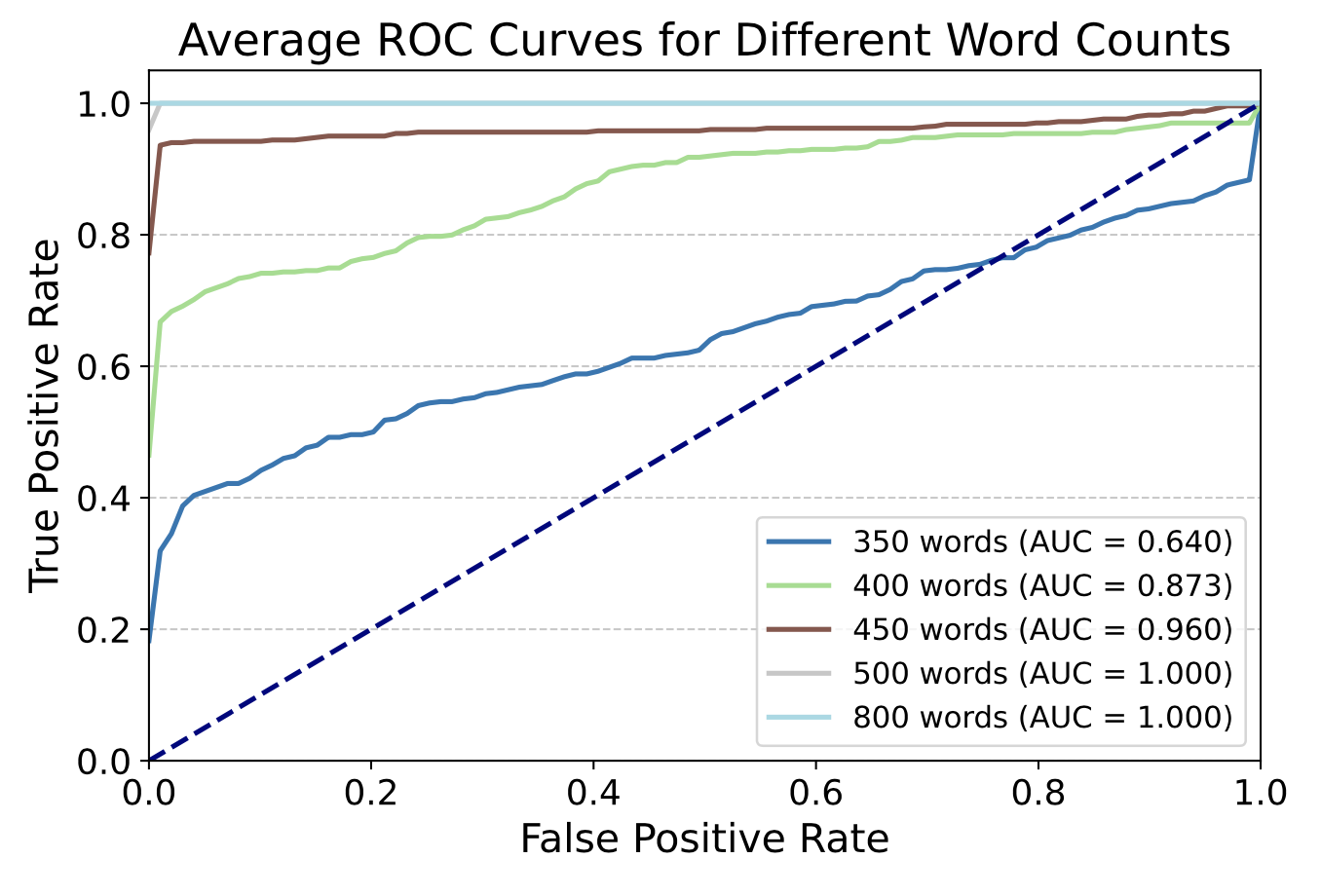}
    \caption{Leakage profile of sharing documents in the RAG scenario, under different document sizes.}
    \label{fig:ragroc}
\end{figure}


\para{Inferring the visited website through semantic cache reuse}
Side-channel information leakage can occur when users on the same host access browsers and use the same LLM plugin. We examined Lumos, an LLM plugin designed to summarize web content and respond to queries. We found that the caching effect occurs when different users access the same web pages and send the same queries, resulting in timing side channels.

Furthermore, we deployed an LLM locally that responds within web pages using Ollama~\cite{ollama-website}, with RedisCache and RedisSementicCache as the back-end cache. We queried the same prompts with different users in different browsers (e.g., Chrome and Edge) and found that there are significant time differences when the cache is hit, and the response time decreases from several seconds to milliseconds. This further proves the possibility of side-channel attacks.

\wenhao{We need to use RAG to demonstrate end-to-end attack on OpenAI's APIs, using image captioning or document asking}

\wenhao{maybe our case is not a typical RAG, and may change the title}
}

%








%

%





\ignore{
\begin{algorithm}[hbt!]
\caption{Semantic Searching Algorithm}\label{alg:semantic}
    \SetKwProg{Fn}{Function}{}{}

    \KwData{

    }

    \KwResult{
    }

    \SetKwFunction{TestSize}{TestSize}
    \SetKwFunction{GetSize}{GetSize}
    
    \Fn{\TestSize{Start, End, A, C, MT}}{
    }

    \Fn{\GetSize{}}{

    }

\end{algorithm}
}







\ignore{\hl{The KV cache can be shared by the same user, within an organization, or even across organizations~\cite{gustanford}. In this section, we focus on a specific scenario involving LLM-based applications. 
Specifically, we examine a document summarization application built on a commodity LLM API service, where all user requests are processed using the same developer’s API key. We provide an end-to-end demonstration of how an adversary could potentially infer processed documents from the application through KV cache side channels.}
\hl{This side channel can inadvertently reveal an organization's interest in highly sensitive content, with substantial privacy and competitive ramifications across various domains. For example, a law firm relying on an LLM-based document processor could unknowingly disclose its involvement in complex litigation or pivotal mergers, tipping off opposing parties about strategic decisions before they become public. Similarly, an investment firm analyzing financial statements might inadvertently signal which companies it views as high-potential opportunities, allowing competitors to anticipate emerging deals or investment moves. These examples illustrate how even subtle timing discrepancies can reveal critical details of an organization's internal activities, underscoring the need for stronger safeguards against such leaks.}








\para{Vulnerable application}
\hl{We implemented the document summarization application with direct summarization~\cite{directsummary} (also known as the stuff approach~\cite{stuffsummary}), using the public \path{Deepseek-chat} model as
the backend LLM API server.}
The application begins by extracting text from user-uploaded PDF files using the \path{pdfplumber} package. 
This text is then formatted into a request and sent to the LLM for  summarization.
In this setup, documents uploaded by users are included in messages under the ``user'' role and sent to the Deepseek model, which returns summaries of the provided documents. Notably, all user inputs are processed under a single Deepseek account, which is typical in such applications. However, this poses a privacy concern because Deepseek's prompt caching~\cite{promptcachingopenai} can inadvertently allow cached content to be reused across different users. As a result, a malicious user could potentially infer which documents other users have processed by exploiting timing side channels.

\para{Characterizing the leakage}
We experimented with 200 documents of various lengths (approximately 12,000, 18,000, and 24,000 tokens), each saved in PDF format and derived from segments of the \path{zero_scrolls} dataset~\cite{shaham2023zeroscrolls}. Our results revealed distinct latencies for cache hits (where documents had been cached) and cache misses (where documents had not been cached), as shown in \autoref{fig:infer_documents}. In particular, responses to cache hits remained consistently fast across different document lengths, while cache misses grew noticeably slower with larger documents.

\para{End-to-end attacks}
In this evaluation, we assume an attacker aims to determine whether a specific document is uploaded by the victim for summarization. The attacker prepares a set of 200 documents of varying lengths (the ``interested'' documents). Meanwhile, the victim submits a total of 200 documents, half of which come from the interested set and half from outside it. This sequence is repeated 5 times, and each time the attacker attempts to distinguish which of the victim's documents belong to the interested set.

Specifically, the victim first submits 100 documents from the interested set. The attacker then probes each of the 200 interested documents once, recording response latencies. Based on a predefined threshold (2.0 seconds in our experiments), the TPR is computed as the fraction of probed documents correctly identified as cache hits (i.e., with latencies below the threshold). Next, the victim uploads 100 additional documents that lie outside the interested set, and the attacker probes the entire set of 200 documents again. The FPR is then calculated as the fraction of documents incorrectly labeled as hits. Our tests produced a TPR of 0.89 and an FPR of 0.05. \hl{The demo for the attack is presented in our website~\cite{Demo}.}

\para{Notes}
For ethical reasons, we did not explore techniques for forcibly evicting cache entries in real-world systems. Such research would require extensive experimentation, potentially violating usage policies or disrupting other users' experience. Without active cache eviction, the timing-based attack primarily operates at the granularity where caches naturally expire due to inactivity---around 5 minutes for systems like OpenAI and Anthropic, as indicated in their documentation~\cite{promptcachingopenai,claudepromptcaching}.







\ignore{
In the current Retrieval-Augmented Generation (RAG) scenario, LLMs generate more specific and accurate responses by knowledge from existing documents or user-uploaded content. In such a scenario, the model will repeatedly use previously accessed documents on one hand, while on the other hand, a small subset of documents will be frequently used in certain specific contexts. Therefore, the KV cache reuse techniques, such as CacheBlend~\cite{yao2024cacheblend} and RAGCache~\cite{jin2024ragcache}, and semantic cache reuse techniques, such as Ollama~\cite{ollama-website} can significantly reduce the response time of the LLM system and increase its throughput. This section presents preliminary findings concerning the timing leakage in the RAG scenario.

\para{Inferring the queried document through KV cache reuse}
We conducted our evaluation on the CacheBlend inference framework~\cite{cacheblend-github}, which is built upon vLLM and supports KV cache reuse. Using the Multi-news dataset~\cite{multinews-github,Fabbri2019MultiNewsAL}, we selected documents with different content at various lengths and obtained ROC curves of response times when users queried a specific document, both with and without the document being cached. We selected 500 documents of different lengths and contents from the dataset through simple data processing. When testing the caching effect of documents of different lengths, we first uploaded one of these documents to the LLM. Then, we repeatedly used this document to query the LLM to obtain a dataset of response times for document cache hits. Additionally, we used the remaining documents, which differ from the target document, to query and obtain a dataset of response times for document cache misses. During the AUC calculation, we labeled the response times for document matches as 1 and those for document misses as 0, and plotted the ROC curves for both. The results, as shown in \autoref{fig:ragroc}, indicate that since document lengths in RAG scenarios are typically in the range of a few hundred words, it becomes apparent whether a document is cached. This could potentially allow for the identification of private documents used by other users.

\begin{figure}[!ht]
    \centering
    \includegraphics[width=0.95\linewidth]{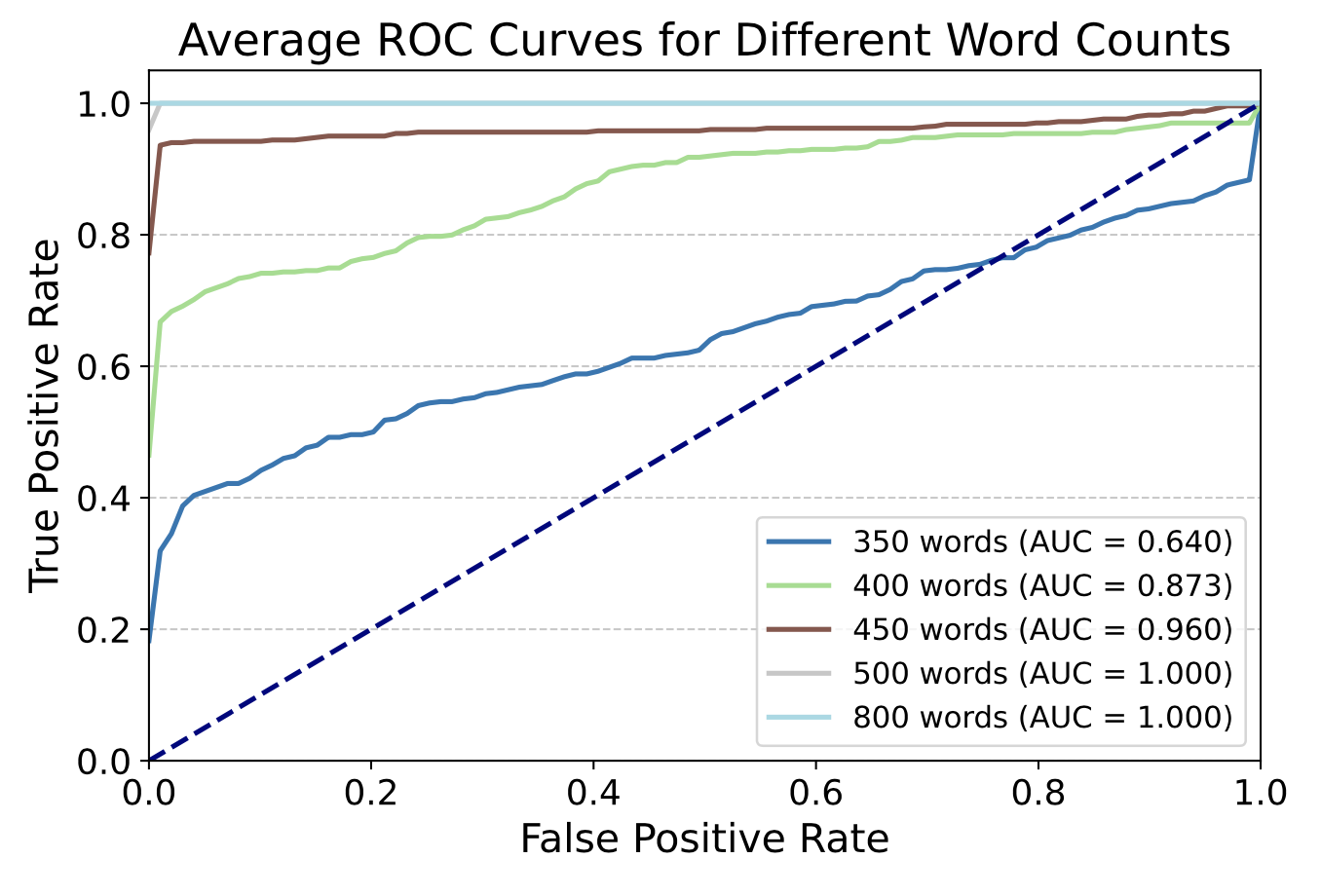}
    \caption{Leakage profile of sharing documents in the RAG scenario, under different document sizes.}
    \label{fig:ragroc}
\end{figure}


\para{Inferring the visited website through semantic cache reuse}
Side-channel information leakage can occur when users on the same host access browsers and use the same LLM plugin. We examined Lumos, an LLM plugin designed to summarize web content and respond to queries. We found that the caching effect occurs when different users access the same web pages and send the same queries, resulting in timing side channels.

Furthermore, we deployed an LLM locally that responds within web pages using Ollama~\cite{ollama-website}, with RedisCache and RedisSementicCache as the back-end cache. We queried the same prompts with different users in different browsers (e.g., Chrome and Edge) and found that there are significant time differences when the cache is hit, and the response time decreases from several seconds to milliseconds. This further proves the possibility of side-channel attacks.

\wenhao{We need to use RAG to demonstrate end-to-end attack on OpenAI's APIs, using image captioning or document asking}

\wenhao{maybe our case is not a typical RAG, and may change the title}
}
}
\subsection{Measurement Study on Commodity LLMs}
\label{subsec:realworld}





\noindent\textbf{KV cache sharing.}
To investigate KV cache sharing in commodity LLM services, we conducted experiments by invoking the APIs provided by these vendors. These APIs support different roles, such as system and user. For the measurement study, we designed requests with system and user prompts of varying lengths and configured them to run in the streaming mode. For this evaluation, we used the \path{zero_scrolls} dataset for generating requests.

\hl{Specifically, we first measured the response latencies by sending initial requests that were likely to miss the cache. Then, we sent identical requests 10 times and measured the median latencies for these subsequent requests.} To maximize the likelihood of co-locating on the same physical machine and ensuring the requests were cached, we conducted continuous tests within the same time period. If we observed lower latencies in the later requests, this indicated the use of caching mechanisms in the LLM services.
{With KV cache sharing, the computation of matched prefix tokens during the prefill phase can be ignored. However, the output generated during the decoding phase still requires computation and inference, which are influenced by parameters such as temperature, introducing randomness.
To verify that the latency reduction was due to KV cache sharing, we deliberately set a high temperature ($0.9$) in the request. This configuration was critical because semantic caching mechanisms typically return identical cached outputs for semantically similar inputs. By introducing substantial randomness in token selection through high temperature, we ensured the LLM would generate diverse responses despite input similarities.}
{We verified whether the LLM produced different responses for each request, with TTFT reductions consistently observed.}
If it did, this strongly indicated that KV cache sharing was supported, enabling a reduction in TTFT while still allowing for diverse outputs.
\hl{To minimize the impact of network latency, we sent requests of varying lengths, ranging from 200 to 2,000 tokens. The time difference between cached and uncached responses typically spanned several hundred milliseconds, making it easy to distinguish between the two. Additionally, we observed when the cache is hit, the TTFT remains consistent, regardless of the request length, whereas when the cache is missed, TTFT increases almost linearly as the length of the request grows.}
As summarized in \autoref{tab:kv_realworld}, most popular LLM service providers support KV cache sharing in specific scenarios.


\begin{table}
    \centering
    \begin{threeparttable}[b]
    \caption{Summary of KV cache sharing in real world LLM serving systems (date: 08/29/2024). }
    \label{tab:kv_realworld}
     {
    \begin{tabular}{C{1.8cm}C{3.0cm}C{2.6cm}}
    \hline 
          \textbf{LLM service}   & \textbf{System prompt sharing} & \textbf{User prompt sharing} \\ \hline  
        GPT-4o-mini  & \Checkmark & \Checkmark \\ \hline 
        Deepinfra & \Checkmark & \Checkmark  \\ \hline 
         Deepseek-chat & \Checkmark & \Checkmark \\ \hline 
        Claude-3.5 & \Checkmark & \Checkmark  \\ \hline 
        Qwen-max & \XSolidBrush & \XSolidBrush  \\ \hline 
        Moonshot & \Checkmark & \Checkmark  \\ \hline 
        Baidu Ernie-8k & \XSolidBrush & \XSolidBrush \\ \hline  
        Google Gemini & \XSolidBrush & \XSolidBrush  \\ \hline  
        Fireworks.ai & \Checkmark & \Checkmark  \\ \hline
        Groq& \XSolidBrush & \XSolidBrush  \\ \hline  
        SiliconFLow & \Checkmark & \Checkmark  \\ \hline  
        
    \end{tabular}
    }
    \end{threeparttable}
\end{table}

\ignore{
\begin{table}
    \centering
    \begin{threeparttable}[b]
    \caption{Summary of KV cache sharing in real world LLM serving systems (date: 08/29/2024). }
    \label{tab:kv_realworld}
     {\small
    \begin{tabular}{C{1.9cm}C{1.9cm}C{1.6cm}C{1.45cm}}
    \hline 
          \textbf{LLM service}   & \textbf{System prompt sharing} & \textbf{User prompt sharing} & \textbf{Document sharing} \\ \hline  
        GPT-4o-mini \tnote{$\dagger$} & \Checkmark & \Checkmark & \XSolidBrush \\ \hline 
        Deepinfra & \Checkmark & \Checkmark & \XSolidBrush \\ \hline 
         Deepseek-chat & \Checkmark & \Checkmark & \XSolidBrush \\ \hline 
        Claude-3.5 & \Checkmark & \Checkmark & \XSolidBrush \\ \hline 
        Qwen-max & \XSolidBrush & \XSolidBrush & \XSolidBrush \\ \hline 
        Moonshot~\cite{qin2024mooncake} & \Checkmark & \Checkmark & \Checkmark \\ \hline 
        Baidu Ernie-8k & \XSolidBrush & \XSolidBrush & \XSolidBrush \\ \hline  
        Google Gemini & \XSolidBrush & \XSolidBrush & \Checkmark \\ \hline  
        Fireworks.ai~\cite{fireworkai}& \Checkmark & \Checkmark & \XSolidBrush \\ \hline
        Groq~\cite{groq}& \XSolidBrush & \XSolidBrush & \XSolidBrush \\ \hline  
        SiliconFLow~\cite{siliconflow}& \Checkmark & \Checkmark & \XSolidBrush \\ \hline  
        
    \end{tabular}
    }
    \begin{tablenotes}
    \item[$\dagger$] \footnotesize We observed a timing difference on 08/29/2024 and reported it to OpenAI. By late December 2024, the timing difference was no longer stable, despite the API indicating that the prompt cache was effective. This may be due to timing obfuscation measures implemented by OpenAI.
    \end{tablenotes}
    \end{threeparttable}
\end{table}
}

\begin{table}
    \centering
     \caption{Native support of semantic caching of popular AI service providers (date: 08/29/2024).}
    \label{tab:llmserviceprovider}
     {
    \begin{tabular}{C{4.6cm}|C{3cm}}
        \hline
        \textbf{Service providers} & \textbf{Semantic cache support} \\ \hline
        Azure OpenAI Service models & \Checkmark \\ \hline
        Amazon Bedrock & \Checkmark \\ \hline
        Google Vertex AI & \XSolidBrush \\ \hline
        Alibaba Elastic Algorithm Service (EAS) of Platform for AI (PAI) & \Checkmark \\ \hline
        
    \end{tabular}
   }
   \vspace{-6pt}
\end{table}

\para{Semantic cache sharing}
We manually reviewed the documentation of public AI service providers to verify  whether they support semantic cache APIs. As shown in \autoref{tab:llmserviceprovider}, semantic caching is supported by major AI platform-as-a-service providers. Even on platforms that do not offer native semantic caching, users can still implement their own solutions or leverage open-source alternatives, such as GPTCache.








\section{Mitigations}


\subsection{Mitigating KV Cache Leakages}


\ignore{
\begin{figure}[!th]
    \centering
    \includegraphics[width=1\linewidth]{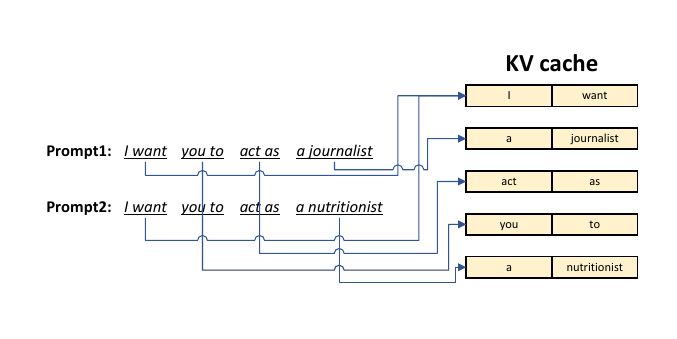}
    \caption{Prompt1 and prompt2 share 2 cache blocks, and each block contains 2 tokens, while the third cache blocks cannot be shared.}
    \label{fig:2token}
\end{figure}
}

\noindent\textbf{Design.}
A straightforward approach to mitigate KV cache leakages is to eliminate any sharing across requests. However, this would negate the computational and cost savings associated with caching. Noting that PSA recovers the request token by token by observing per-token timing differences, we explore the effect of a simple mitigation strategy that the prefix cache can only be shared in units of at least $K$ tokens ($K = 2, 3, 4$ etc.).
In SGLang, this could be achieved by modifying the radix tree structure used to manage the KV cache.
This approach reduces the likelihood of KV cache sharing, but it is unlikely to significantly impact performance.

\para{Evaluation}
To evaluate the effectiveness of the mitigation and reduce the cost of querying the LLM, we conducted a simulation-based experiment. First, we built the classifiers that detect the hits and misses of $K$ tokens for each value of $K$ ($K = 1, 2, 3, 4$), following the method outlined in \autoref{subsec:evalpsa}. Since the timing differences become more pronounced as $K$ increases, we reduced the number of samples (i.e., $n$) in multi-sampling, and obtained the corresponding TPRs and FPRs for each classifier. Then we used an \textit{oracle} to simulate the classifiers by randomly sampling a number and determining whether it fell within the classification range.
In this evaluation, we used the same repetitive trials method, dataset and fine-tuned model as described in \autoref{subsec:evalpsa}. The next-token predictor was modified to predict the next $K$ tokens. 
\autoref{tab:chunks_sim} presents the token recovery rate and the average number of queries per recovered token for $K = 1, 2, 3$ and $4$. \hl{The results show that increasing $K$ still yields a notable recovery rate, but the attack bottleneck shifts fundamentally. Larger $K$ amplifies timing differences between cache hits and misses, reducing false positives. At the same time, the predictor’s search space grows exponentially, making predictor accuracy the new bottleneck. For tokens that are successfully recovered, the average number of queries decreases with larger $K$ values, since the predictor can more easily identify tokens in simpler prompts. However, the overall recovery rates decline because the predictor struggles with harder tokens at higher $K$. As a result, although queries per recovered token drop, the total cost of recovering an entire prompt rises significantly.}

}

\begin{table}
    \centering
    \centering
    \caption{Token recovery results under different numbers of minimum shared tokens.}
    \label{tab:chunks_sim}
     {
    \begin{tabular}{C{0.3 cm} | C{1.5 cm} | C{1.3 cm} | C{2.1  cm} | C{1.4  cm}}
    \hline 
         \textbf{$K$}    & \textbf{Recovery rate} & \textbf{Accuracy} & \textbf{\#queries per \newline recovered token } & \textbf{\#queries \newline per token }   \\ \hline  
        1 & 91.5\% & 97.9\% & 118.58 & 215.41  \\ \hline
        2 & 81.0\% & 98.8\% & 108.89 & 286.79  \\ \hline
        3 & 67.5\% & 98.5\% & 88.30  & 337.28 \\ \hline
        4 & 49.0\% & 98.0\% & 62.55  & 470.82 \\ \hline
    \end{tabular}
    }
    \vspace{-8pt}
\end{table}


\subsection{Mitigating Semantic Cache Leakages}

\noindent\textbf{Design.}
As investigated in \autoref{subsec:evalpna}, private attributes have a significant impact on the semantic similarity between requests. As a result, the PNA infers private attributes by probing whether a semantically similar request is cached. To mitigate this leakage, we propose a strategy that involves identifying and anonymizing the private attributes present in the requests. This approach not only prevents the leakage of private attributes but also increases the potential for sharing requests across users.
\begin{figure}
    \centering
    \includegraphics[width=0.9\linewidth]{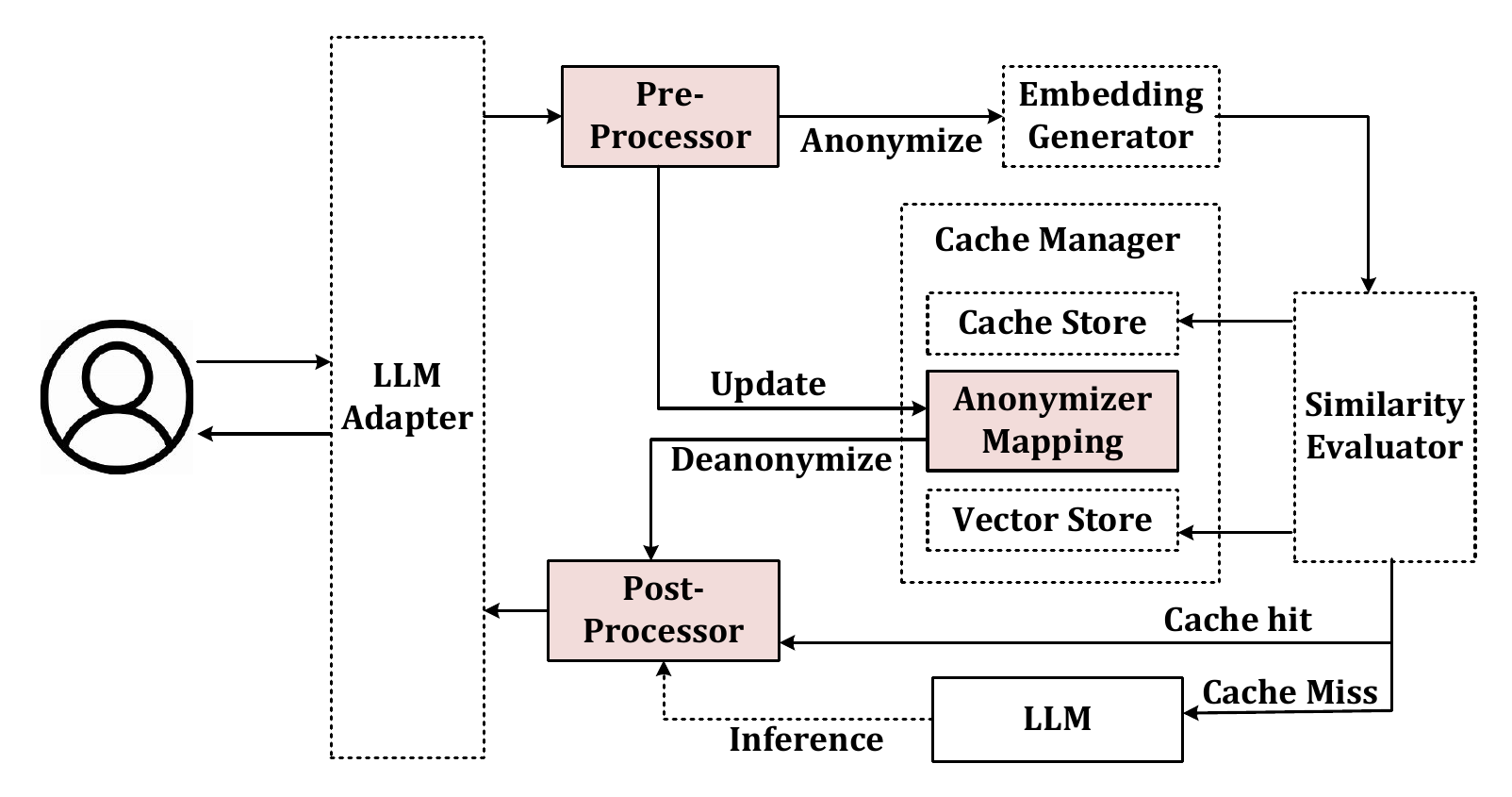}
    \caption{Mitigating semantic cache leakages. The shaded components are customized as part of our mitigation.}
    \label{fig:Semantic_defense}
    \vspace{-8pt}
\end{figure}
As shown in \autoref{fig:Semantic_defense}, we integrate a custom pre-processor and post-processor into the GPTCache framework. The pre-processor is designed to identify private attributes within the requests, 
and replace them with anonymized identifiers. In this approach, we selectively de-identify Personally Identifiable Information (PII) attributes, such as names, email addresses, phone numbers, credit card numbers, and IP addresses, while ensuring that no essential information needed for the LLMs is removed.

To facilitate reuse, the cache manager maintains a mapping structure that stores the anonymized identifier alongside its corresponding private attribute in a key-value format. The post-processor then identifies the anonymized identifiers in the response and replaces them with the private attributes by referencing this mapping. This ensures that the user receives an accurate response. 


\para{Evaluation}
In our prototype implementation, we used the Presidio tool~\cite{presidio-github} to automatically identify private attributes. For performance evaluation, we used the evaluation dataset released by Presidio, which includes sentences containing private information. Specifically, we randomly sampled 1,000 sentences from the dataset and fed them into both the original GPTCache and the enhanced GPTCache. Then we measured the average delay introduced by the pre-processor and post-processor. 
The results show that the anonymization process adds an average delay of approximately 6 ms, while GPTCache's response latency for a semantic cache hit without anonymization is around 0.14 s. Thus, de-identification introduces only about 4\% additional overhead, which has a minimal impact on GPTCache's overall performance.
\section{Related Works}
\label{sec:related}

\para{Prompt extraction attacks with adversarial prompts}
Most existing research focuses on stealing system prompts from LLMs by eliciting the previous prompts, typically through direct output or translation. 
Earlier studies involved manually constructing attacker prompts~\cite{perez2022ignore, zhang2023prompts}, while more recent work, such as PLeak, leverages output feedback from a given prompt and introduces an incremental search algorithm to optimize prompt retrieval~\cite{hui2024pleak}. 
While these approaches exploit the model's vulnerability to adversarial prompts, our proposed attacks take advantage of timing differences introduced in LLM service deployment. As such, our attacks do not rely on the specific details of any particular LLM.






\para{Side channel attacks on LLM}
Debenedetti et al. propose system-level side channels within the deep learning lifecycle, such as training data filtering, input preprocessing, output monitoring, and query filtering. These side channels can potentially be exploited to infer the training data or the requests~\cite{debenedetti2023privacy}. 
LLM keystroking attacks~\cite{weiss2024your} 
are a type of packet analysis based side channels. These attacks exploit the length of response tokens, assuming the attacker has access to encrypted network packets. 
Besides, Carlini et al. investigated the privacy risks of timing attacks in multi-turn interactions with LLM chatbots \cite{carlini2024remote}. More recently, Soleimani et al. exposed network side-channel vulnerabilities in speculative decoding-optimized LLM services \cite{soleimani2025wiretapping}. By comparison, we are the \textit{first} to study the timing leaks introduced by LLM serving system optimizations, rather than relying on output data or token packet sizes to recover requests.
\ignore{
Inspired by our work, a follow-up study by Gu et al. reports a comprehensive measurement analysis of prompt caching in real-world LLMs~\cite{gustanford}, detecting prompt caching in 8 out of 17 LLM service providers. More recently, Zheng et al. conducted a similar study on LLM timing side channels~\cite{zheng2024inputsnatch}, which however does not feature our optimized search strategy for efficient request recovery.}

\para{Micro-architectural side channel attacks on deep learning systems}
Numerous studies have explored methods for extracting deep learning models or structures, as well as fingerprinting these models, by exploiting various side channels, such as CPU~\cite{duddu2018stealing, yan2020cache, rakin2022deepsteal, gongye2020reverse, shukla2024stealing}, GPU~\cite{wei2020leaky}, FPGA~\cite{zhang2021stealing}, power and magnetic channels~\cite{maia2022can, horvath2023barracuda}, and PCIe traffic~\cite{zhu2021hermes}. In comparison, our work focuses on leaking private prompts rather than stealing model parameters. Additionally, our approach does not rely on the micro-architectural or power characteristics of specific hardware; instead, it exploits timing leaks inherent in LLM systems. As a result, our attacks are applicable across CPU, GPU, and FPGA platforms, provided they utilize KV cache or semantic cache sharing techniques.











\section{Conclusions}
\label{sec:conclusion}

LLM inference is a resource-intensive task, prompting studies focused on reducing inference latency. These optimizations often involve the use of various caches. When multiple users share the LLM system, these optimizations can lead to interference between users. This paper examines the side channels created by such interference, identifying two types of leaks: one in the KV cache and another in the semantic cache. We urge LLM system providers to recognize this emerging threat and prioritize security in their design choices.

\bibliographystyle{IEEEtran}

\bibliography{ref}

\begin{thebibliography}{10}
\providecommand{\url}[1]{#1}
\csname url@samestyle\endcsname
\providecommand{\newblock}{\relax}
\providecommand{\bibinfo}[2]{#2}
\providecommand{\BIBentrySTDinterwordspacing}{\spaceskip=0pt\relax}
\providecommand{\BIBentryALTinterwordstretchfactor}{4}
\providecommand{\BIBentryALTinterwordspacing}{\spaceskip=\fontdimen2\font plus
\BIBentryALTinterwordstretchfactor\fontdimen3\font minus \fontdimen4\font\relax}
\providecommand{\BIBforeignlanguage}[2]{{%
\expandafter\ifx\csname l@#1\endcsname\relax
\typeout{** WARNING: IEEEtran.bst: No hyphenation pattern has been}%
\typeout{** loaded for the language `#1'. Using the pattern for}%
\typeout{** the default language instead.}%
\else
\language=\csname l@#1\endcsname
\fi
#2}}
\providecommand{\BIBdecl}{\relax}
\BIBdecl

\bibitem{chatgpt}
``Chatgpt | openai,'' \url{https://openai.com/chatgpt/}, 2024.

\bibitem{perplexity}
``Perplexity ai,'' \url{https://www.perplexity.ai/}, 2024.

\bibitem{copilot}
``Github copilot · your ai pair programmer,'' \url{https://github.com/features/copilot/}, 2024.

\bibitem{yao2022zeroquant}
Z.~Yao, R.~Yazdani~Aminabadi, M.~Zhang, X.~Wu, C.~Li, and Y.~He, ``Zeroquant: Efficient and affordable post-training quantization for large-scale transformers,'' \emph{Advances in Neural Information Processing Systems}, vol.~35, pp. 27\,168--27\,183, 2022.

\bibitem{wei2022outlier}
X.~Wei, Y.~Zhang, X.~Zhang, R.~Gong, S.~Zhang, Q.~Zhang, F.~Yu, and X.~Liu, ``Outlier suppression: Pushing the limit of low-bit transformer language models,'' \emph{Advances in Neural Information Processing Systems}, vol.~35, pp. 17\,402--17\,414, 2022.

\bibitem{xiao2023smoothquant}
G.~Xiao, J.~Lin, M.~Seznec, H.~Wu, J.~Demouth, and S.~Han, ``Smoothquant: Accurate and efficient post-training quantization for large language models,'' in \emph{International Conference on Machine Learning}.\hskip 1em plus 0.5em minus 0.4em\relax PMLR, 2023, pp. 38\,087--38\,099.

\bibitem{wang2024model}
W.~Wang, W.~Chen, Y.~Luo, Y.~Long, Z.~Lin, L.~Zhang, B.~Lin, D.~Cai, and X.~He, ``Model compression and efficient inference for large language models: A survey,'' \emph{arXiv preprint arXiv:2402.09748}, 2024.

\bibitem{dao2022flashattention}
T.~Dao, D.~Fu, S.~Ermon, A.~Rudra, and C.~R{\'e}, ``Flashattention: Fast and memory-efficient exact attention with io-awareness,'' \emph{Advances in Neural Information Processing Systems}, vol.~35, pp. 16\,344--16\,359, 2022.

\bibitem{kwon2023efficient}
W.~Kwon, Z.~Li, S.~Zhuang, Y.~Sheng, L.~Zheng, C.~H. Yu, J.~Gonzalez, H.~Zhang, and I.~Stoica, ``Efficient memory management for large language model serving with pagedattention,'' in \emph{Proceedings of the 29th Symposium on Operating Systems Principles (SOSP)}, 2023, pp. 611--626.

\bibitem{zheng2023efficiently}
L.~Zheng, L.~Yin, Z.~Xie, J.~Huang, C.~Sun, C.~H. Yu, S.~Cao, C.~Kozyrakis, I.~Stoica, J.~E. Gonzalez \emph{et~al.}, ``Efficiently programming large language models using {SGLang},'' \emph{arXiv preprint}, 2023.

\bibitem{langchain-prompttemplate}
``Prompt template,'' \url{https://python.langchain.com.cn/docs/modules/model_io/prompts/prompt_templates/}, 2024.

\bibitem{claudepromptcaching}
``Prompt caching with claude,'' \url{https://www.anthropic.com/news/prompt-caching}, 2024.

\bibitem{bang2023gptcache}
F.~Bang, ``Gptcache: An open-source semantic cache for llm applications enabling faster answers and cost savings,'' in \emph{Proceedings of the 3rd Workshop for Natural Language Processing Open Source Software (NLP-OSS 2023)}, 2023, pp. 212--218.

\bibitem{langchain-website}
``Langchain,'' \url{https://www.langchain.com/}, 2024.

\bibitem{wuknow}
G.~Wu, Z.~Zhang, Y.~Zhang, W.~Wang, J.~Niu, Y.~Wu, and Y.~Zhang, ``I know what you asked: Prompt leakage via kv-cache sharing in multi-tenant llm serving,'' in \emph{32nd Annual Network and Distributed System Security Symposium, {NDSS} 2025, San Diego, California, USA}, 2025.

\bibitem{zheng2024inputsnatch}
X.~Zheng, H.~Han, S.~Shi, Q.~Fang, Z.~Du, Q.~Guo, and X.~Hu, ``Inputsnatch: Stealing input in llm services via timing side-channel attacks,'' \emph{arXiv preprint arXiv:2411.18191}, 2024.

\bibitem{gustanford}
C.~Gu, X.~L. Li, R.~Kuditipudi, P.~Liang, and T.~Hashimoto, ``Stanford cs 191w senior project: Timing attacks on prompt caching in language model apis,'' 2024.

\bibitem{vaswani2017attention}
A.~Vaswani, ``Attention is all you need,'' \emph{Advances in Neural Information Processing Systems}, 2017.

\bibitem{anythingllm}
``Anythingllm: The all-in-one desktop \& docker ai application with built-in rag, ai agents, and more.'' \url{https://github.com/Mintplex-Labs/anything-llm}, 2024.

\bibitem{kurth2020netcat}
M.~Kurth, B.~Gras, D.~Andriesse, C.~Giuffrida, H.~Bos, and K.~Razavi, ``Netcat: Practical cache attacks from the network,'' in \emph{2020 IEEE Symposium on Security and Privacy (SP)}.\hskip 1em plus 0.5em minus 0.4em\relax IEEE, 2020, pp. 20--38.

\bibitem{schwarz2019netspectre}
M.~Schwarz, M.~Schwarzl, M.~Lipp, J.~Masters, and D.~Gruss, ``Netspectre: Read arbitrary memory over network,'' in \emph{Computer Security--ESORICS 2019: 24th European Symposium on Research in Computer Security, Luxembourg, September 23--27, 2019, Proceedings, Part I 24}.\hskip 1em plus 0.5em minus 0.4em\relax Springer, 2019, pp. 279--299.

\bibitem{naghibijouybari2018rendered}
H.~Naghibijouybari, A.~Neupane, Z.~Qian, and N.~Abu-Ghazaleh, ``Rendered insecure: Gpu side channel attacks are practical,'' in \emph{Proceedings of the 2018 ACM SIGSAC conference on computer and communications security}, 2018, pp. 2139--2153.

\bibitem{zhou2016vulnerable}
Z.~Zhou, W.~Diao, X.~Liu, Z.~Li, K.~Zhang, and R.~Liu, ``Vulnerable gpu memory management: towards recovering raw data from gpu,'' \emph{arXiv preprint arXiv:1605.06610}, 2016.

\bibitem{taneja2023hot}
H.~Taneja, J.~Kim, J.~J. Xu, S.~Van~Schaik, D.~Genkin, and Y.~Yarom, ``Hot pixels: Frequency, power, and temperature attacks on $\{$GPUs$\}$ and arm $\{$SoCs$\}$,'' in \emph{32nd USENIX Security Symposium (USENIX Security 23)}, 2023, pp. 6275--6292.

\bibitem{samsung}
``Samsung bans staff’s ai use after spotting chatgpt data leak,'' \url{https://www.bloomberg.com/news/articles/2023-05-02/samsung-bans-chatgpt-and-other-generative-ai-use-by-staff-after-leak}.

\bibitem{syspromptvalue}
``System prompts in large language models,'' \url{https://promptengineering.org/system-prompts-in-large-language-models/}, 2024.

\bibitem{lammaindex-github}
``Llamaindex is a data framework for your llm applications,'' \url{https://github.com/run-llama/llama_index}, 2024.

\bibitem{jpmorgan}
``Jpmorgan rolls out in-house genai-based chatbot to employees,'' \url{https://www.financedirectoreurope.com/news/jpmorgan-rolls-out-ai-based-chatbot/}, 2024.

\bibitem{openaiapi}
``Text generation and prompting,'' \url{https://platform.openai.com/docs/guides/text?api-mode=responses}, 2024.

\bibitem{zheng2023judging}
L.~Zheng, W.-L. Chiang, Y.~Sheng, S.~Zhuang, Z.~Wu, Y.~Zhuang, Z.~Lin, Z.~Li, D.~Li, E.~P. Xing, H.~Zhang, J.~E. Gonzalez, and I.~Stoica, ``Judging llm-as-a-judge with mt-bench and chatbot arena,'' 2023.

\bibitem{DBLP:journals/corr/abs-2303-17564}
\BIBentryALTinterwordspacing
S.~Wu, O.~Irsoy, S.~Lu, V.~Dabravolski, M.~Dredze, S.~Gehrmann, P.~Kambadur, D.~S. Rosenberg, and G.~Mann, ``Bloomberggpt: {A} large language model for finance,'' \emph{CoRR}, vol. abs/2303.17564, 2023. [Online]. Available: \url{https://doi.org/10.48550/arXiv.2303.17564}
\BIBentrySTDinterwordspacing

\bibitem{systemprompt-huggingface}
``System prompt leakage dataset,'' \url{https://huggingface.co/datasets/gabrielchua/system-prompt-leakage}, 2024.

\bibitem{mcinnes2018umap}
L.~McInnes, J.~Healy, and J.~Melville, ``Umap: Uniform manifold approximation and projection for dimension reduction,'' \emph{arXiv preprint arXiv:1802.03426}, 2018.

\bibitem{flush_cache}
S.~Group, ``flush cache,'' \url{https://github.com/sgl-project/sglang/blob/25e5d589e39b3b605296395e4f9c96ec42f09055/python/sglang/srt/server.py#L164}, 2024.

\bibitem{Demo}
``llm side-channel demo,'' \url{https://sites.google.com/view/early-bird-catches-the-leak}, 2025.

\bibitem{MedQuAD-website}
``Medquad,'' \url{https://huggingface.co/datasets/lavita/MedQuAD}, 2019.

\bibitem{sbert_model}
S.~Transformers, ``all-mpnet-base-v2,'' \url{https://huggingface.co/sentence-transformers/all-mpnet-base-v2}, 2024.

\bibitem{stuffsummary}
``Ai document summarization,'' \url{https://www.ibm.com/architectures/hybrid/genai-document-summarization}, 2024.

\bibitem{shaham2023zeroscrolls}
U.~Shaham, M.~Ivgi, A.~Efrat, J.~Berant, and O.~Levy, ``Zeroscrolls: A zero-shot benchmark for long text understanding,'' \emph{arXiv preprint arXiv:2305.14196}, 2023.

\bibitem{yao2024cacheblend}
J.~Yao, H.~Li, Y.~Liu, S.~Ray, Y.~Cheng, Q.~Zhang, K.~Du, S.~Lu, and J.~Jiang, ``Cacheblend: Fast large language model serving with cached knowledge fusion,'' \emph{arXiv preprint arXiv:2405.16444}, 2024.

\bibitem{lmcache}
``Lmcache: Supercharge your llm with the fastest kv cache layer,'' \url{https://github.com/LMCache/LMCache}, 2025.

\bibitem{yang2025kvshare}
H.~Yang, R.~Zhang, M.~Huang, W.~Wang, Y.~Tang, Y.~Li, Y.~Liu, and D.~Zhang, ``Kvshare: An llm service system with efficient and effective multi-tenant kv cache reuse,'' \emph{arXiv preprint arXiv:2503.16525}, 2025.

\bibitem{anonymous2025semsharekv}
\BIBentryALTinterwordspacing
Anonymous, ``{SemShareKV}: Efficient {KVC}ache sharing for semantically similar prompts via token-level {LSH} matching,'' in \emph{Submitted to ACL Rolling Review - July 2025}, 2025, under review. [Online]. Available: \url{https://openreview.net/forum?id=B5xRER6OKT}
\BIBentrySTDinterwordspacing

\bibitem{presidio-github}
``Presidio,'' \url{https://github.com/microsoft/presidio}, 2022.

\bibitem{perez2022ignore}
F.~Perez and I.~Ribeiro, ``Ignore previous prompt: Attack techniques for language models,'' \emph{arXiv preprint arXiv:2211.09527}, 2022.

\bibitem{zhang2023prompts}
Y.~Zhang and D.~Ippolito, ``Prompts should not be seen as secrets: Systematically measuring prompt extraction attack success,'' \emph{arXiv preprint arXiv:2307.06865}, 2023.

\bibitem{hui2024pleak}
B.~Hui, H.~Yuan, N.~Gong, P.~Burlina, and Y.~Cao, ``Pleak: Prompt leaking attacks against large language model applications,'' \emph{arXiv preprint arXiv:2405.06823}, 2024.

\bibitem{debenedetti2023privacy}
E.~Debenedetti, G.~Severi, N.~Carlini, C.~A. Choquette-Choo, M.~Jagielski, M.~Nasr, E.~Wallace, and F.~Tram{\`e}r, ``Privacy side channels in machine learning systems,'' in \emph{33rd USENIX Security Symposium}, 2024.

\bibitem{weiss2024your}
R.~Weiss, D.~Ayzenshteyn, G.~Amit, and Y.~Mirsky, ``What was your prompt? a remote keylogging attack on ai assistants,'' \emph{arXiv preprint arXiv:2403.09751}, 2024.

\bibitem{carlini2024remote}
N.~Carlini and M.~Nasr, ``Remote timing attacks on efficient language model inference,'' \emph{arXiv preprint arXiv:2410.17175}, 2024.

\bibitem{soleimani2025wiretapping}
M.~Soleimani, G.~Jia, I.~Gim, S.-s. Lee, and A.~Khandelwal, ``Wiretapping llms: Network side-channel attacks on interactive llm services,'' \emph{Cryptology ePrint Archive}, 2025.

\bibitem{duddu2018stealing}
V.~Duddu, D.~Samanta, D.~V. Rao, and V.~E. Balas, ``Stealing neural networks via timing side channels,'' \emph{arXiv preprint arXiv:1812.11720}, 2018.

\bibitem{yan2020cache}
M.~Yan, C.~W. Fletcher, and J.~Torrellas, ``Cache telepathy: Leveraging shared resource attacks to learn $\{$DNN$\}$ architectures,'' in \emph{29th USENIX Security Symposium (USENIX Security 20)}, 2020, pp. 2003--2020.

\bibitem{rakin2022deepsteal}
A.~S. Rakin, M.~H.~I. Chowdhuryy, F.~Yao, and D.~Fan, ``Deepsteal: Advanced model extractions leveraging efficient weight stealing in memories,'' in \emph{2022 IEEE symposium on security and privacy (SP)}.\hskip 1em plus 0.5em minus 0.4em\relax IEEE, 2022, pp. 1157--1174.

\bibitem{gongye2020reverse}
C.~Gongye, Y.~Fei, and T.~Wahl, ``Reverse-engineering deep neural networks using floating-point timing side-channels,'' in \emph{2020 57th ACM/IEEE Design Automation Conference (DAC)}.\hskip 1em plus 0.5em minus 0.4em\relax IEEE, 2020, pp. 1--6.

\bibitem{shukla2024stealing}
S.~Shukla, M.~Alam, P.~Mitra, and D.~Mukhopadhyay, ``Stealing the invisible: Unveiling pre-trained cnn models through adversarial examples and timing side-channels,'' \emph{arXiv preprint arXiv:2402.11953}, 2024.

\bibitem{wei2020leaky}
J.~Wei, Y.~Zhang, Z.~Zhou, Z.~Li, and M.~A. Al~Faruque, ``Leaky dnn: Stealing deep-learning model secret with gpu context-switching side-channel,'' in \emph{2020 50th Annual IEEE/IFIP International Conference on Dependable Systems and Networks (DSN)}.\hskip 1em plus 0.5em minus 0.4em\relax IEEE, 2020, pp. 125--137.

\bibitem{zhang2021stealing}
Y.~Zhang, R.~Yasaei, H.~Chen, Z.~Li, and M.~A. Al~Faruque, ``Stealing neural network structure through remote fpga side-channel analysis,'' \emph{IEEE Transactions on Information Forensics and Security}, vol.~16, pp. 4377--4388, 2021.

\bibitem{maia2022can}
H.~T. Maia, C.~Xiao, D.~Li, E.~Grinspun, and C.~Zheng, ``Can one hear the shape of a neural network?: Snooping the gpu via magnetic side channel.'' in \emph{USENIX Security Symposium}, 2022, pp. 4383--4400.

\bibitem{horvath2023barracuda}
P.~Horvath, L.~Chmielewski, L.~Weissbart, L.~Batina, and Y.~Yarom, ``Barracuda: Bringing electromagnetic side channel into play to steal the weights of neural networks from nvidia gpus,'' \emph{arXiv preprint arXiv:2312.07783}, 2023.

\bibitem{zhu2021hermes}
Y.~Zhu, Y.~Cheng, H.~Zhou, and Y.~Lu, ``Hermes attack: Steal {DNN} models with lossless inference accuracy,'' in \emph{30th USENIX Security Symposium (USENIX Security 21)}, 2021.

\end{thebibliography}

\begin{IEEEbiography}[{\includegraphics[width=1in,height=1.25in,clip]
{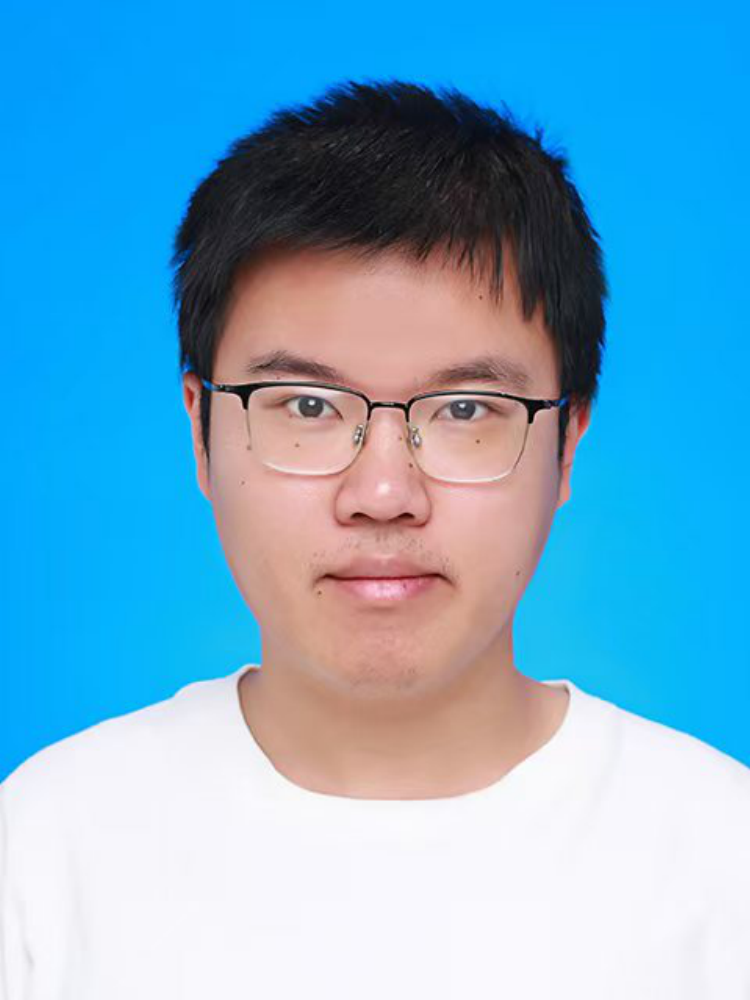}}] {Linke Song} is currently pursuing the Ph.D. degree with Institute of Information Engineering, Chinese Academy of Sciences. His research interests include system security and confidential computing.

\end{IEEEbiography}

\begin{IEEEbiography}[{\includegraphics[width=1in,height=1.25in,clip]
{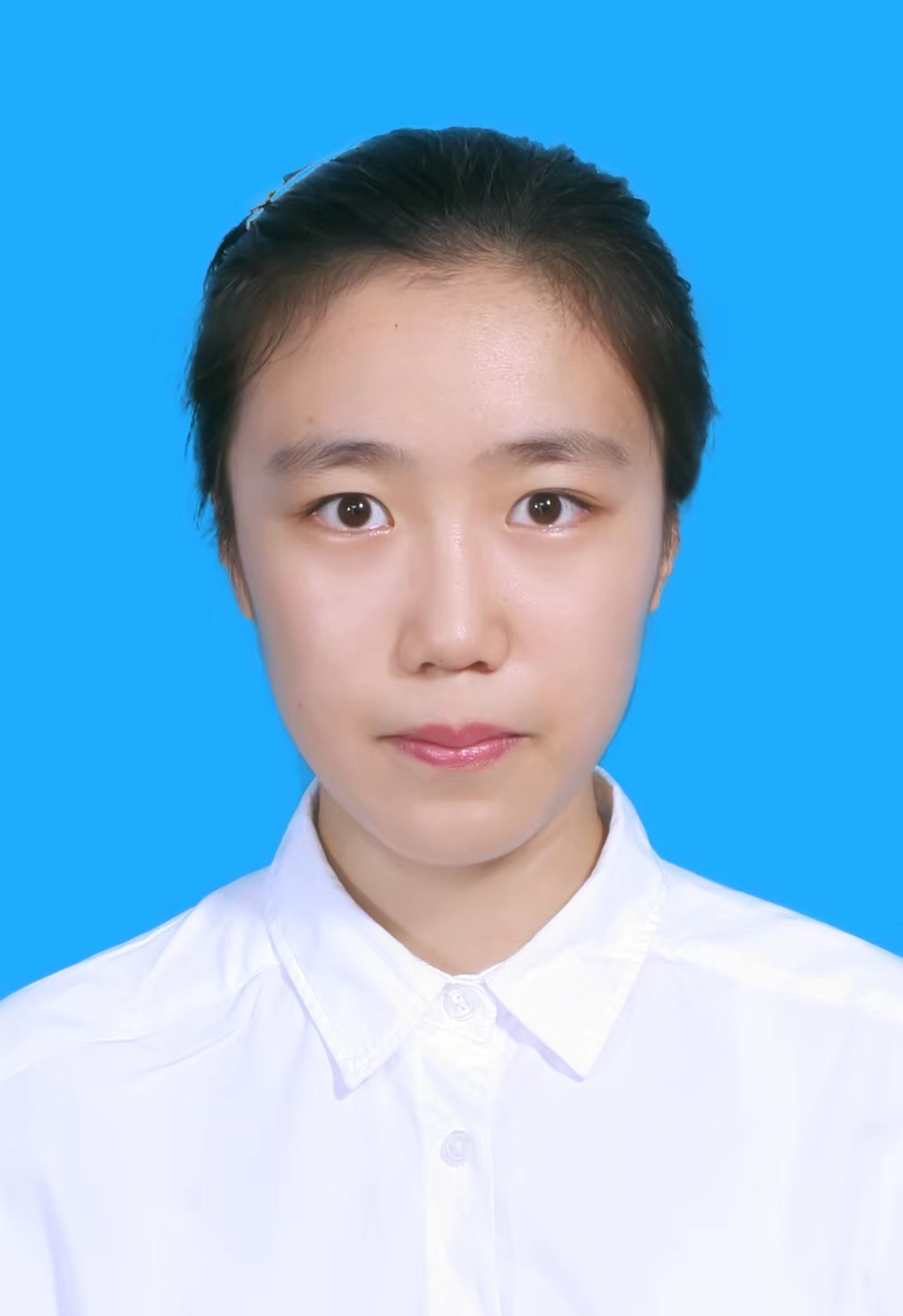}}] {Zixuan Pang} is currently pursuing her Master's degree at the University of Science and Technology of China. Her research interests focus on system security and confidential computing.

\end{IEEEbiography}

\begin{IEEEbiography}[{\includegraphics[width=1in,height=1.25in,clip]
{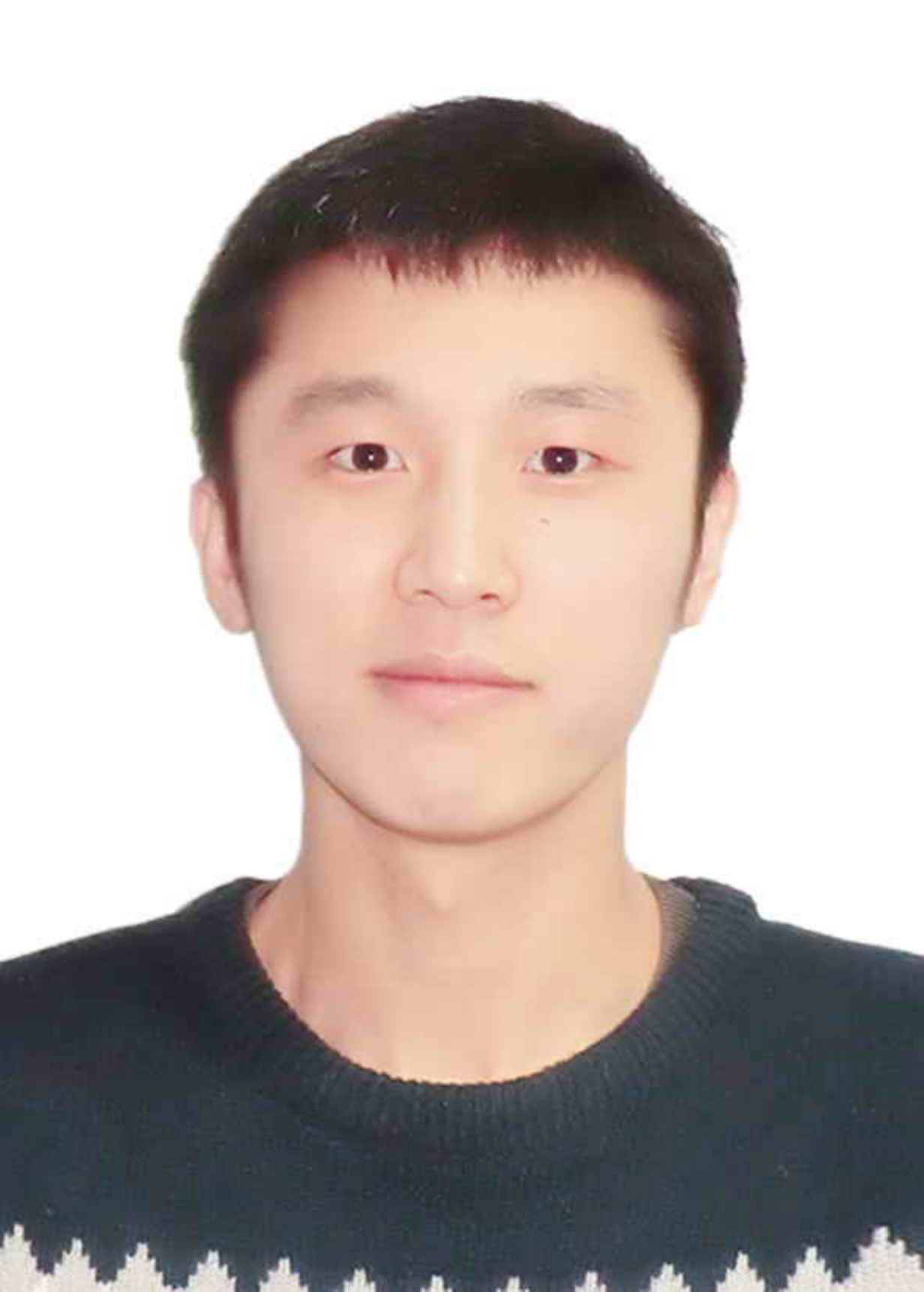}}] {Wenhao Wang} received the B.S. degree from Ocean University of China in 2009, and the Ph.D. degree from University of Chinese Academy of Sciences in 2015. He is an Associate Professor in Institute of Information Engineering, Chinese Academy of Sciences. His research interests include system security, confidential computing and side channels.
\end{IEEEbiography}

\begin{IEEEbiography}[{\includegraphics[width=1in,height=1.25in,clip]{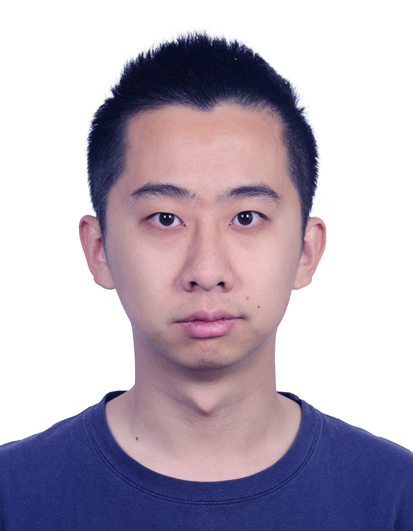}}] {Zihao Wang} received his Ph.D. degree from Indiana University Bloomington in 2025. He is currently a Research Fellow at Nanyang Technological University. His research interests are in trustworthy machine learning, including safety, security, and privacy.
\end{IEEEbiography}

\begin{IEEEbiography}[{\includegraphics[width=1in,height=1.25in,clip]{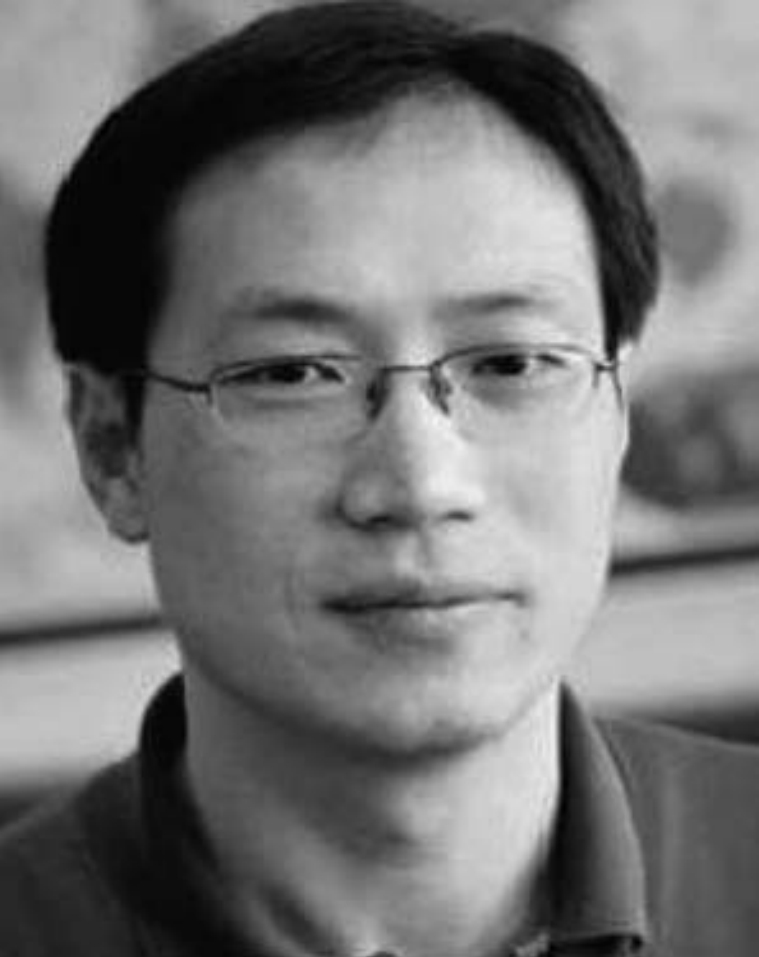}}] {XiaoFeng Wang} received the PhD degree in computer engineering from Carnegie Mellon University in 2004. His research interests include cloud and mobile security, and data and health informatics security. Dr. Wang is currently the Chair of ACM Special Interest Group on Security, Audit and Control (SIGSAC), and was also TPC Co-Chair of the ACM Conference on Computer and Communications Security (CCS), the ACM’s flagship security and privacy conference, during 2018 and 2019. He is a Fellow of the ACM, IEEE and AAAS.
\end{IEEEbiography}

\begin{IEEEbiography}[{\includegraphics[width=1in,height=1.25in,clip]{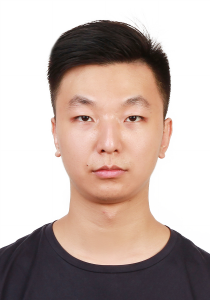}}] {Hongbo Chen} is a Ph.D. candidate at Indiana University Bloomington. He received his bachelor’s degree from Xi’an Jiaotong University in 2018. He mainly focuses
on system security research, especially on AI for security and confidential computing.
\end{IEEEbiography}

\begin{IEEEbiography}[{\includegraphics[width=1in,height=1.25in,clip]{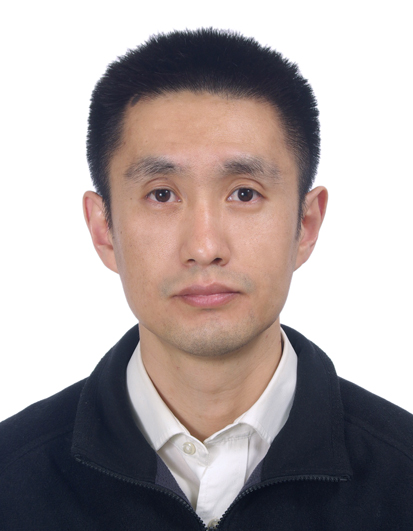}}] {Wei Song} received the Ph.D. degree in computer science from the  University of Manchester, Manchester, U.K., in 2011. He is currently an  Associate Professor with the Institute of Information Engineering, CAS.
His current research focuses on the security enhancement of computer architectures, such as the defenses for cache side channel and control-flow hijacking attacks.
\end{IEEEbiography}

\begin{IEEEbiography}[{\includegraphics[width=1in,height=1.25in,clip]{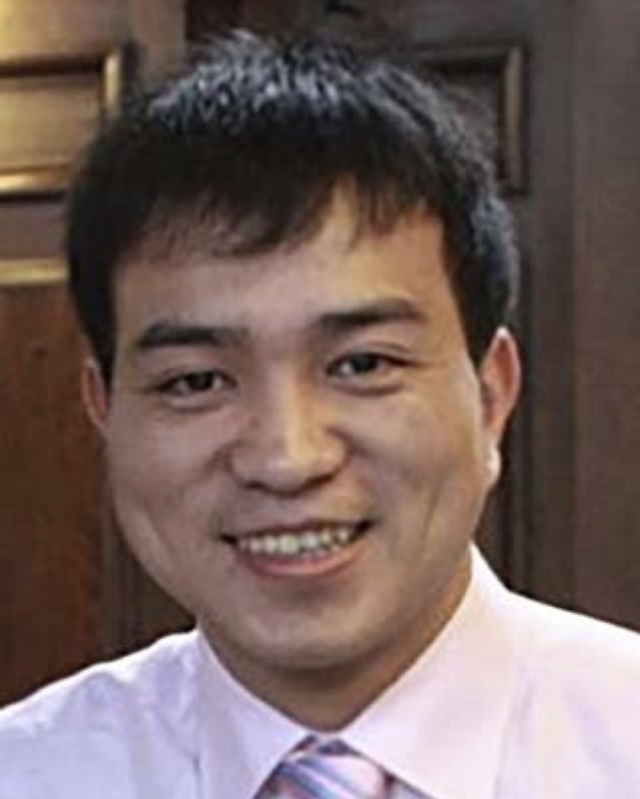}}] {Yier Jin} received the Ph.D. degree in electrical engineering from Yale University, in 2012. Currently, he is a Professor with the University of Science and Technology of China and an Adjacent Professor with the University of Florida. His research interests include hardware security, embedded systems design
and security, trusted hardware intellectual property (IP) cores, hardware software co-design for modern computing systems, security analysis on the Internet
of Things (IoT), and wearable devices with particular emphasis on information integrity and privacy protection in the IoT era. Dr. Jin was a recipient of the DoE Early CAREER Award in 2016 and the ONR Young Investigator Award in 2019.
\end{IEEEbiography}

\begin{IEEEbiography}[{\includegraphics[width=1in,height=1.25in,clip]{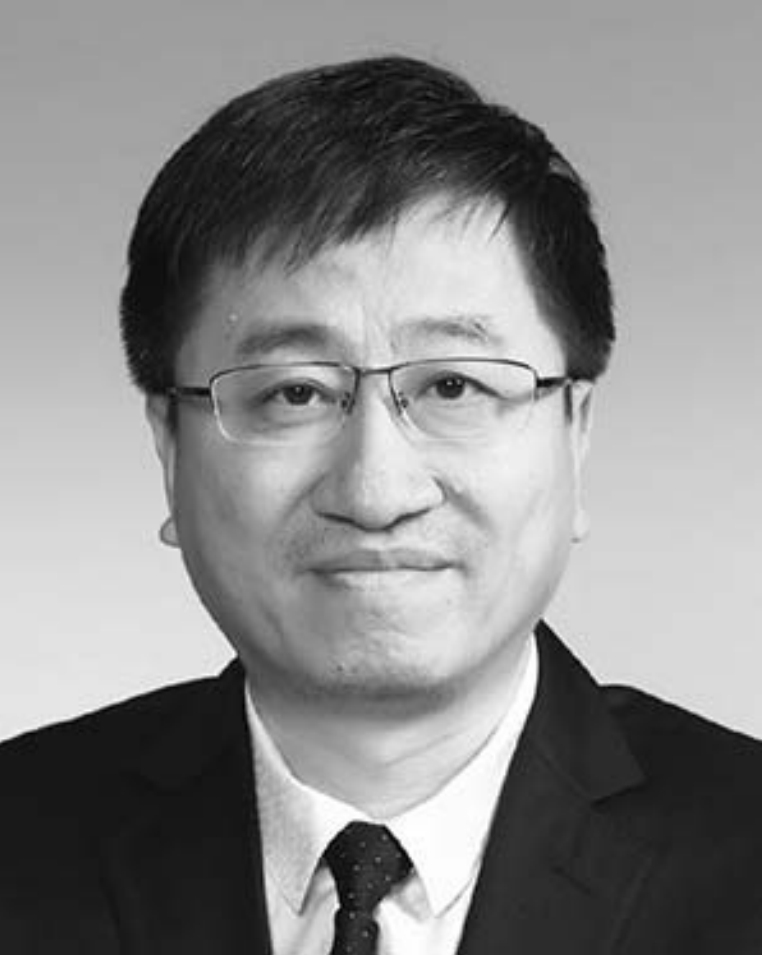}}] {Dan Meng} received the BS, MS, and PhD degrees from the Harbin Institute of Technology, Harbin, Heilongjiang, China. His research interests include high performance computer architecture and cyber security.
\end{IEEEbiography}

\begin{IEEEbiography}[{\includegraphics[width=1in,height=1.25in,clip]{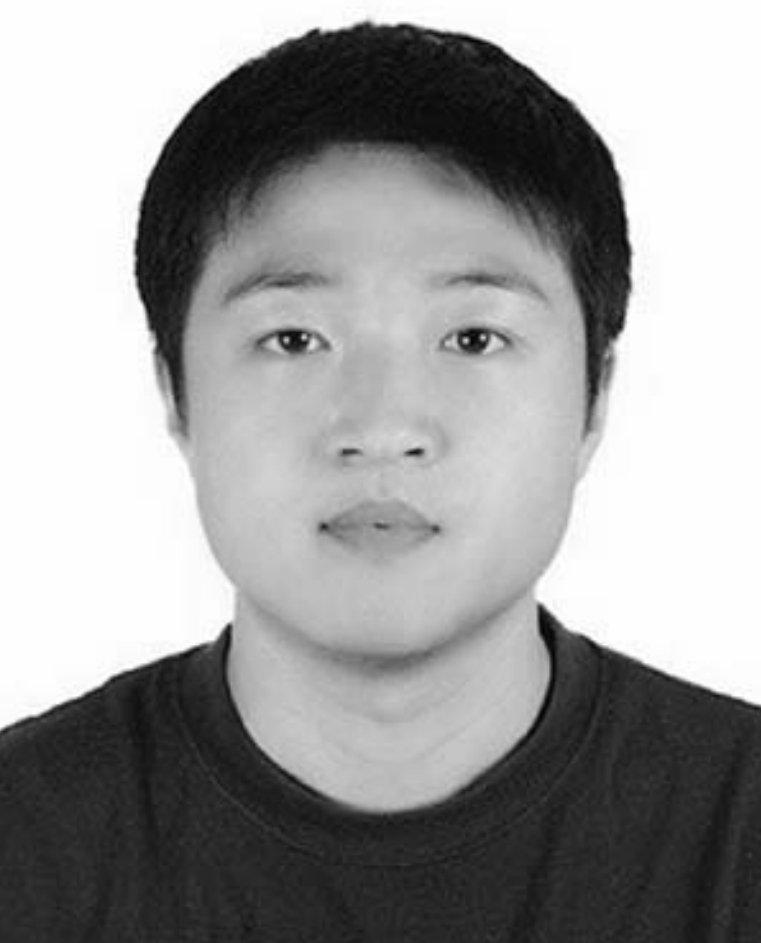}}] {Rui Hou} received the BS and MS degrees in computer architecture from the Harbin Institute of Technology, China, in 2001 and in 2003 respectively, and the PhD degree in computer architecture from the Institute of Computing Technology (ICT), Chinese Academy of Sciences (CAS), Beijing, China, in 2007. He is a professor and director with the State Key Laboratory of of Cyberspace Security Defense, Institute of Information Engineering, Chinese Academy of Sciences. He published more than 40 papers in international conferences and journals, and got more than 50 patents. His current research interests include computer architecture, processor security, data center server architecture and AI security.
\end{IEEEbiography}

 





\end{document}